\documentclass[notitlepage,nofootinbib,
superscriptaddress,showpacs,preprintnumbers,amsmath,amssymb,prd,onecolumn,12 point]{revtex4-2}

\usepackage{graphicx}
\usepackage{dcolumn}
\usepackage{bm}
\usepackage{hyperref}
\usepackage[mathlines]{lineno}
\usepackage{float}
\usepackage[T1]{fontenc}
\usepackage{amsmath}
\usepackage{subfigure}
\DeclareUnicodeCharacter{2212}{\ensuremath{-}}
 \DeclareUnicodeCharacter{202F}{FIX ME!!!!}
   \DeclareUnicodeCharacter{2212}{-}

\begin{document}


\title{Unification and Texture Universality: The Essence of Hermiticity}

\author{Pralay Chakraborty}%
 \email{pralay@gauhati.ac.in}
\affiliation{Department of Physics, Gauhati University, Guwahati, Assam -781014, India}%

\author{Subhankar Roy}
\email{subhankar@gauhati.ac.in}
 \affiliation{Department of Physics, Gauhati University, Guwahati, Assam -781014, India}



\date{\today}

\begin{abstract}

A unified framework sheltering the type-I Dirac seesaw, based on the $\Delta(27)$ group supplemented by other cyclic symmetries is proposed, preserving the naturalness of Yukawa couplings. It helps to visualize mass matrices within the lepton and quark sectors in terms of \textit{three universal parameters}: $\Sigma_1$, $\Sigma_2$, and $\Sigma_3$, highlighting that the down-type quark and light neutrino mass matrices exhibit a \textit{Hermitian} texture. Several phenomenological aspects, such as lepton flavour violation and nonunitarity of the lepton mixing matrix are explored. The stability of the proposed texture under renormalization group evolution is also investigated.

\end{abstract}

\maketitle


\newcolumntype{P}[1]{>{\centering\arraybackslash}p{#1}}

\section{Introduction}
\label{introduction}

The quark and lepton sectors exhibit significant differences rather than being similar. We see that the top quark mass lies around $173$ GeV. In contrast, the masses of the three neutrino mass eigenstates are presumed to lie within the eV scale. In addition, there is a significant difference between the mixing patterns of quarks and leptons. Specifically, in the quark sector, the mixing matrix, known as the Cabibbo-Kobayashi-Maskawa (CKM) matrix: $V_{CKM}$\,\cite{Cabibbo:1963yz, Kobayashi:1973fv}, is dictated mostly by the Cabibbo angle, $\theta_C\sim 13^\circ$\,\cite{ParticleDataGroup:2024cfk}, whereas, within the lepton sector, the smallest mixing angle, which resides in the Pontecorvo-Maki-Nakagawa-Sakata (PMNS) matrix: $U_{PMNS}$\,\cite{Maki:1962mu}, is the reactor angle, $\theta_{13}\sim 9^\circ$\,\cite{nufit}. In other words, $V_{CKM}\sim I$, whereas $U_{PMNS}$ deviates significantly from the identity matrix.

In spite of these prominent differences, the unified theories always leave scope to treat both lepton and quark sectors on an equal footing\,\cite{Georgi:1974sy, Carena:1994bv, Ananthanarayan:1994qt, Leontaris:1987nq}. The experimental observation that $\theta_{13}\sim \mathcal{O}(\theta_C)$ is interesting. Either this happens to be a simple a numerical coincidence, or may resemble a strong signature of unification. Both bottom-up and top-down studies are found in the literature in this context\,\cite{Roy:2020vtm, Roy:2014nua, Roy:2015cza, Ding:2012wh, Boucenna:2012xb, Branco:2014zza, Pati:1974yy,Elias:1975kf,Elias:1977bv, Blazek:2003wz,Dent:2007eu, Antusch:2009gu,Marzocca:2011dh,Antusch:2012fb,Antusch:2013ti,Antusch:2013rxa,Antusch:2013jca,Charles:2004jd}. Several unification schemes have been studied starting from discrete flavour symmetries\,\cite{Barranco:2010we, GonzalezCanales:2009zz, deMedeirosVarzielas:2017sdv, Koide:2003rx, King:2006np, King:2013hoa, Vien:2023xyq, Vien:2023zid, Vien:2022sxh, Vien:2021eog, Hernandez:2021mxo, Vien:2021xfp, Vien:2021ciw, Vien:2020uzf, Vien:2019eju, Vien:2019zhs, Vien:2019lso, Vien:2016qbb, Gupta:2015iku, Gupta:2011zzg, Randhawa:1999hi, Gupta:2012fsl, Berezhiani:2024fsw, Garces:2018nar}. In Refs.\,\cite{King:2013hj, Chen:2023mwt, Bonilla:2014xla, Morisi:2011pt}, how the masses of charged leptons and quarks can be correlated is exhibited. The above mentioned works either adopt Fritzsch-like or general textures within the quark sector. Nevertheless, they always stick to Majorana nature of neutrino.

In the present work, however, the motivation differs slightly from the existing ones. In general, when unified frameworks are studied, the parameter $\theta_C$ or the Wolfenstein parameter, $\lambda=\sin\theta_{C}$\,\cite{Wolfenstein:1983yz}, plays a leading role. We adhere to a distinct and independent strategy, where we try to visualize the quark and lepton mass matrices in terms of three unification parameters: $\Sigma_1$, $\Sigma_2$, and $\Sigma_3$.
In our approach, we emphasize the naturalness of the Yukawa sector by adhering to the choice: $y\sim\mathcal{O}(1)$.

The unification scheme that assumes neutrinos to be purely Dirac type is not very common. The present work is characterized by this feature. We adopt the type-I Dirac seesaw mechanism\,\cite{CentellesChulia:2017koy,  Borboruah:2024lli, Borah:2024gql, Borah:2017dmk, Singh:2024imk, Mahapatra:2023oyh, Chen:2022bjb, Goswami:2025jde} to explain the origin of neutrino masses and mixing. The mass matrices are more fundamental than the mixing matrices. Respecting this key concept, the present work explores the realization of mass matrices within both of these alienated sectors and investigates whether the patterns or textures are somehow similar. We find that a general three-dimensional Hermitian mass matrix texture plays a very important role in achieving universality.

The plan of the paper is outlined as follows: In section (\ref{sec2}), we discuss an effective unified theory to derive the quark and lepton mass matrix textures. In this regard, we discuss the phenomenological viability of the obtained textures and ensure the naturalness of the Yukawa couplings. In section (\ref{sec6}), we study some phenomenological constrains of the proposed model. In section (\ref{sec7}), we study the stability of the textures under renormalization group running. In section (\ref{sec8}), we discuss the salient facets of the study and conclude.

\section{The Model \label{sec2}}

 We start with the $\Delta(27)$ symmetry group\,\cite{Vien:2016tmh, CarcamoHernandez:2016piw, Vien:2022sxh,Vien:2020hzy, CarcamoHernandez:2018iel, Branco:1983tn, deMedeirosVarzielas:2006fc, Ma:2007wu, deMedeirosVarzielas:2012ylr, Bhattacharyya:2012pi, Ferreira:2012ri, Ma:2013xqa, Nishi:2013jqa, deMedeirosVarzielas:2013xas, Harrison:2014jqa, CarcamoHernandez:2018djj, Bjorkeroth:2019csz, Bazzocchi:2009qg, Varzielas:2012nn, Aranda:2013gga, Abbas:2014ewa, Abbas:2015zna, Bjorkeroth:2015uou}. While the quark masses are explained by the Higgs mechanism\,\cite{Higgs:1964pj, Englert:1964et, Guralnik:1964eu}, in the neutrino sector, we incorporate the type-I Dirac seesaw mechanism. In this regard, we extend the field content of the Standard Model\,\cite{Glashow:1961tr, Weinberg:1967tq, Salam:1968rm} by introducing right-handed neutrinos and additional scalar fields. Furthermore, we add two fermion triplets, $N_L$ and $N_R$. The transformation properties of the extended field content are summarised in Table\,(\ref{rotated-field-table}).

The $SU(2)_L \times  \Delta (27) \times Z_{7} \times Z_{5} \times  Z_4 \times Z_3$ invariant Lagrangian is constructed in the following way,

\begin{align}
- \mathcal{L}_Y &=  \mathcal{L}_{\text{quark}}+\mathcal{L}_{\text{lepton}},
\end{align}

where,

\begin{align}
\mathcal{L}_{\text{quark}} &= \frac{y_{u}}{\Lambda^7}(\overline{Q}_{l_{L}}\psi)_{1_{00}}\tilde{H} u_{R_{1_{00}}} \xi^6  + \frac{y_{c}}{\Lambda^4}(\overline{Q}_{l_{L}}\psi)_{1_{20}}\tilde{H} c_{R_{1_{10}}} \xi^3 + \frac{y_{t}}{\Lambda^2}(\overline{Q}_{l_{L}}\psi_{1_{10}})\tilde{H}t_{R_{1_{20}}} \xi +\nonumber\\
& \frac{x_1^d }{\Lambda}(\overline{Q}_{l_{L}}\eta)_{3^*_{S_1}}H d_{R_3}+\frac{x_2^d }{\Lambda}(\overline{Q}_{l_{L}}\eta)_{3^*_{S_2}}H d_{R_3}+\frac{\tilde{x_3}^d }{\Lambda}(\overline{Q}_{l_{L}}\eta)_{3^*_{A}}H d_{R_3}+\frac{y_1^d }{\Lambda^2}(\overline{Q}_{l_{L}}\chi)_{3^*_{S_1}}\nonumber\\
&H \xi d_{R_3}+\frac{y_2^d }{\Lambda^2}(\overline{Q}_{l_{L}}\chi)_{3^*_{S_2}}H \xi d_{R_3}+\frac{\tilde{y_3}^d }{\Lambda^2}(\overline{Q}_{l_{L}}\chi)_{3^*_{A}}H \xi d_{R_3}+\frac{z_1^d }{\Lambda^2}(\overline{Q}_{l_{L}}\kappa)_{3^*_{S_1}}H \sigma  d_{R_3}+\nonumber\\
&\frac{z_2^d }{\Lambda^2}(\overline{Q}_{l_{L}}\kappa)_{3^*_{S_2}}H \sigma d_{R_3}+\frac{\tilde{z_3}^d }{\Lambda^2}(\overline{Q}_{l_{L}}\kappa)_{3^*_{A}}H \sigma d_{R_3},
\end{align}

and

\begin{align}
\mathcal{L}_{\text{lepton}} &= \frac{y_{e}}{\Lambda^{10}} (\overline{D}_{l_{L}}\psi)_{1_{00}}H e_{R_{1_{00}}} \xi^9+ \frac{y_{\mu}}{\Lambda^7}(\overline{D}_{l_{L}}\psi)_{1_{20}}H\mu_{R_{1_{10}}} \xi^6 + \frac{y_{\tau}}{\Lambda^4}(\overline{D}_{l_{L}}\psi)_{1_{10}}H\tau_{R_{1_{20}}} \xi^3 \nonumber\\
&+\frac{x_1^\nu }{\Lambda}(\overline{D}_{l_{L}}\eta)_{3^*_{S_1}}\tilde{H} N_{R_{3}}+\frac{x_2^\nu }{\Lambda}(\overline{D}_{l_{L}}\eta)_{3^*_{S_2}}\tilde{H} N_{R_3}+\frac{\tilde{x_3}^\nu }{\Lambda}(\overline{D}_{l_{L}}\eta)_{3^*_{A}}\tilde{H} N_{R_3}+\frac{y_1^\nu }{\Lambda^2}(\overline{D}_{l_{L}}\nonumber\\
&\chi)_{3^*_{S_1}}\tilde{H} \xi N_{R_3}+\frac{y_2^\nu }{\Lambda^2}(\overline{D}_{l_{L}}\chi)_{3^*_{S_2}}\tilde{H} \xi N_{R_3}+ \frac{\tilde{y_3}^\nu }{\Lambda^2}(\overline{D}_{l_{L}}\chi)_{3^*_{A}}\tilde{H} \xi N_{R_3}+\frac{z_1^\nu }{\Lambda^2}(\overline{D}_{l_{L}}\kappa)_{3^*_{S_1}}\tilde{H}\nonumber\\
& \sigma N_{R_3}+\frac{z_2^\nu }{\Lambda^2}(\overline{D}_{l_{L}}\kappa)_{3^*_{S_2}}\tilde{H} \sigma N_{R_3}+\frac{\tilde{z_3}^\nu }{\Lambda^2}(\overline{D}_{l_{L}}\kappa)_{3^*_{A}}\tilde{H} \sigma N_{R_3}+ \frac{y_{p_1}}{\Lambda}  (\overline{N}_L \kappa)_{3^*_{S_1}}\rho  \nu_{R_3}+\nonumber\\
&\frac{y_{p_2}}{\Lambda}  (\overline{N}_L \kappa)_{3^*_{S_2}}\rho \nu_{R_3}+\frac{y_{p_3}}{\Lambda}(\overline{N}_L\kappa )_{3^*_{A}}\rho \nu_{R_3}+  \frac{y_{s_1}}{\Lambda}  (\overline{N}_L \phi)_{3^*_{S_1}} \sigma  N_{R_3}+\frac{y_{s_2}}{\Lambda}  (\overline{N}_L \phi)_{3^*_{S_2}}\nonumber\\
&\sigma  N_{R_3}+\frac{y_{s_3}}{\Lambda}  (\overline{N}_L \phi)_{3^*_{A}} \sigma  N_{R_3}+h.c.
\label{chap5:Yukawa Lagrangian M1}
\end{align}
where, $\Lambda$ is the cut-off scale of the theory. The auxiliary groups $Z_5$, $Z_4$ and $Z_3$ are introduced to eliminate certain undesired terms that are allowed by $\Delta(27)$. The group $Z_7$ is introduced to forbid all possible Majorana terms. The terms restricted by the cyclic symmetries are given in \ref{appendix c}. The product rules under the $\Delta(27)$ group are given in \ref{appendix a}. To obtain the desired texture, we assume the following: $\tilde{x_3}^d=i\,x_3^d$, $\tilde{y_3}^d=i\,y_3^d$, $\tilde{z_3}^d=i\,z_3^d$, $\tilde{x_3}^\nu=i\,x_3^\nu$, $\tilde{y_3}^\nu=i\,y_3^\nu$ and $\tilde{z_3}^\nu=i\,z_3^\nu$.

 \begin{table}
\centering
\begin{tabular*}{\textwidth}{@{\extracolsep{\fill}} ccccccc}
\hline\hline
\textbf{Fields} & $SU(2)_L$ & $\Delta(27)$ & $Z_7$ & $Z_5$ & $Z_4$ & $Z_3$ \\
\hline\hline
$\overline{D}_{l_{L}}$ & 2 & 3 & $4$ & 0 & 1 & 1 \\
$\overline{Q}_{l_{L}}$ & 2 & 3 & $3$ & 0 & 1 & 1 \\
$(e_R, \mu_R, \tau_R)$ & 1 & $(1_{00},1_{10},1_{20})$ & $(3,3,3)$ & (1, 4, 2) & 1 & (1, 1, $\omega^2$) \\
$(u_R, c_R, t_R)$ & 1 & $(1_{00},1_{10},1_{20})$ & $(4,4,4)$ & (4, 2, 4) & 1 & 1 \\
$d_R$ & 1 & 3 & $4$ & 0 & 1 & 1 \\
$H$ & 2 & $1_{00}$ & $0$ & 0 & 1 & 1 \\
$\nu_R$ & 1 & 3 & $3$ & 1 & $i$ & 1 \\
$\overline{N}_L$ & 1 & 3 & $4$ & 3 & 1 & $\omega$ \\
$N_R$ & 1 & 3 & $3$ & 0 & 1 & 1 \\
$\psi$ & 1 & $3^*$ & $0$ & 0 & 1 & 1 \\
$\eta$ & 1 & $3$ & $0$ & 0 & 1 & 1 \\
$\chi$ & 1 & 3 & $0$ & 4 & 1 & $\omega^2$ \\
$\kappa$ & 1 & 3 & $0$ & 0 & $-i$ & $\omega^2$ \\
$\sigma$ & 1 & $1_{00}$ & $0$ & 0 & $i$ & $\omega$ \\
$\xi$ & 1 & $1_{00}$ & $0$ & 1 & 1 & $\omega$ \\
$\phi$ & 1 & 3 & $0$ & 2 & $-i$ & $\omega$ \\
$\rho$ & 1 & $1_{00}$ & $0$ & 1 & 1 & 1 \\
\hline
\end{tabular*}
\caption{Transformation properties of fields under $SU(2)_L \times \Delta(27) \times Z_7 \times Z_5 \times Z_4 \times Z_3$.}
\label{rotated-field-table}
\end{table}

To obtain the proposed texture, the vacuum expectation values\,(vev) for the scalar fields are chosen as: $\langle H \rangle_{0}=v_{H}$, $\langle \psi \rangle_{0}=v_\psi(1,1,1)^{T}$, $\langle \eta \rangle_{0}=v_{\eta}(0,1,0)^{T}$, $\langle \chi \rangle_{0}=v_{\chi}(1,0,0)^{T}$, $\langle \kappa \rangle_{0}=v_{\kappa}(0,0,1)^{T}$, $\langle \phi \rangle_{0}=v_{\phi}(0,0,1)^{T}$, $\langle\sigma\rangle_{0}=v_{\sigma}$, $\langle\rho\rangle_{0}=v_{\rho}$, and $\langle\xi\rangle_{0}=v_{\xi}$, respectively. The $SU(2)_L \times \times \Delta (27) \times Z_{7} \times Z_{5} \times Z_4 \times Z_3$ invariant scalar potential along with the optimization conditions is discussed in \ref{appendix b}.

We obtain similar textures for charged lepton $(M_l)$ and up-type $(M_u)$ quarks as shown below,

\begin{align}
		M_{l}&=v_{H} v_\psi \begin{bmatrix}
			\frac{y_{e} v_\xi^9}{\Lambda^{10}} & \frac{ y_{\mu} v_\xi^6}{\Lambda^7} & \frac{y_{\tau} v_\xi^3}{\Lambda^4} \\
			\frac{y_{e} v_\xi^9}{\Lambda^{10}} & \omega\,\frac{ y_{\mu} v_\xi^6}{\Lambda^7} & \omega^2\, \frac{y_{\tau} v_\xi^3}{\Lambda^4} \\
			\frac{y_{e} v_\xi^9}{\Lambda^{10}} & \omega^2\, \frac{ y_{\mu} v_\xi^6}{\Lambda^7} & \omega \,\frac{y_{\tau} v_\xi^3}{\Lambda^4} \\
		\end{bmatrix} \quad \text{and} \quad M_u=v_{H} v_\psi \begin{bmatrix}
			\frac{y_{u} v_\xi^6}{\Lambda^{7}} & \frac{ y_{c} v_\xi^3}{\Lambda^4} & \frac{y_{t} v_\xi}{\Lambda^2} \\
			\frac{y_{u} v_\xi^6}{\Lambda^{7}} & \omega\,\frac{ y_{c} v_\xi^3}{\Lambda^4} & \omega^2\, \frac{y_{t} v_\xi}{\Lambda^2} \\
			\frac{y_{u} v_\xi^6}{\Lambda^{7}} & \omega^2\, \frac{ y_{c} v_\xi^3}{\Lambda^4} & \omega \,\frac{y_{t} v_\xi}{\Lambda^2} \\
		\end{bmatrix},
	\end{align}

where, $\omega= e^{2 \pi i/3}$. We diagonalize $M_l$ and $M_u$ as $M_{l}^{diag}=U_{l_{L}}^{\dagger}M_{l}U_{l_R}$ and $M_{u}^{diag}=V_{u_{L}}^{\dagger}M_{u}V_{u_R}$ respectively. For our model, the left-handed diagonalising matrices $U_{l_{L}}$ and $V_{l_{L}}$ are exactly equal and is expressed as shown below,
	\begin{equation}
		U_{l_{L}}=V_{u_{L}}=\frac{1}{\sqrt{3}} \begin{bmatrix}
			1 &  1 & 1\\
			1 & \omega & \omega^2\\
		1   & \omega^2 & \omega\\
		\end{bmatrix}.
		\label{UL}
	\end{equation}

The right-handed diagonalising matrices $U_{l_R}$ and $V_{u_R}$ are simply $3\times 3$ identity matrices.

The diagonalized charged leptons and up-type quarks mass matrices $M_l^{\text{diag}}$ and $M_u^{\text{diag}}$ takes the following form:

 \begin{align}
		M_{l}^{\text{diag}}&= \sqrt{3} v_{H} v_\psi \begin{bmatrix}
			 \frac{y_{e} v_\xi^9}{\Lambda^{10}} & 0 & 0 \\
			0 &   \frac{ y_{\mu} v_\xi^6}{\Lambda^7} & 0 \\
			0 & 0 &   \frac{y_{\tau} v_\xi^3}{\Lambda^4}\\
		\end{bmatrix} \quad \text{and} \quad M_u^{\text{diag}}= \sqrt{3} v_{H} v_\psi \begin{bmatrix}
			 \frac{y_{u} v_\xi^6}{\Lambda^{7}} & 0 & 0 \\
			0 &   \frac{ y_{c} v_\xi^3}{\Lambda^4} & 0 \\
			0 & 0 &   \frac{y_{t} v_\xi}{\Lambda^2}\\
		\end{bmatrix}.
	\end{align}
	
It is important to mention that the up-type quarks and charged lepton mass hierarchies can be explained through the Froggatt–Nielsen (FN) mechanism\,\cite{Froggatt:1978nt}, in which the quarks and fermion mass hierarchies emerge from higher-dimensional operators suppressed by higher powers of cut off  scale $\Lambda$. In the present work, we find that the  charged lepton and up-type quark masses satisfy the following relation:
	
\begin{eqnarray}
m_e : m_\mu : m_\tau &=& \Omega^6 y_e : \Omega^3 y_\mu : y_\tau,\nonumber \quad \text{and} \quad m_u : m_c : m_t = \Omega^5 y_u : \Omega^2 y_c : y_t,
\end{eqnarray}

where, $\Omega=\frac{v_\xi}{\Lambda}$.

The above relations hold good for the observed mass hierarchies of charged leptons and up-type quarks: 

\begin{align}
m_e : m_\mu : m_\tau &\approx 0.00028 : 0.0594 : 1,\nonumber \quad \text{and} \quad m_u : m_c : m_t \approx 0.00112 : 0.022 : 1,
\end{align}

provided $\Omega\approx0.35$, $y_e\approx0.15$, $y_\mu\approx0.14$, $y_u\approx0.21$, $y_c\approx0.18$ and $y_\tau, y_t\approx 1$.

We derive the down-type quarks mass matrix as shown in the following,

\begin{eqnarray}
M_d&=&\begin{bmatrix}
 2 y_1^d \Sigma_1 &  (z_2^d-iz_3^d) \Sigma_2  &  (x_2^d+i x_3^d) \Sigma_3\\
 (z_2^d+iz_3^d) \Sigma_2 & 2 x_1^d \Sigma_3 &  (y_2^d-i y_3^d)\Sigma_2 \\
  (x_2^d-i x_3^d) \Sigma_3 & (y_2^d+i y_3^d)\Sigma_2 & 2 z_1^d \Sigma_2 \\
\end{bmatrix},
\end{eqnarray}

where, $\Sigma_1=\frac{1}{2 \Lambda ^2} v_H v_\xi v_\chi$, $\Sigma_2=\frac{1}{2 \Lambda ^2}v_H v_\sigma v_\kappa$ and $\Sigma_3=\frac{1}{2 \Lambda}v_H v_\eta$.

We highlight that parameters $\Sigma_1$, $\Sigma_2$ and $\Sigma_3$  also appear in the neutrino sector. The neutrino mass matrix is obtained from the following,

\begin{equation}
M_\nu = B M^{-1} D,
\end{equation}

where, \begin{align}
B &= \begin{bmatrix}
\frac{ y_1^\nu}{\Lambda ^2} v_H v_\xi v_\chi & \frac{z_2^\nu}{2 \Lambda ^2}v_H v_\sigma v_\kappa-i \frac{z_3^\nu}{2 \Lambda ^2}v_H v_\sigma v_\kappa   &  \frac{x_2^\nu}{2 \Lambda}v_H v_\eta +i \frac{x_3^\nu}{2 \Lambda}v_H v_\eta \\
\frac{z_2^\nu}{2 \Lambda ^2}v_H v_\sigma v_\kappa+i \frac{z_3^\nu}{2 \Lambda ^2}v_H v_\sigma v_\kappa & \frac{x_1^\nu}{\Lambda}v_H v_\eta & \frac{y_2^\nu}{2 \Lambda^2}v_H v_\chi v_\xi -i \frac{y_3^\nu}{2 \Lambda^2}v_H v_\chi v_\xi &\\
\frac{x_2^\nu}{2 \Lambda}v_H v_\eta -i \frac{x_3^\nu}{2 \Lambda}v_H v_\eta & \frac{y_2^\nu}{2 \Lambda^2}v_H v_\chi v_\xi +i \frac{y_3^\nu}{2 \Lambda^2}v_H v_\chi v_\xi &  \frac{z_1^\nu}{ \Lambda^2}v_H v_\kappa v_\sigma\\
\end{bmatrix},\\
M&=\begin{bmatrix}
 0 & \frac{y_{s_2} v_\phi v_\sigma}{2\Lambda}-\frac{y_{s_3} v_\phi v_\sigma}{2\Lambda} & 0 \\
\frac{y_{s_2} v_\phi v_\sigma}{2\Lambda}+\frac{y_{s_3} v_\phi v_\sigma}{2\Lambda} & 0 & 0 \\
 0 & 0 &\frac{y_{s_1}}{\Lambda} v_\phi v_\sigma \\
\end{bmatrix},\\ D&=\begin{bmatrix}
 0 & \frac{y_{p_2} v_\kappa v_\rho}{2 \Lambda}-\frac{y_{p_3} v_\kappa v_\rho}{2 \Lambda} & 0 \\
 \frac{y_{p_2} v_\kappa v_\rho}{2 \Lambda}+\frac{y_{p_3} v_\kappa v_\rho}{2 \Lambda} & 0 & 0 \\
 0 & 0 &\frac{y_{p_1} v_\kappa v_\rho}{\Lambda} \\
\end{bmatrix}.
\label{BMD}
\end{align}

 By considering the Yukawa couplings $\sim \mathcal{O}(1)$, we take the following choices: $\frac{y_{p_2}}{y_{p_3}}\approx\frac{ y_{s_2}}{y_{s_3}}\,\,\text{and}\,\,\frac{y_{p_1}}{y_{p_3}}\approx\frac{y_{s_1}}{y_{s_3}}$. With this, we obtain the final neutrino mass matrix as shown in the following,

\begin{eqnarray}
M_\nu&=& \gamma^\nu  \begin{bmatrix}
 2 y_1^\nu \Sigma_1 &  (z_2^\nu-iz_3^\nu) \Sigma_2  &  (x_2^\nu+i x_3^\nu) \Sigma_3\\
 (z_2^\nu+iz_3^\nu) \Sigma_2 & 2 x_1^\nu \Sigma_3 &  (y_2^\nu-i y_3^\nu)\Sigma_2 \\
  (x_2^\nu-i x_3^\nu) \Sigma_3 & (y_2^\nu+i y_3^\nu)\Sigma_2 & 2 z_1^\nu \Sigma_2 \\
\end{bmatrix}
\end{eqnarray}

where, $\gamma^\nu=\frac{y_{p_3} v_\kappa v_\rho}{y_{s_3} v_\phi v_\sigma}$.

It is interesting to note that both quark and neutrino sectors are correlated through common parameters $\Sigma_1$, $\Sigma_2$ and $\Sigma_3$. The parameter $\gamma^\nu$ appearing in the neutrino sector dictates the smallness of neutrino masses as compared to quarks. We highlight that the proposed model manifests a promising relation: $U_{PMNS} = V_{uL}^{\dagger} U_\nu$.

The obtained texture for down type quark and neutrino is Hermitian takes the following form,

\begin{equation}
M_{d/\nu} = \begin{bmatrix}
a & r_1 e^{i \theta_1} & r_2 e^{i\theta_2}\\
r_1 e^{-i\theta_1} & b & r_3 e^{i\theta_3}\\
r_2 e^{-i\theta_2} & r_3 e^{-i\theta_3} & c\\
\end{bmatrix}.
\label{Texture}
\end{equation}

The mass matrices obtained from the symmetry framework carrying several parameters involving Yukawa couplings, vevs, and the cut-off scale, resemble the exact form of the Hermitian texture $M_{d/\nu}$ expressed in terms of $a, b, c, r_1, r_2, r_3, \theta_1, \theta_2 \, \text{and}\, \theta_3$ as presented in Eq.\,(\ref{Texture}). It is worth mentioning that we maintain the independence of the texture parameters while expressing them in terms of the model parameters. The given model contains several scalar fields, which play a important role in maintaining the independence of the texture parameters.

In the next section, we shall check the consistency of the textures with experimental data.

\subsection{Numerical Analysis \label{sec3}}

From the symmetry framework, we obtain universal textures for both the lepton and quark sectors. The charged leptons and up-type quarks are governed by a common texture $M_{u/l}$, while a \textit{Hermitian} texture $M_{d/\nu}$ governs the neutrinos and down-type quarks. The $M_u$ and $M_l$ are special complex matrices and we identify the related diagonalising matrices in Eq.\,(\ref{UL}). In this section, we examine whether the presently available experimental data from both the quark and neutrino sectors support the proposed Hermitian texture structure. The obtained matrices $M_d$ and $M_\nu$ being hermitian, require unitary matrices to get diagonalised. We see $M^{diag}_d = V_{d_L} M_d V_{d_L}^\dagger$; $M^{diag}_\nu = U_{\nu_{L}} M_\nu U_{\nu_{L}}^\dagger$. We finally identify CKM and PMNS matrices as : $V_{CKM}= V_{u_{L}}^{\dagger} V_{d_{L}}$ and $U_{PMNS}= U_{l_{L}}^{\dagger} U_{\nu_{L}}$. For further study, we stick to the standard parametrization of both the $V_{CKM}$ and $U_{PMNS}$ as per the Particle Data Group\,(PDG)\,\cite{ParticleDataGroup:2024cfk} displayed below,

\begin{figure}
  \centering
    \subfigure[]{\includegraphics[width=0.49\textwidth]{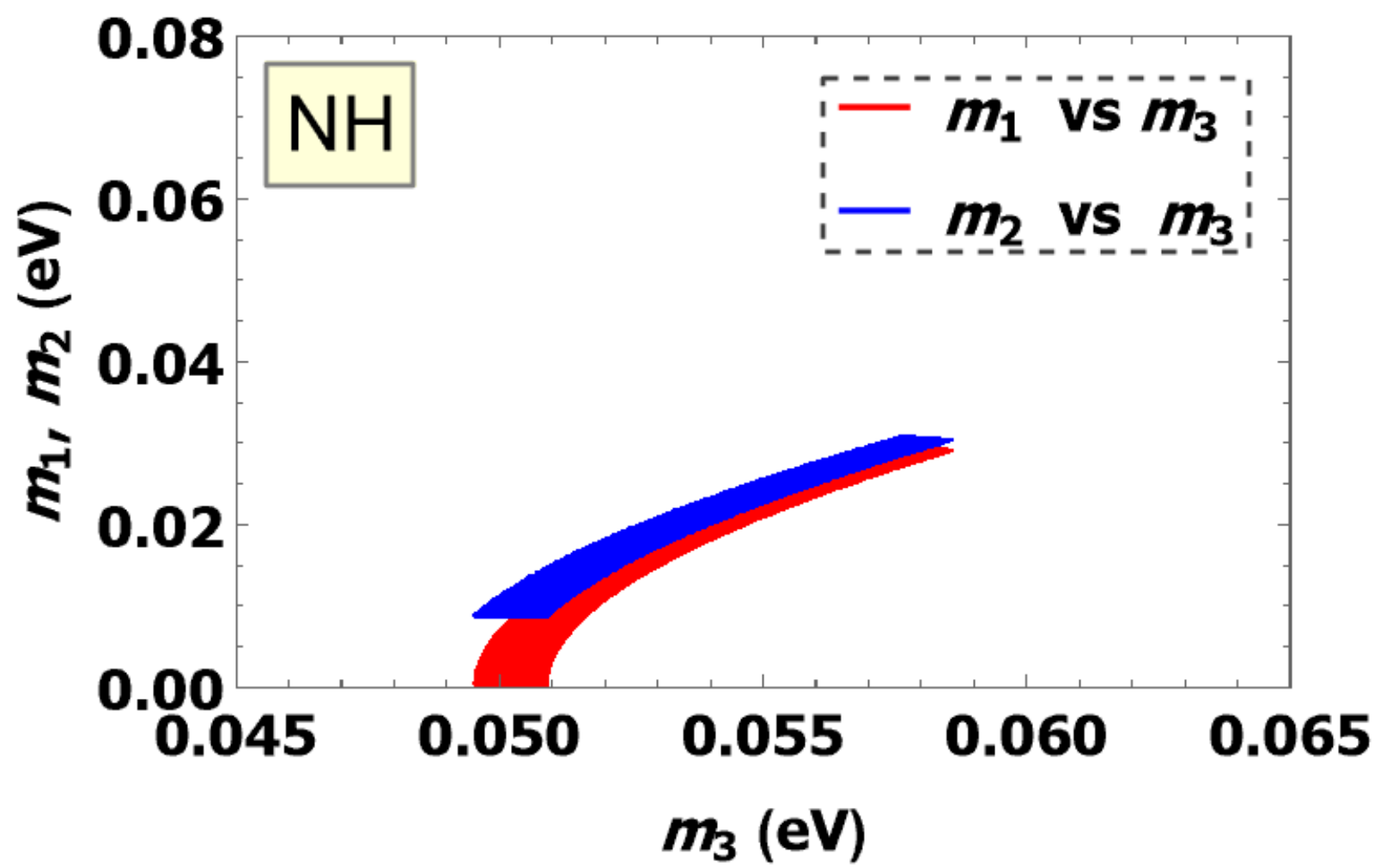}\label{fig:1a}} 
    \subfigure[]{\includegraphics[width=0.49\textwidth]{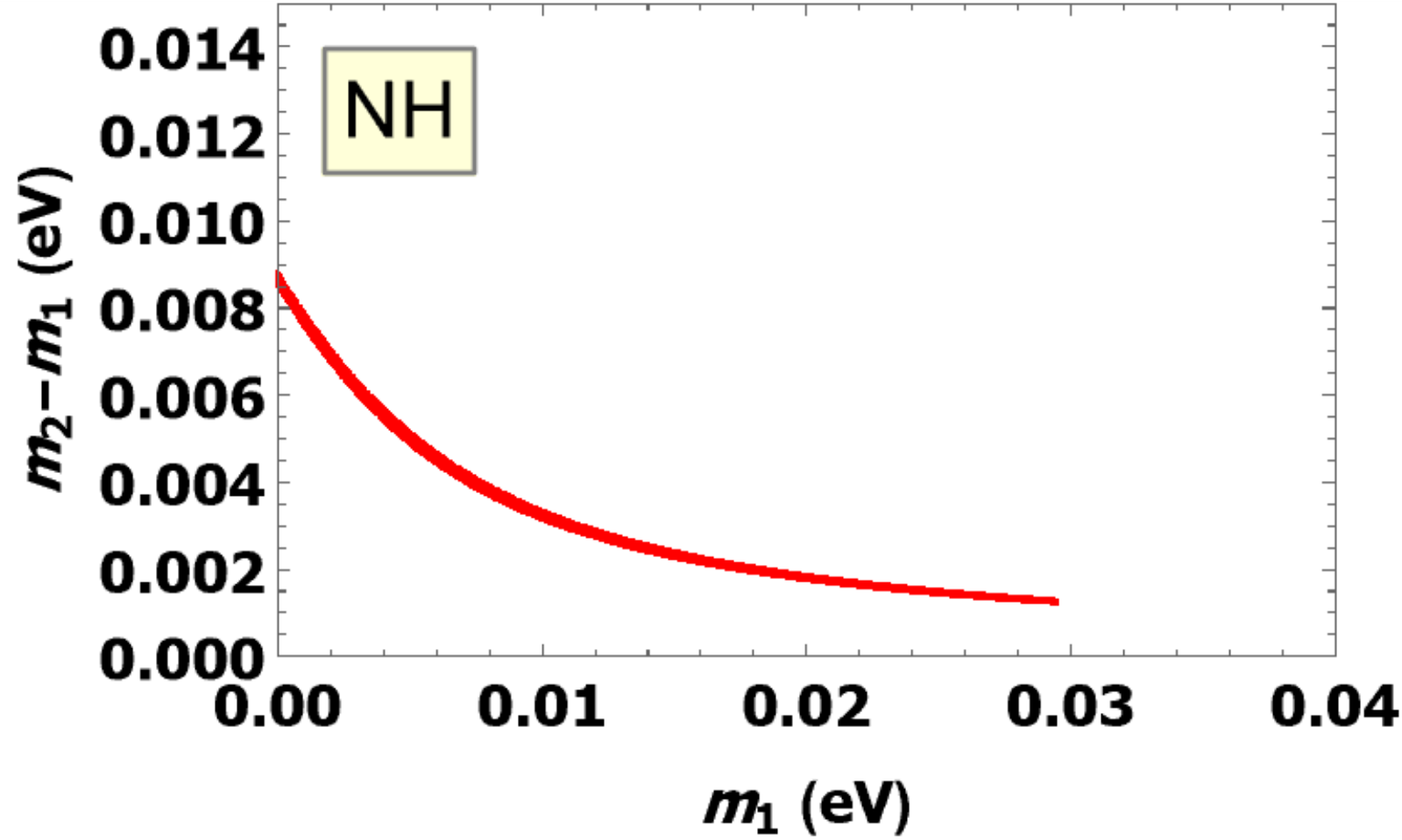}\label{fig:1b}}\\
    \subfigure[]{\includegraphics[width=0.49\textwidth]{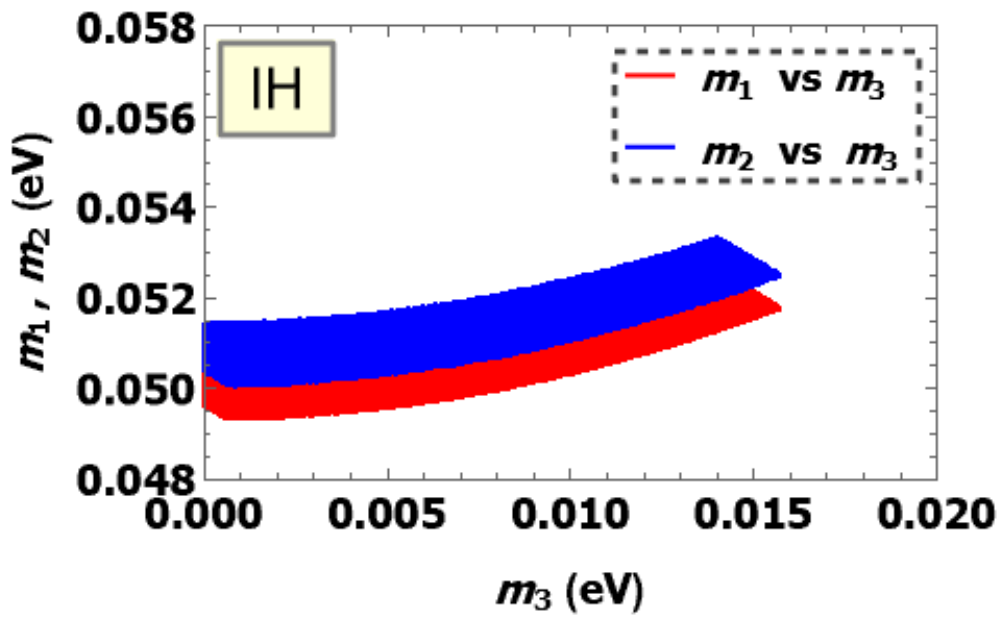}\label{fig:1c}} 
    \subfigure[]{\includegraphics[width=0.49\textwidth]{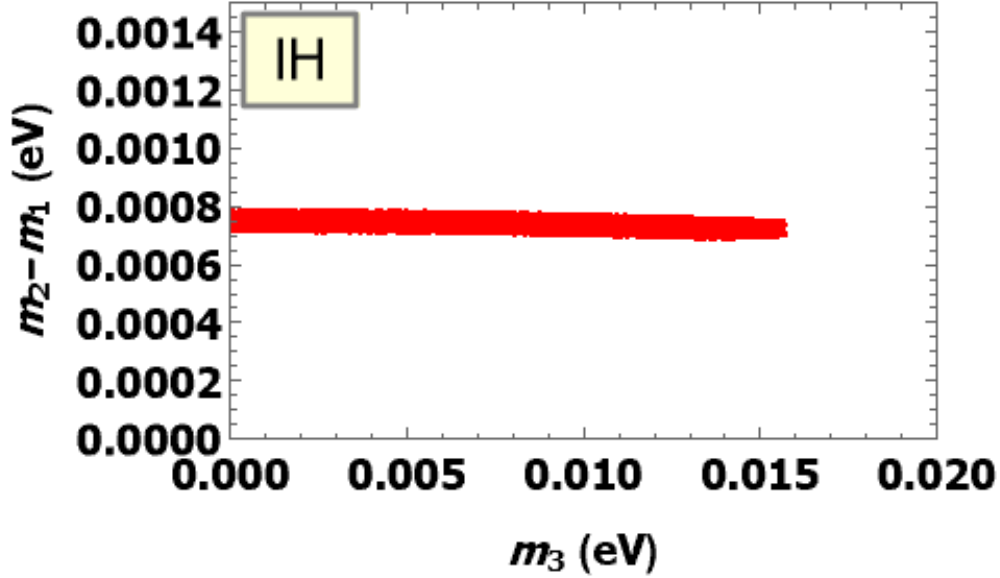}\label{fig:1d}} 
   \caption{ The scatterplots between (a) $m_3$ vs $m_1$ and $m_2$ for NH, (b) $m_2 - m_1$ vs $m_1$ for IH, (c) $m_3$ vs $m_1$ and $m_2$ for IH, (d) $m_2 - m_1$ vs $m_3$ for IH. It seems there is a apparent overlapping between $m_1$ and $m_2$ for both cases. To clarify this, we plot $m_2 -m_1$ and we find that this difference never becomes exactly zero.}
\label{fig:mass parameters}
\end{figure}

\begin{align}
V_{CKM} /U_{PMNS} &\sim  \begin{bmatrix}
1 & 0 & 0\\
0 & c_{23} & s_{23}\\
0 & - s_{23} & c_{23}
\end{bmatrix}\times \begin{bmatrix}
c_{13} & 0 & s_{13}\,e^{-i\delta}\\
0 & 1 & 0\\
-s_{13} e^{i\delta} & 0 & c_{13}
\end{bmatrix} \times \begin{bmatrix}
c_{12} & s_{12} & 0\\
-s_{12} & c_{12} & 0\\
0 & 0 & 1
\end{bmatrix},
\end{align}

where, $s_{ij}=\sin\theta_{ij}$ and $c_{ij}=\cos\theta_{ij}$. Needless to mention that the above expression reflects the parametrised decomposition of the general form, and by no means the associated angles are same for both sectors.  

\begin{figure}
  \centering
    \subfigure[]{\includegraphics[width=0.49\textwidth]{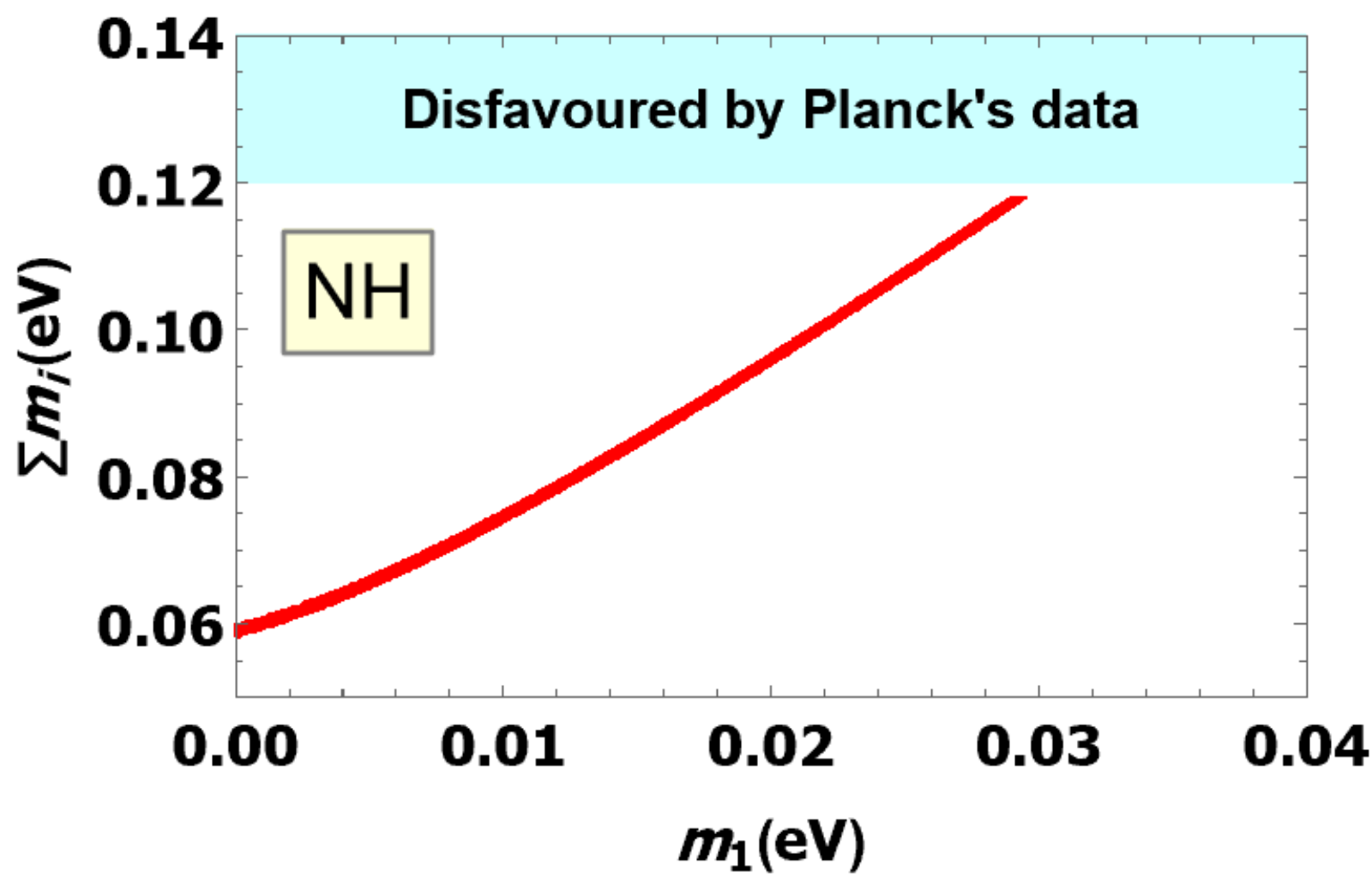}\label{fig:2a}} 
    \subfigure[]{\includegraphics[width=0.49\textwidth]{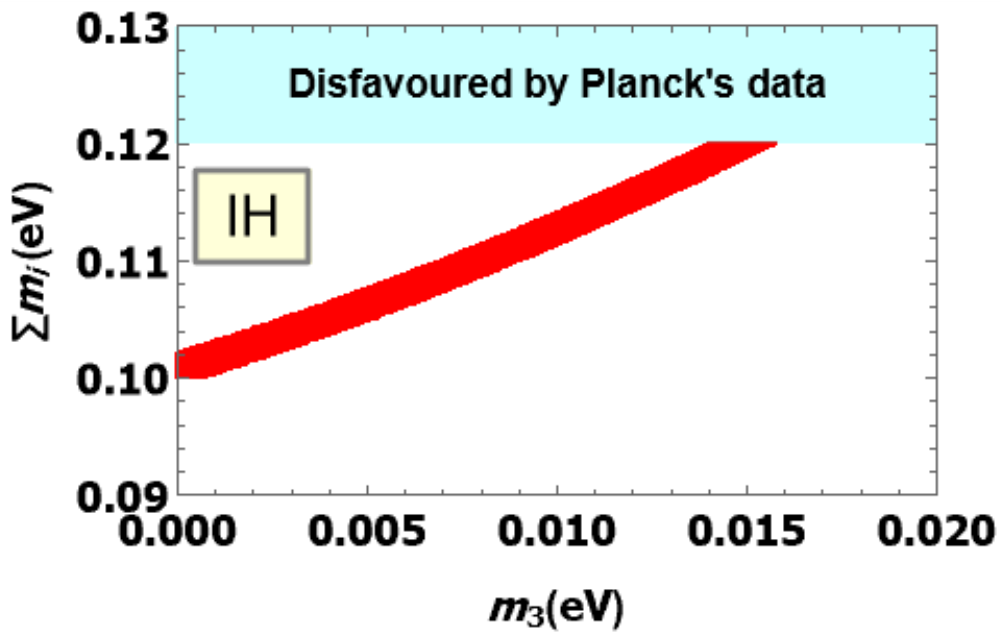}\label{fig:2b}}  
   \caption{ The correlation plots between (a) $m_1$ vs $\Sigma m_i$ for NH, (b) $m_3$ vs $\Sigma m_i$ for IH. The sum of three neutrino mass eigenvalues is consistent with Cosmological observation.}
\label{fig:sum}
\end{figure} 

The present work involves nine real texture parameters and seven physical parameters. These are three complex diagonalizing conditions:

 \begin{align}
  M_{d/\nu_{12}}^{\text{diag}}&=0,\quad M_{d/\nu_{13}}^{\text{diag}}=0,\quad M_{d/\nu_{23}}^{\text{diag}}=0
  \label{diagonalization}
 \end{align}
 
appearing as transcendental equations. We first start with the quark sector. It is well known that the CKM parameters are precisely determined. In this regard, we fix the three mixing angles $\theta_{12}^q=12.78^\circ$, $\theta_{13}^q=0.213^\circ$, $\theta_{23}^q=2.39^\circ$ and the Dirac CP phase $\delta^q=65.71^\circ$ \cite{ParticleDataGroup:2024cfk}. For simplicity, we stick to their best-fit values and ignore the uncertainties. We solve the three complex diagonalization conditions as given in Eq.\,(\ref{diagonalization}) numerically, and solve six texture parameters $r_1^q, r_2^q, r_3^q, \theta_1^q, \theta_2^q$ and $\theta_3^q$. We find that the three mass eigenvalues and other texture parameters\,($r_1^q, r_2^q, r_3^q, \theta_1^q, \theta_2^q$, $\theta_3^q$) can be expressed in terms of $a^q$, $b^q$, and $c^q$. We set $a^q=1.56$\,GeV, $b^q=1.37$\,GeV and $c^q=1.35$\,GeV and find $r_1^q=1.403$\,GeV, $r_2^q=1.39$\,GeV, $r_3^q=1.33$\,GeV, $\theta_1^q=122.69^\circ$, $\theta_2^q=-123.28^\circ$ and $\theta_3^q=116.96^\circ$ respectively. Further, we obtain the down-type quark masses as: $m_d = 0.0047$ GeV, $m_s = 0.093$ GeV, and $m_b = 4.183$ GeV respectively\,\cite{ParticleDataGroup:2024cfk}.

However, in the neutrino sector, the parameter space is quite broad. Thus, instead of fixing the mixing angles $\theta_{ij}^\nu$ and the Dirac CP phase $\delta^\nu$ at specific values, we generate a sufficient number of random data points for these parameters within their respective $3\sigma$ bounds. For this purpose, we use the global fit data provided by NuFIT 6.1\,\cite{nufit}, which incorporates the recent results from the JUNO experiment\,\cite{JUNO:2025gmd}. We solve the three complex diagonalization conditions numerically, as discussed earlier to determine the six texture parameters $r_1^\nu, r_2^\nu, r_3^\nu, \theta_1^\nu, \theta_2^\nu$, and $\theta_3^\nu$. These parameters, along with the three mass eigenvalues, are expressed in terms of $a_\nu$, $b_\nu$, and $c_\nu$. The mass eigenvalues are computed such that the numerical values of the free parameters $a_\nu$, $b_\nu$, and $c_\nu$ remain consistent with the experimental bounds on the two mass-squared differences, $\Delta m_{21}^2$ and $\Delta m_{31}^2$\,\cite{nufit}. In addition, the sum of the three neutrino masses, $\sum m_i$, aligns with the cosmological constraint $\sum m_i \leq 0.12~\text{eV}$\,\cite{Planck:2018vyg}. The analysis is performed for both the normal hierarchy\,(NH) and the inverted hierarchy\,(IH) of neutrino masses.

\begin{figure}
  \centering
    \subfigure[]{\includegraphics[width=0.49\textwidth]{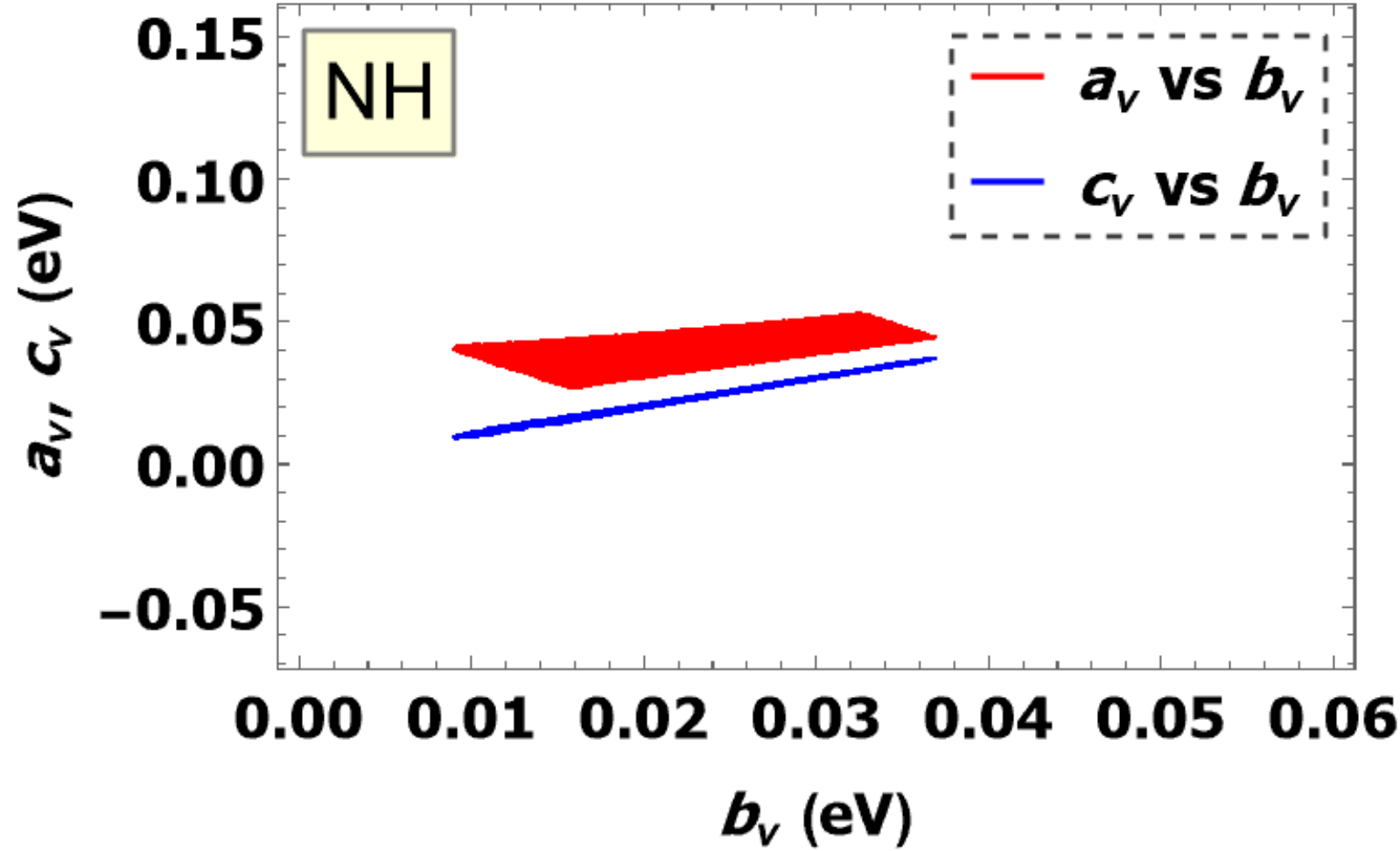}\label{fig:4a}} 
    \subfigure[]{\includegraphics[width=0.49\textwidth]{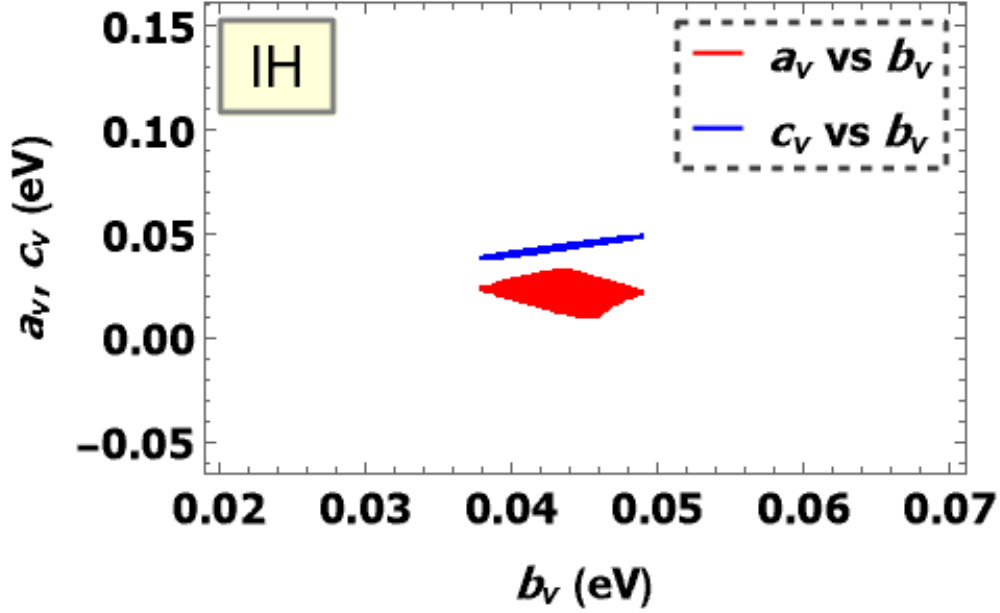}\label{fig:4b}}
     \subfigure[]{\includegraphics[width=0.49\textwidth]{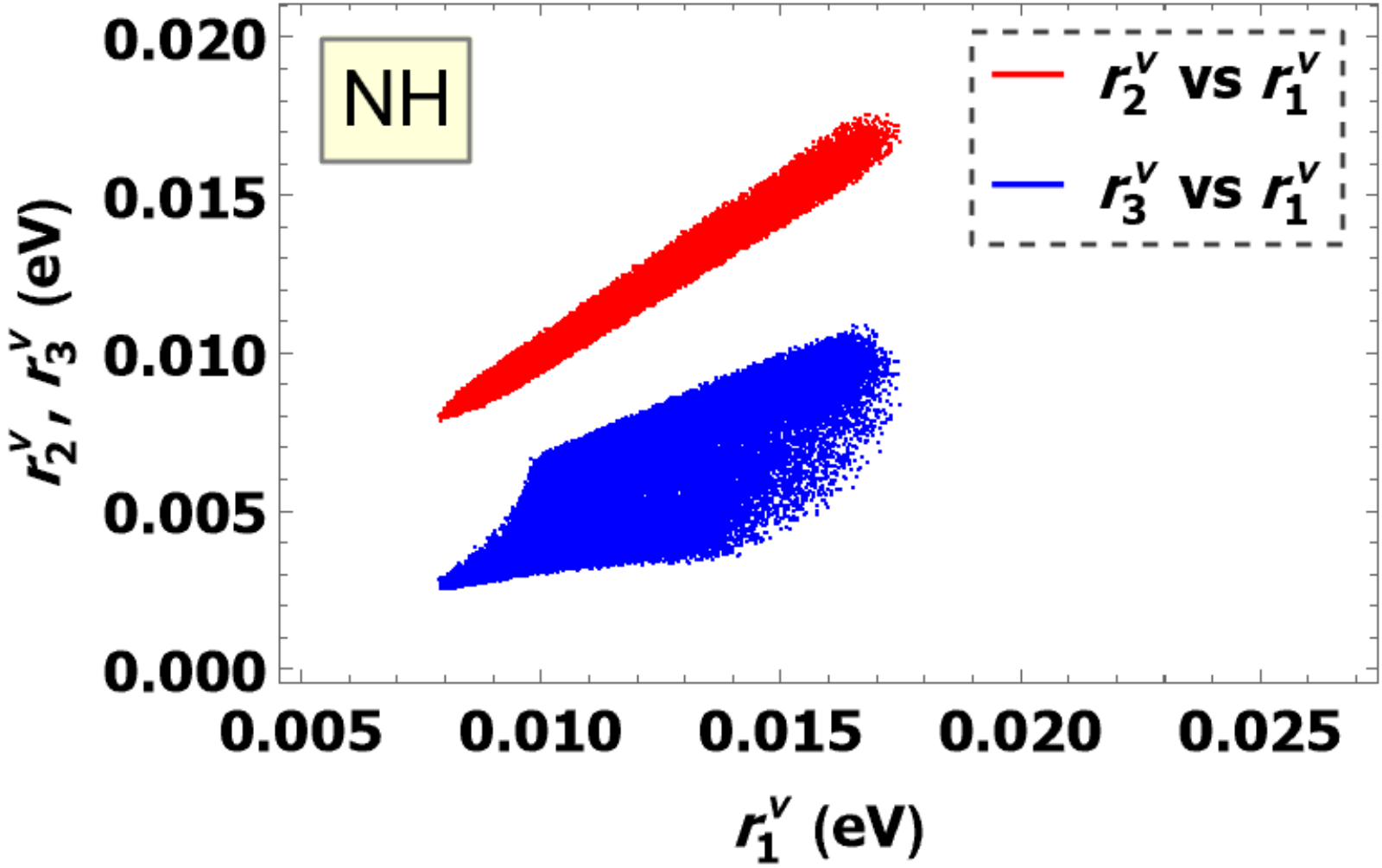}\label{fig:4a}} 
    \subfigure[]{\includegraphics[width=0.49\textwidth]{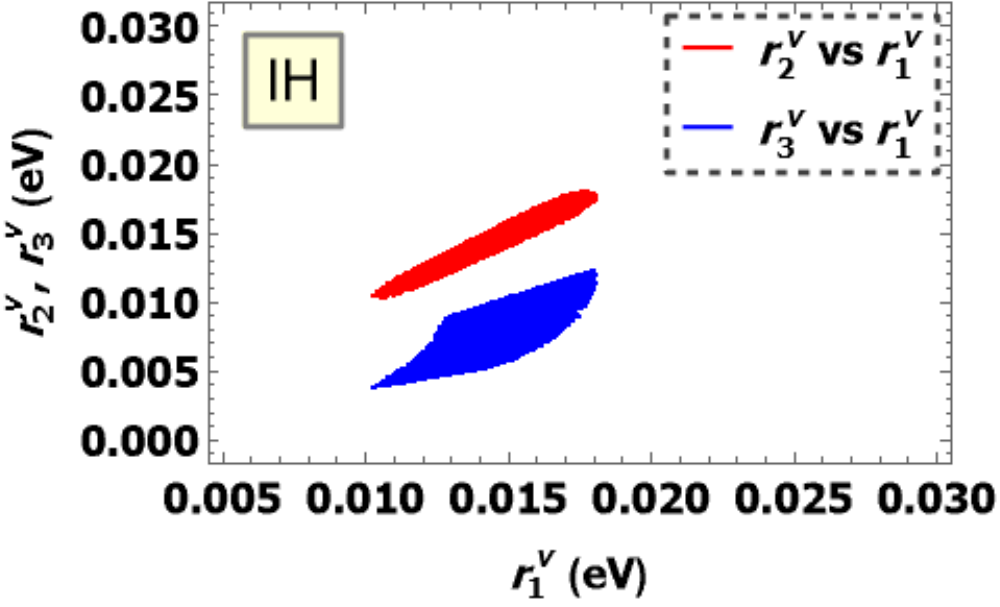}\label{fig:4b}}
     \subfigure[]{\includegraphics[width=0.49\textwidth]{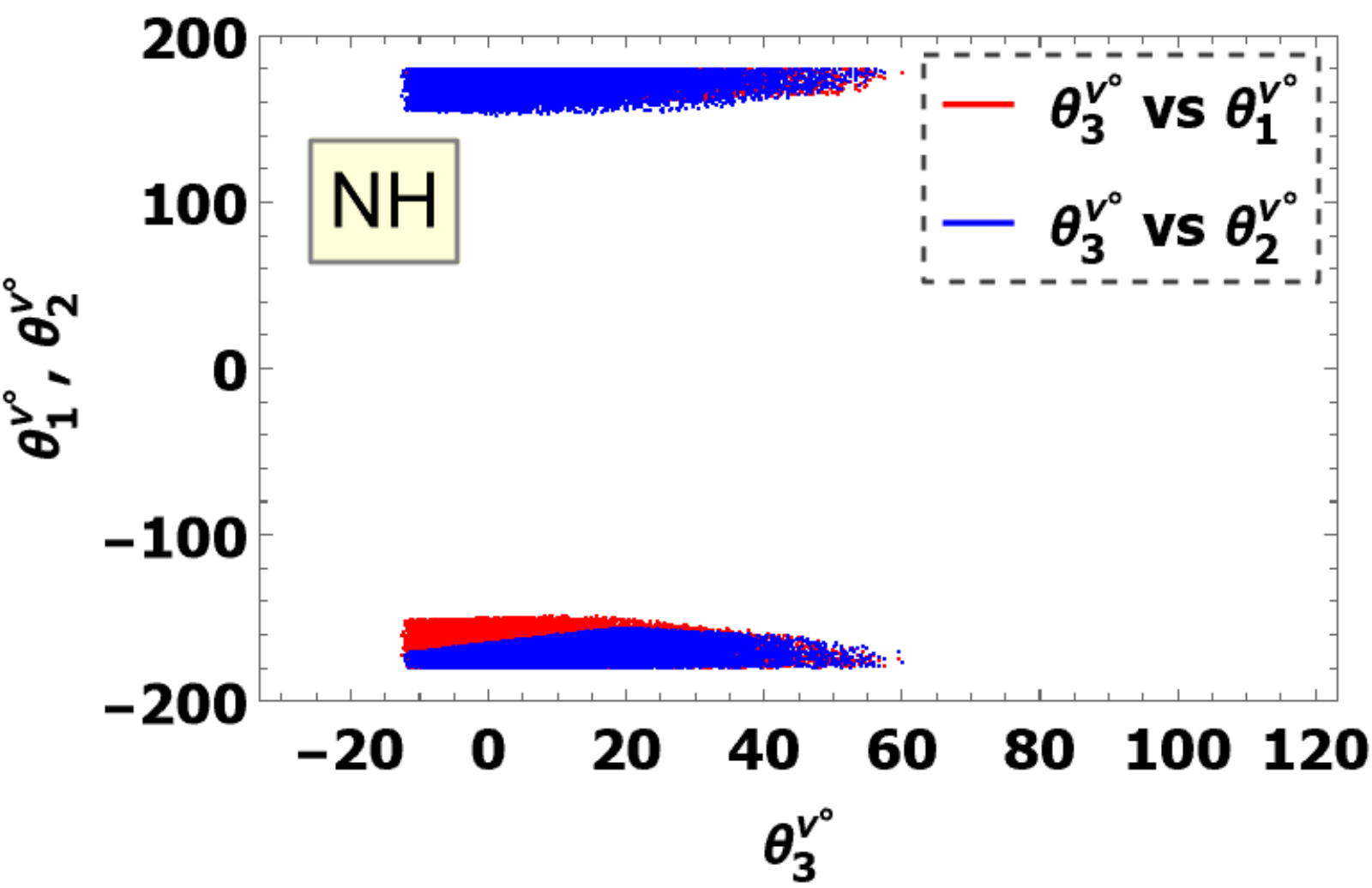}\label{fig:4a}} 
    \subfigure[]{\includegraphics[width=0.49\textwidth]{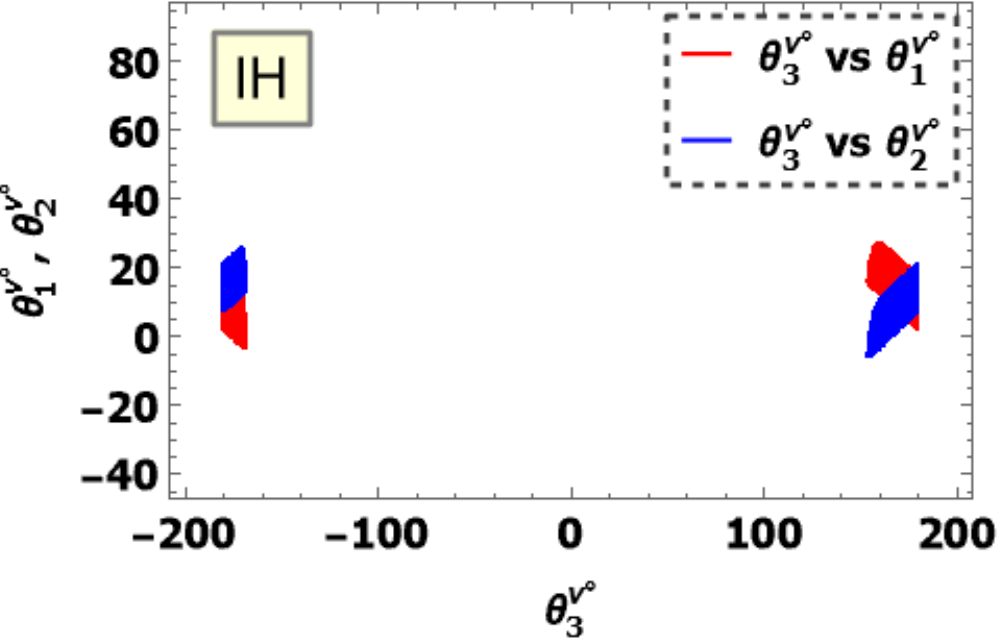}\label{fig:4b}}
   \caption{ The variation of the texture parameters for both NH and IH. It is seen that there exists some apparent overlapping region in the parameter space. These regions hint towards additional correlations in the mass matrix.}
\label{fig:texture parameters}
\end{figure}

\begin{table}
\centering
\begin{tabular*}{\textwidth}{@{\extracolsep{\fill}} ccccc}
\hline
\hline
Texture & Min Value &  Max Value & Min Value & Max Value\\
Parameters & (NH) & (NH) & (IH) & (IH)\\
\hline
\hline
$a_\nu /\text{eV}$ & 0.026 & 0.052 & 0.010 & 0.033\\
\hline
$b_\nu /\text{eV}$ & 0.009 & 0.037 & 0.038 & 0.049 \\
\hline
$c_\nu /\text{eV}$ & 0.009 & 0.037  & 0.038 & 0.049\\
\hline
$r_1^\nu /\text{eV}$ & 0.0079 & 0.017  & 0.010 &0.018\\
\hline
$r_2^\nu /\text{eV}$ & 0.0079 & 0.017  & 0.010 & 0.018\\
\hline
$r_3^\nu /\text{eV}$ & 0.0025 & 0.011  & 0.004 & 0.012\\
\hline
$\theta_1^\nu/^\circ$ & -180 & 180   & -3.58 & 27.09\\
\hline
$\theta_2^\nu/^\circ$ & -180 & 180  & -5.99 & 25.53 \\
\hline
$\theta_3^\nu/^\circ$ & -12.62 & 60.10  & -180 & 180\\
\hline
\end{tabular*}
\caption{Shows the maximum and minimum values of the texture parameters for neutrino sector for both NH and IH.}
\label{values of texture parameters}
\end{table}

\begin{figure}
  \centering
    \subfigure[]{\includegraphics[width=0.49\textwidth]{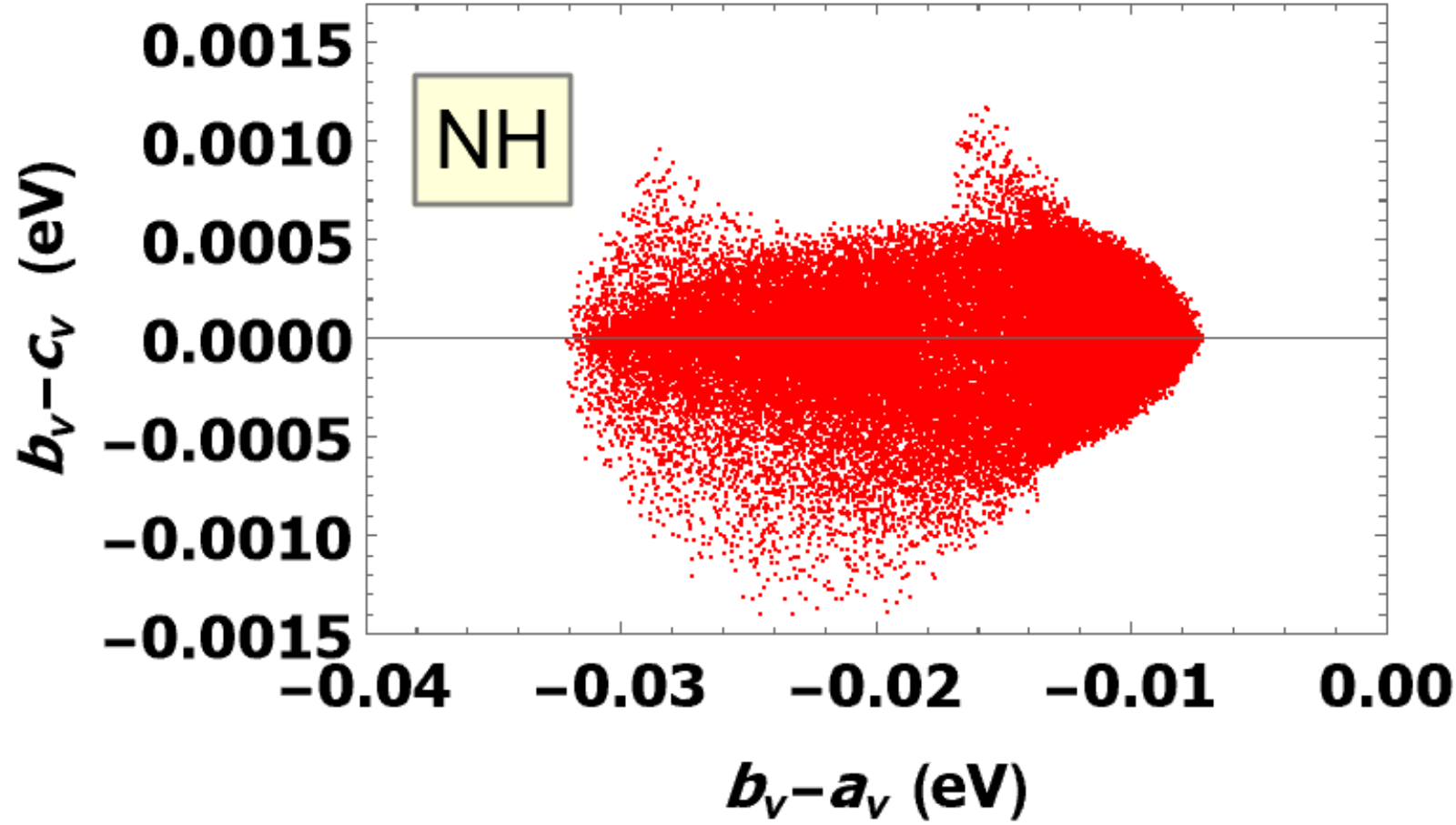}\label{fig:2a}} 
    \subfigure[]{\includegraphics[width=0.49\textwidth]{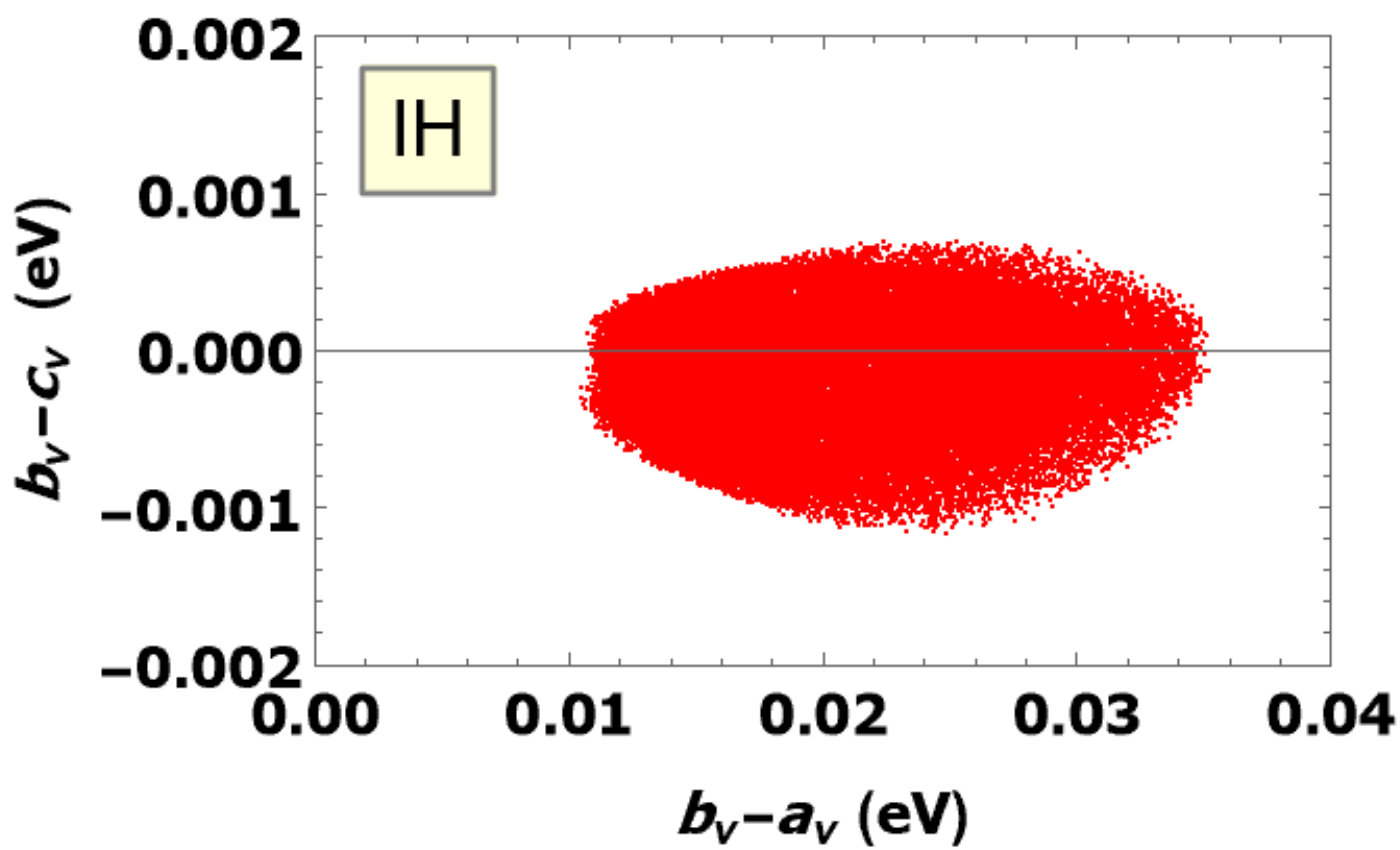}\label{fig:2b}}  
   \caption{ The correlation plots between ($b_\nu - c_\nu$) vs ($b_\nu - a_\nu$) for both NH and IH. It is seen that for both cases there exists the possibility of $b_\nu \simeq c_\nu$ which reduce the number of parameters.}
\label{fig:correlation1}
\end{figure}

\begin{figure}
  \centering
    \subfigure[]{\includegraphics[width=0.49\textwidth]{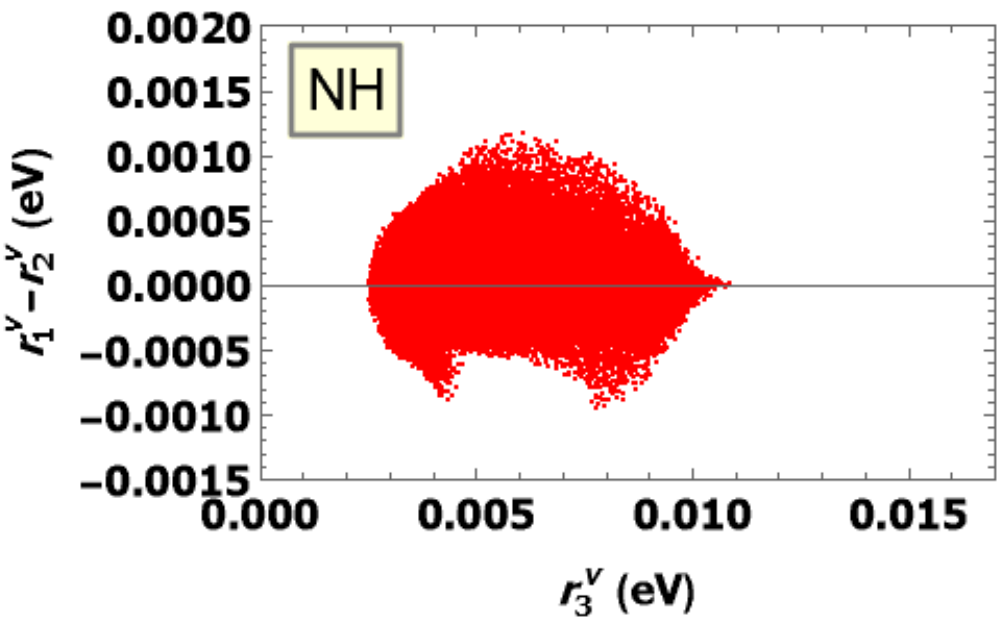}\label{fig:2a}} 
    \subfigure[]{\includegraphics[width=0.49\textwidth]{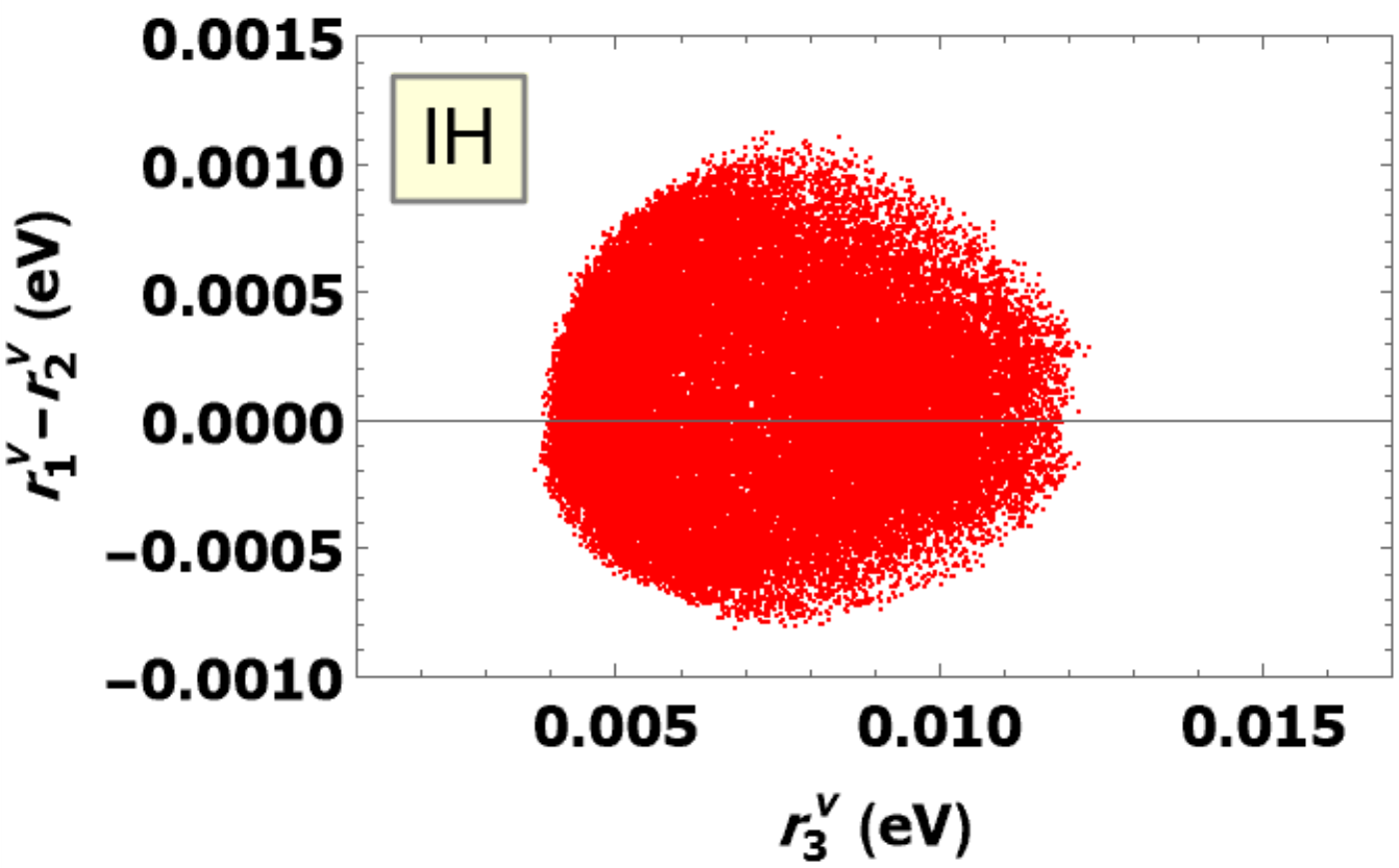}\label{fig:2b}}  
   \caption{ The correlation plots between ($r_1^\nu - r_2^\nu$) vs $r_3^\nu$ for both NH and IH. It is seen that for both cases there exists the possibility of $r_1^\nu \simeq r_2^\nu$ which reduce the number of parameters.}
\label{fig:correlation2}
\end{figure}

We obtain the neutrino mass spectrum for both hierarchies. For NH, the mass spectra of $m_1$, $m_2$, and $m_3$ are obtained as follows: $6.75 \times 10^{-7}\, \text{eV} < m_1 < 0.029\,\text{eV}$, $0.008\,\text{eV}< m_2 < 0.031\,\text{eV}$, and $0.049\,\text{eV}< m_3 < 0.058\,\text{eV}$ (see Fig.\,(\ref{fig:1a})). Here, we observe an apparent overlap in the spectra of $m_1$ and $m_2$. This may seem misleading, as it could suggest an exact degeneracy of $m_1$ and $m_2$ within the overlapping region. To clarify this, we graphically analyse $m_2 - m_1$ in Fig.\,(\ref{fig:1b}) and verify that this difference never becomes exactly zero. On the other hand, the proposed texture supports the possibility of exact inverted hierarchy\,(IH): $0.049\,\text{eV} < m_1 < 0.052\,\text{eV}$, $0.050\,\text{eV} < m_2 < 0.053\,\text{eV}$, and $7.61 \times 10^{-7}\,\text{eV}< m_3 < 0.015\,\text{eV}$ (see Fig.\,(\ref{fig:1c})). In similar fashion, we graphically analyse $m_2 - m_1$ to verify that this difference is non-zero (see Fig.\,(\ref{fig:1d})). It is to be mentioned that diagonalization conditions were satisfied within a numerical tolerance of $\mathcal{O}(10^{-15})$ for the present work.

\begin{figure}
  \centering
    \subfigure[]{\includegraphics[width=0.49\textwidth]{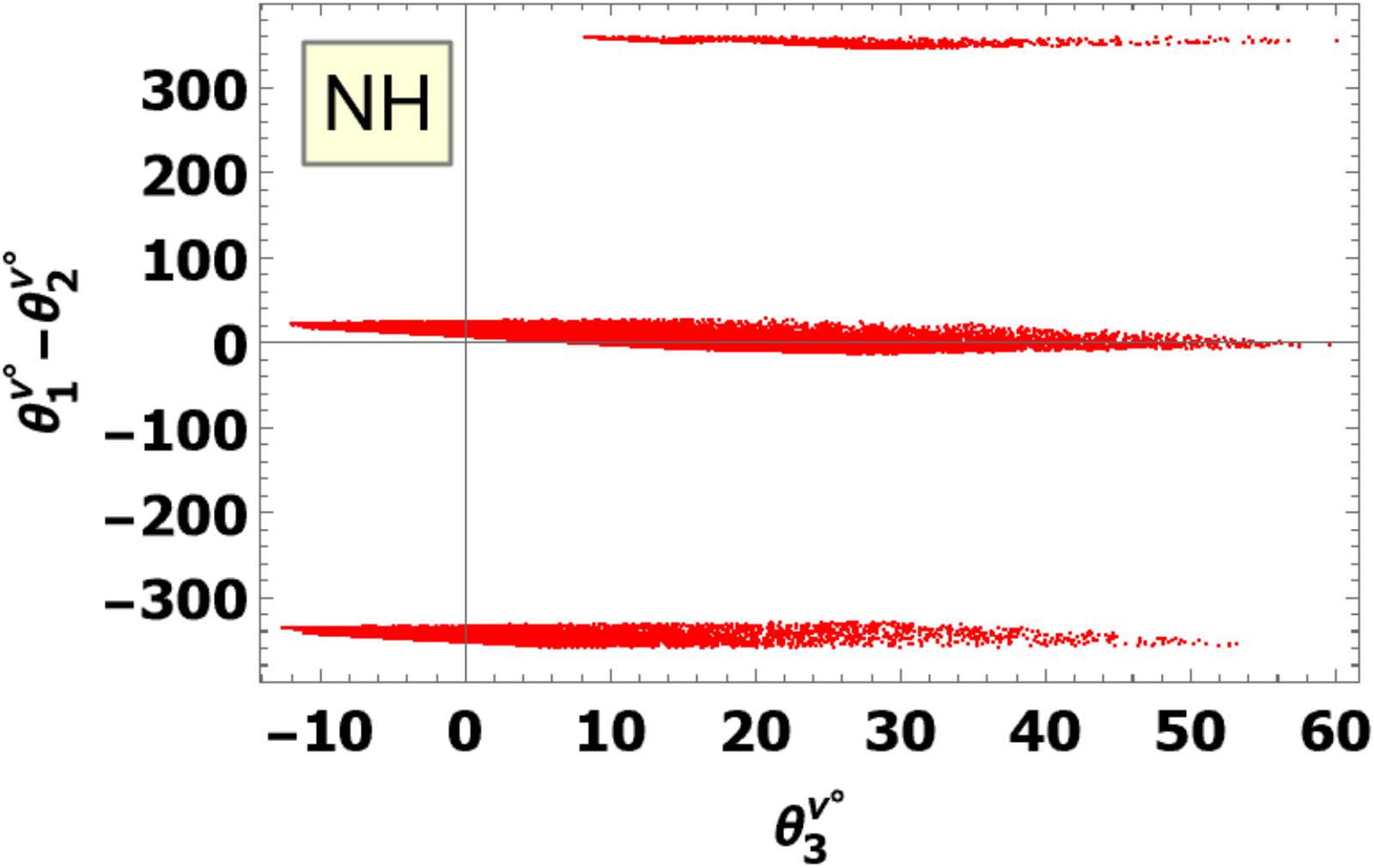}\label{cor1}} 
    \subfigure[]{\includegraphics[width=0.49\textwidth]{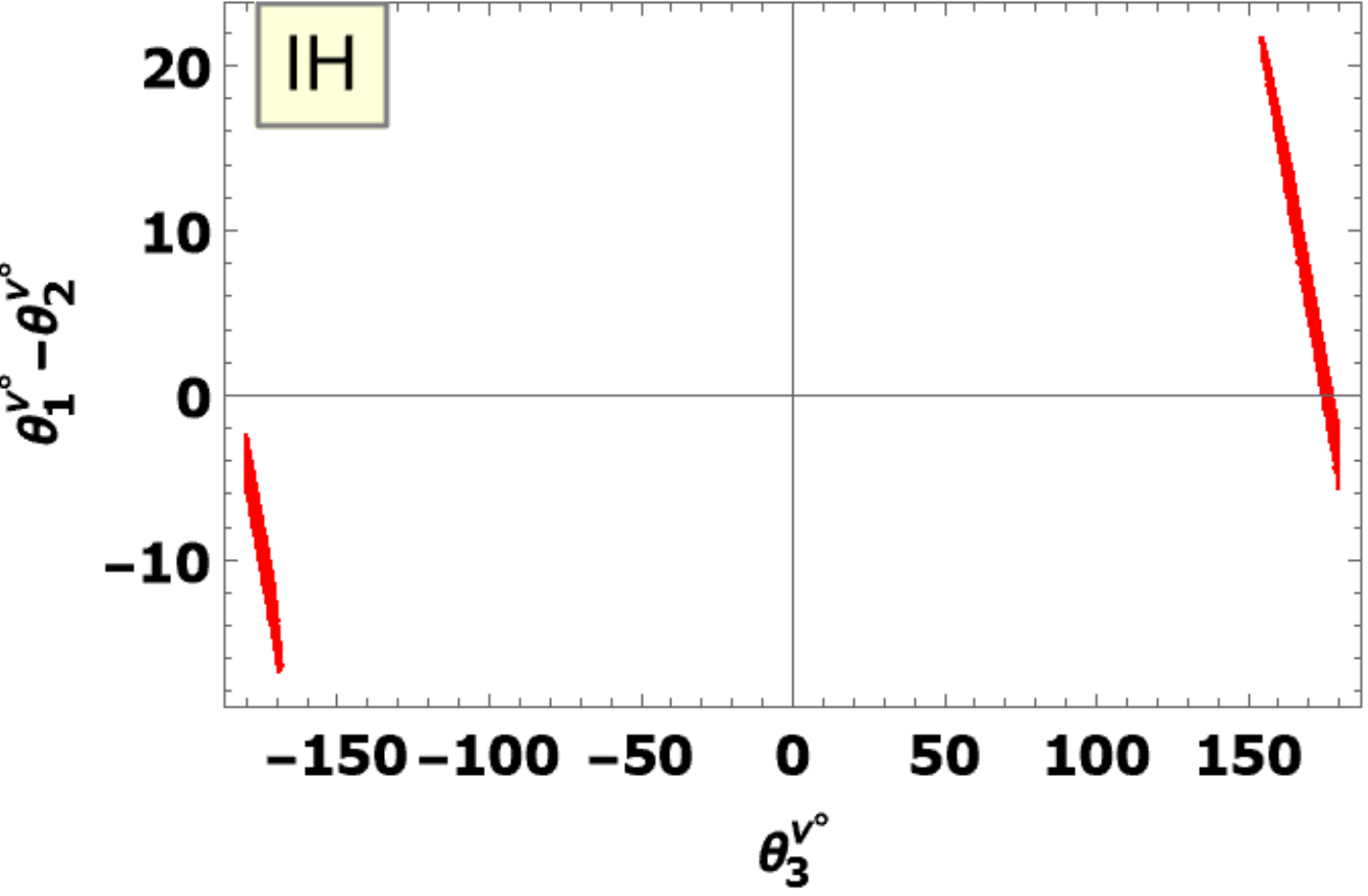}\label{fig:2b}}  
   \caption{The correlation plots between ($\theta_1^\nu - \theta_2^\nu$) vs $\theta_3^\nu$ for both NH and IH. It is seen that for both cases there exists the possibility of $\theta_1^\nu \simeq \theta_2^\nu$.}
\label{fig:correlation3}
\end{figure}

\begin{figure}
  \centering
    \subfigure[]{\includegraphics[width=0.49\textwidth]{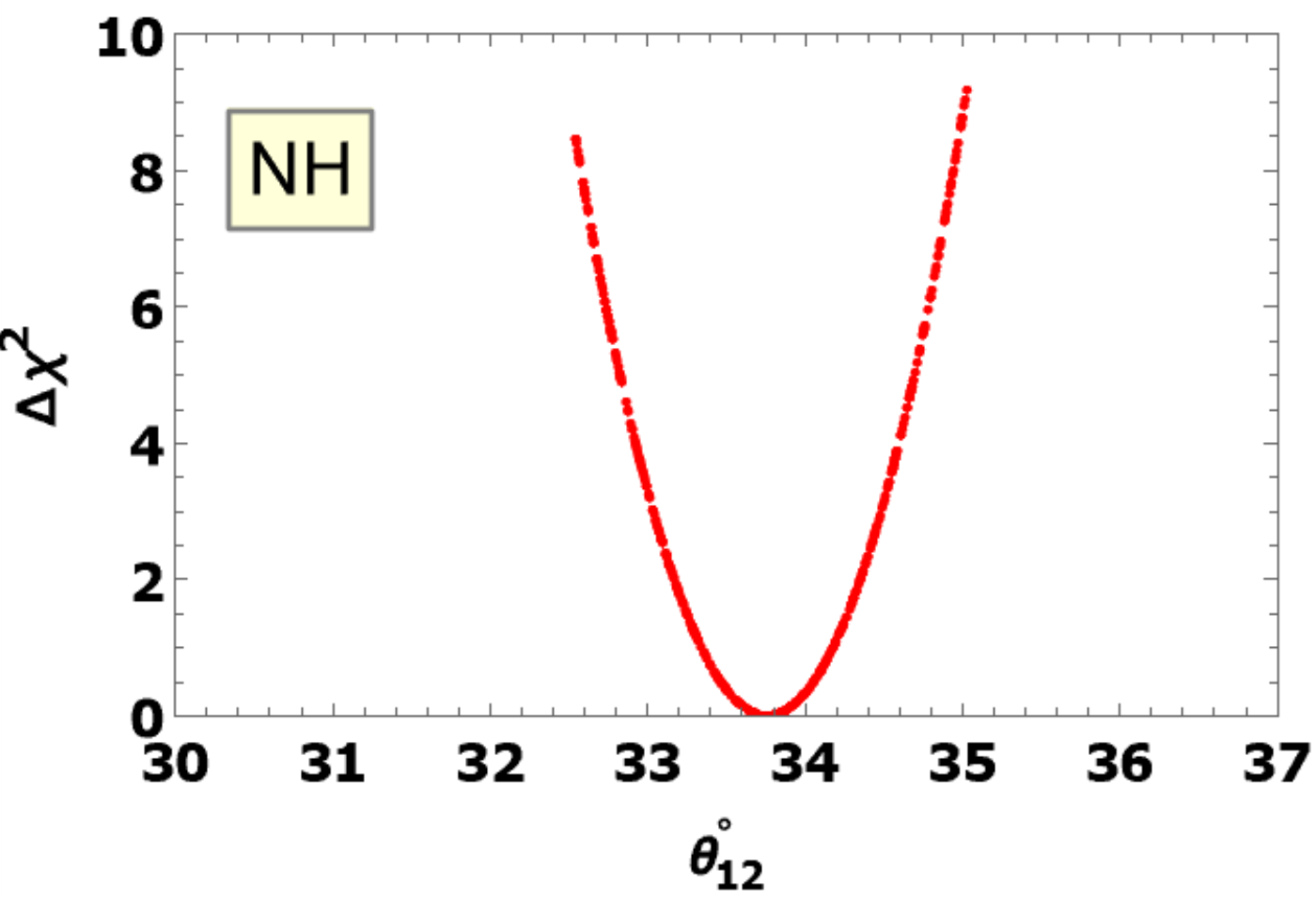}\label{fig:8a}} 
    \subfigure[]{\includegraphics[width=0.49\textwidth]{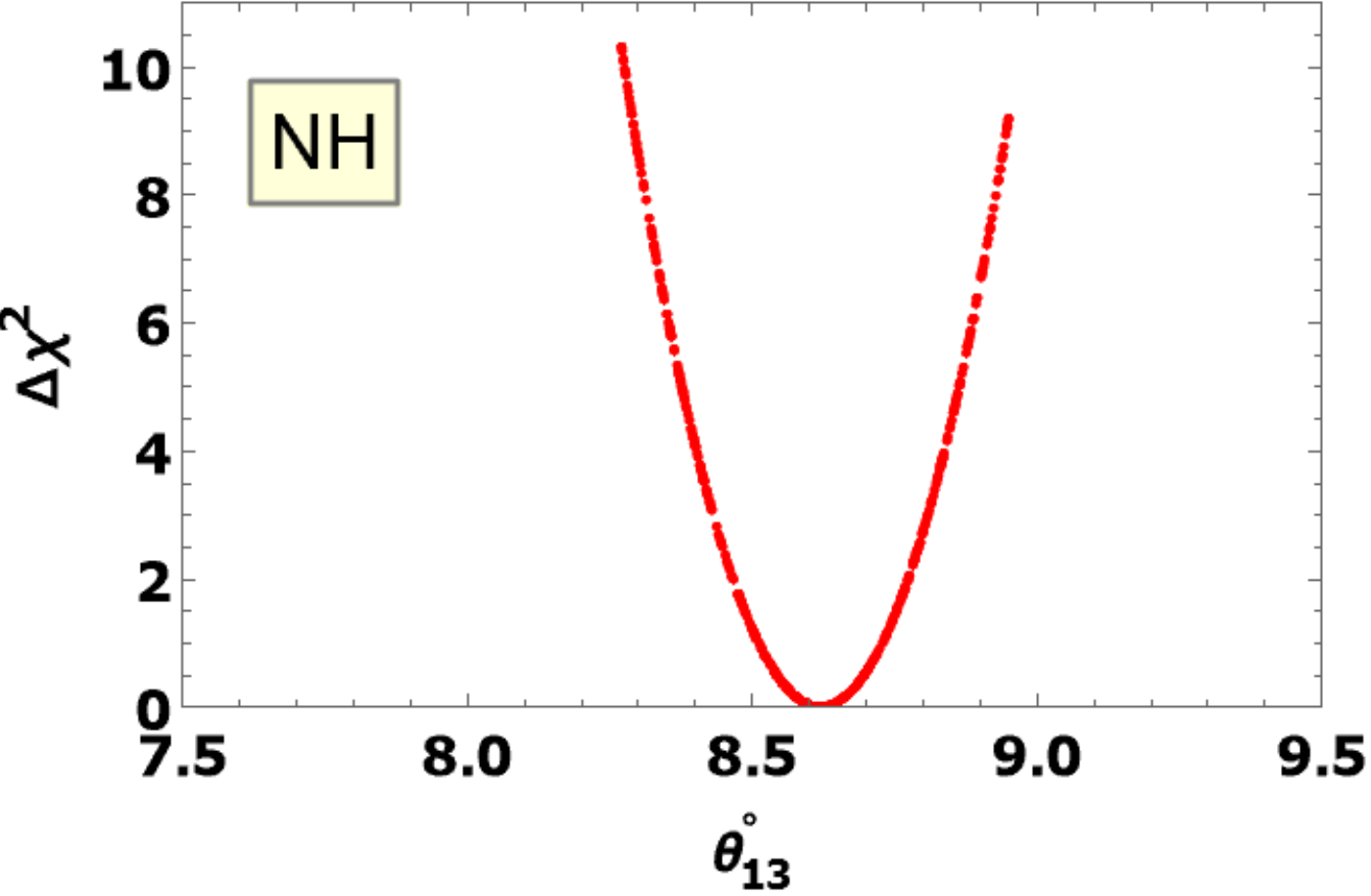}\label{fig:8b}}
     \subfigure[]{\includegraphics[width=0.49\textwidth]{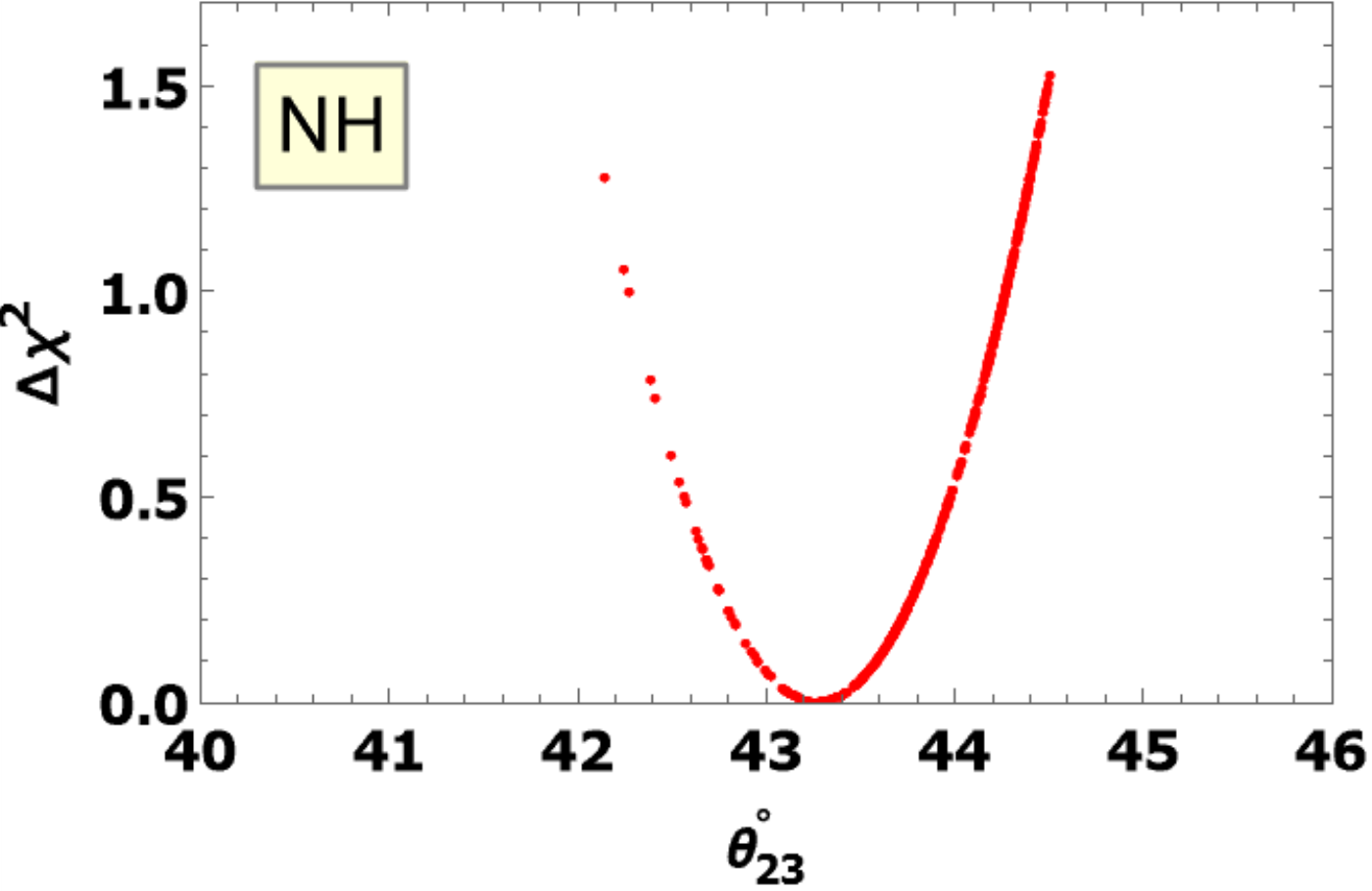}\label{fig:8c}} 
    \subfigure[]{\includegraphics[width=0.49\textwidth]{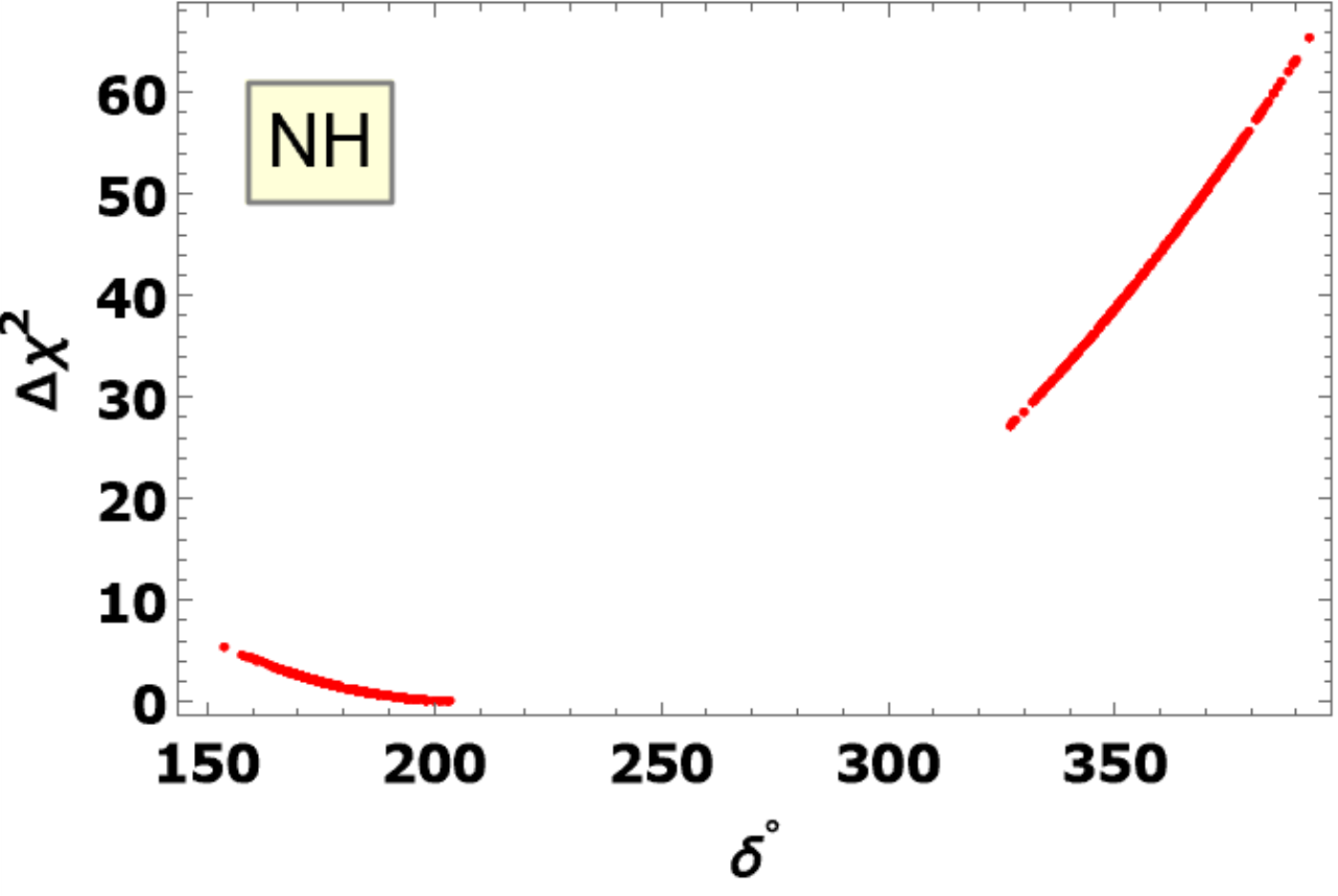}\label{fig:8d}}
     \subfigure[]{\includegraphics[width=0.49\textwidth]{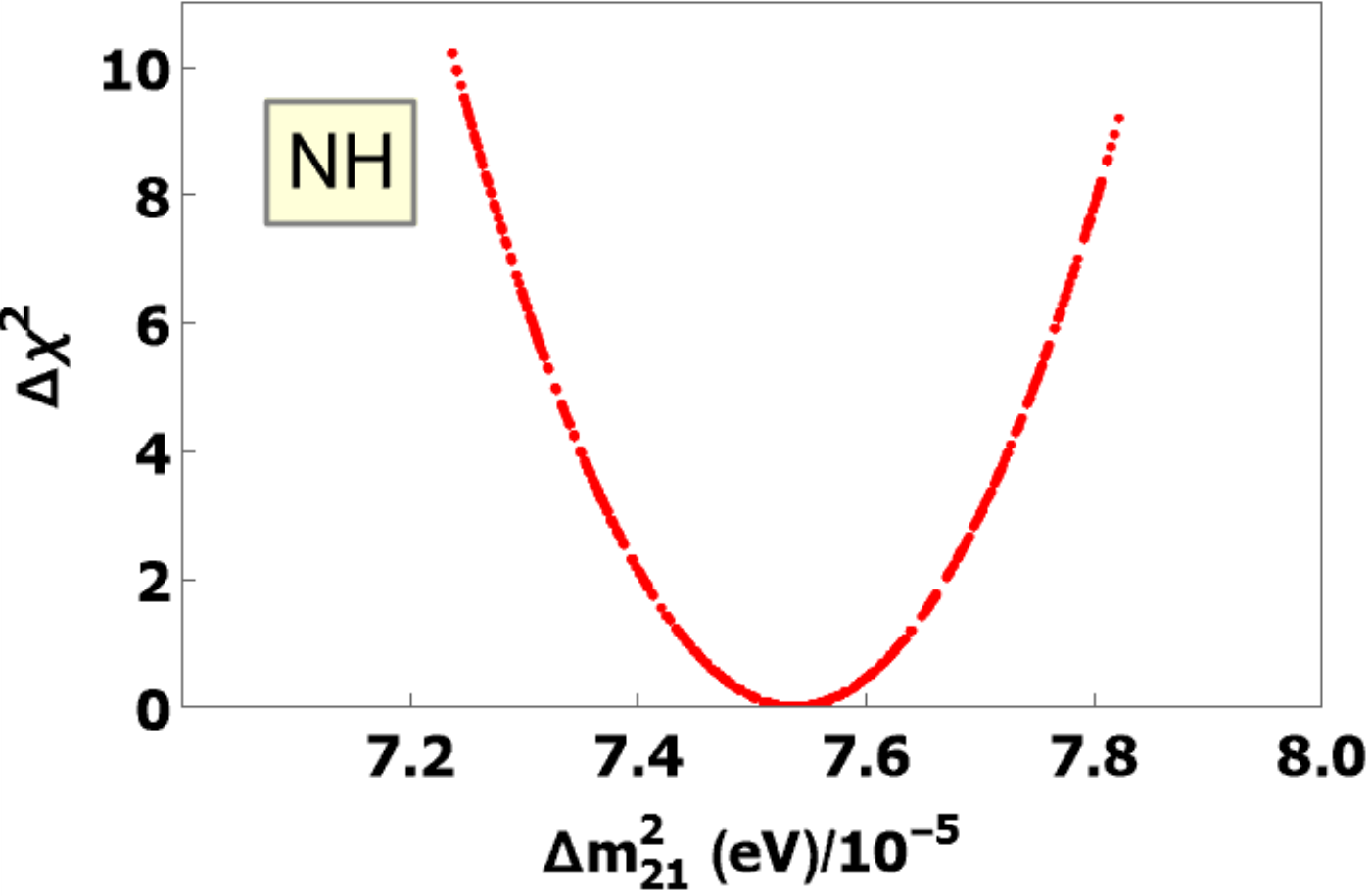}\label{fig:8e}} 
    \subfigure[]{\includegraphics[width=0.49\textwidth]{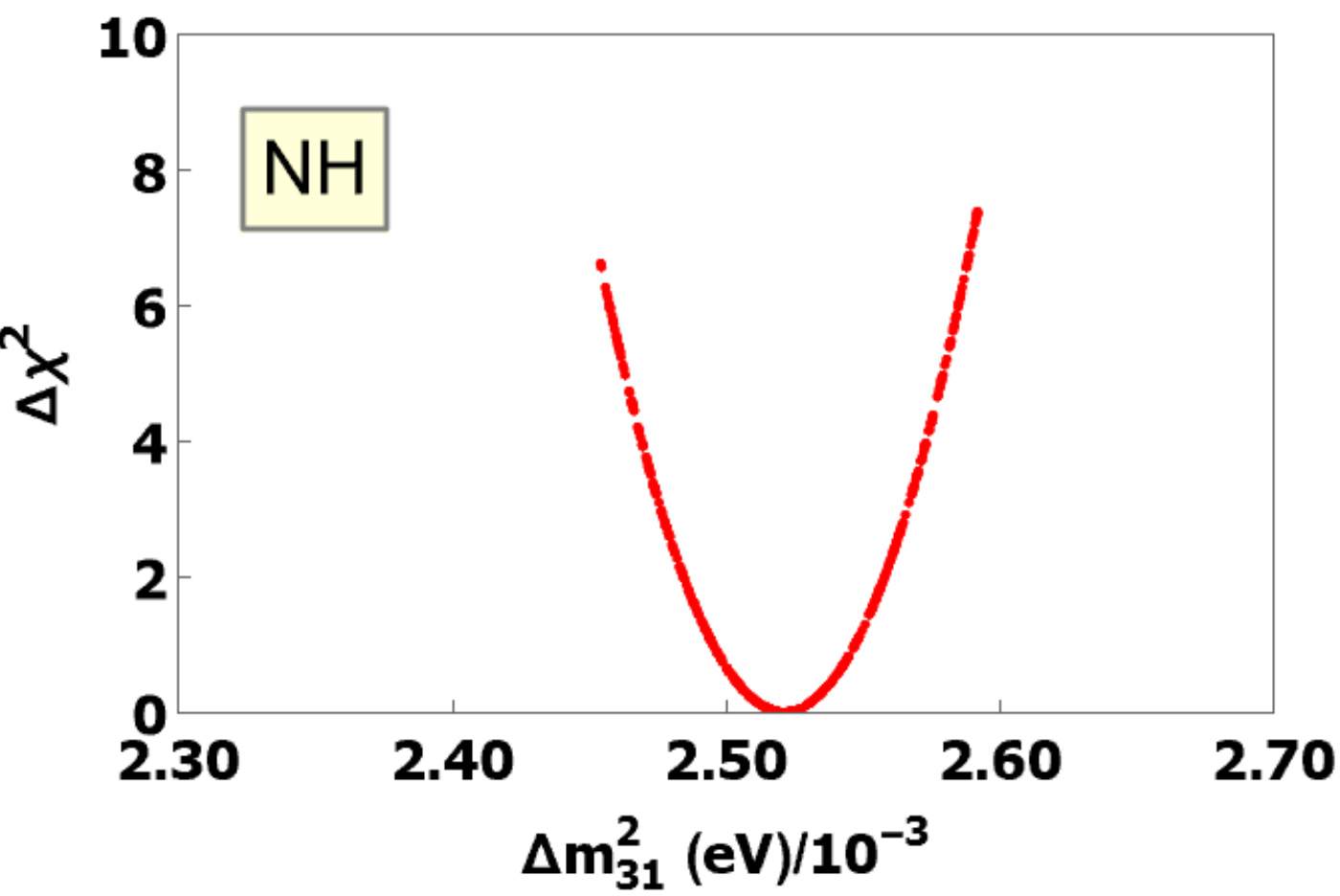}\label{fig:8f}}
   \caption{Variation of $ \chi^2$ as a function of the studied parameters for NH. The reduced texture predictions for the physical parameters are compared with the corresponding experimental $1\sigma$ best-fit values within the allowed confidence interval. The reduced texture predicts sharp bound for $\theta_{23}$ in the interval $(42.14^\circ-44.50^\circ)$. On the other hand, for $\delta$, we see two constrained bounds $(153.60^\circ-203.40^\circ)$ and $(326.90^\circ-392.80^\circ)$ respectively.}
\label{fig:chinh}
\end{figure}

\begin{figure}
  \centering
    \subfigure[]{\includegraphics[width=0.49\textwidth]{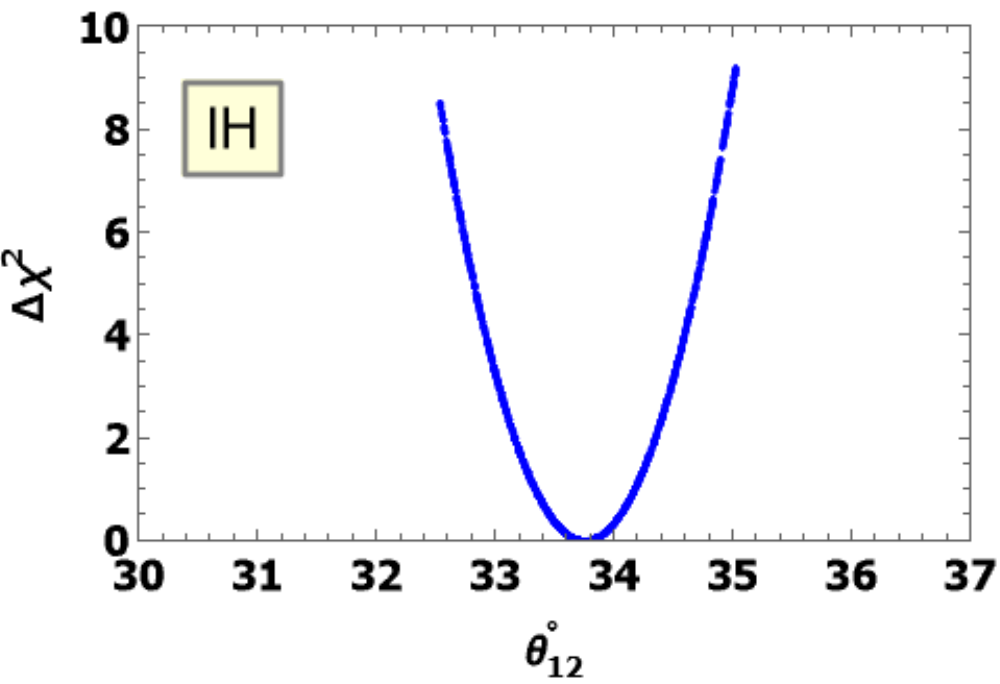}\label{fig:8a}} 
    \subfigure[]{\includegraphics[width=0.49\textwidth]{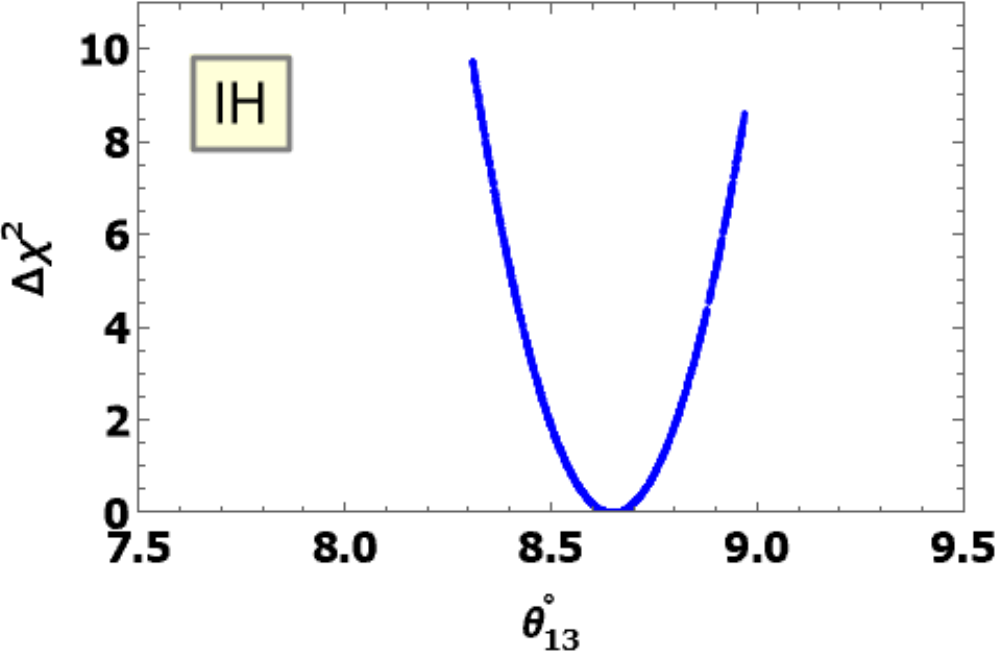}\label{fig:8b}}
     \subfigure[]{\includegraphics[width=0.49\textwidth]{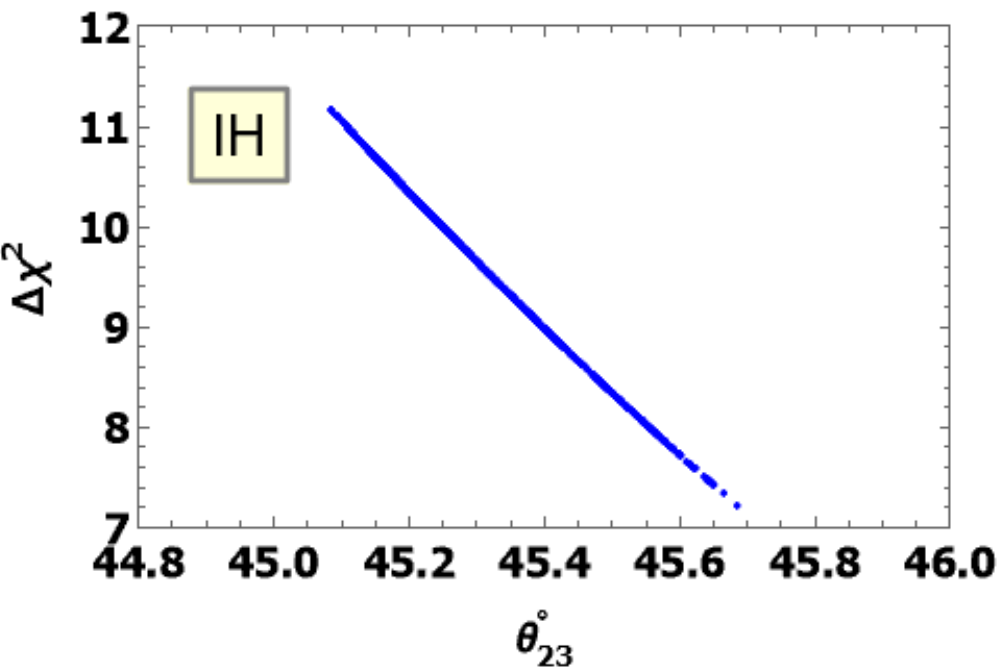}\label{fig:8c}} 
    \subfigure[]{\includegraphics[width=0.49\textwidth]{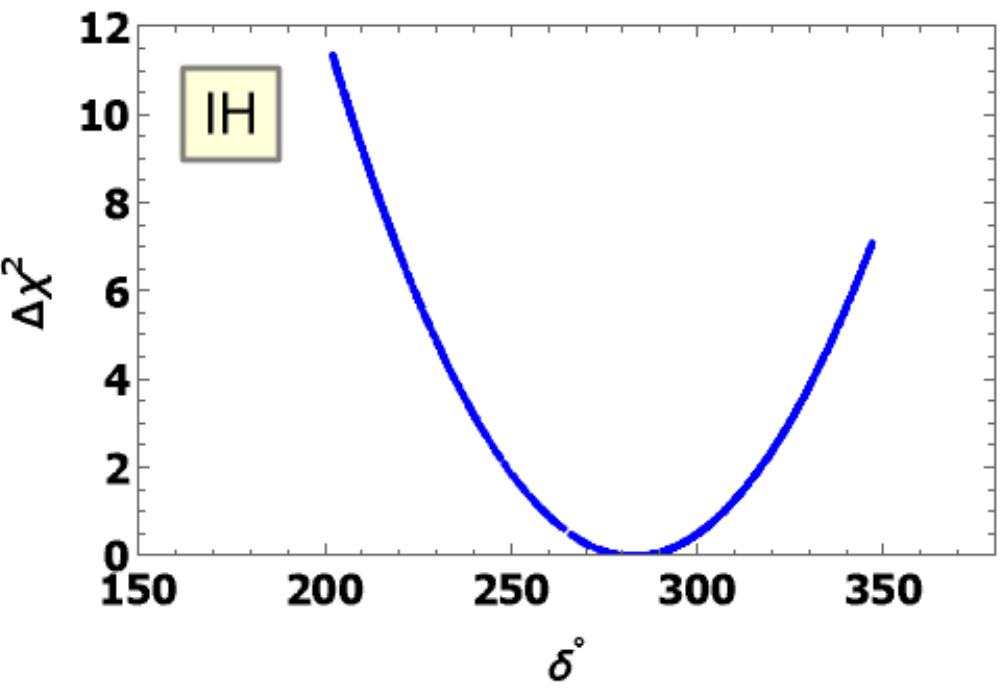}\label{fig:8d}}
     \subfigure[]{\includegraphics[width=0.49\textwidth]{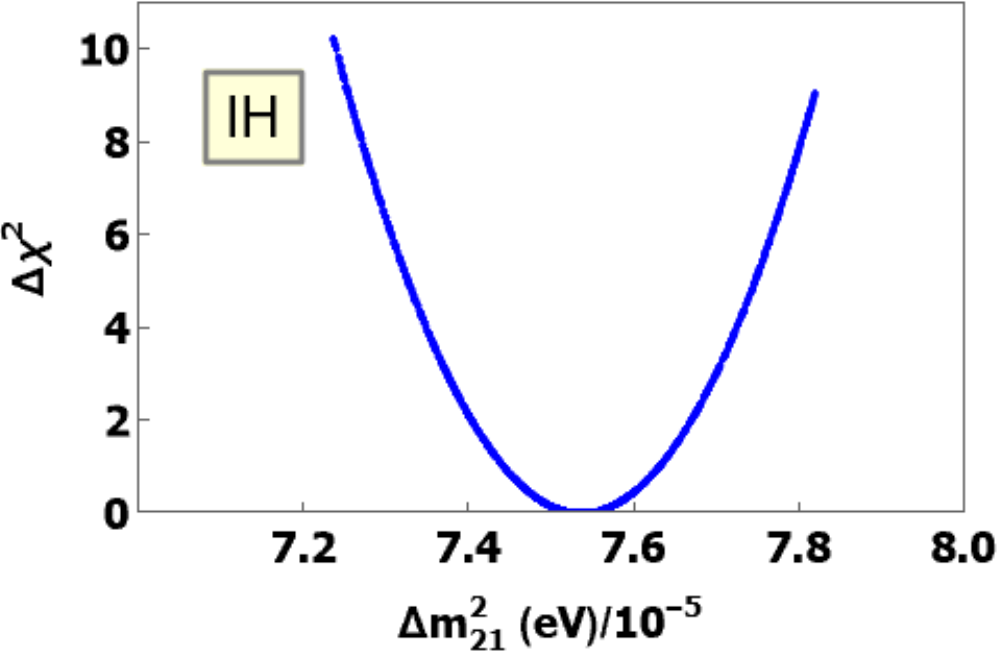}\label{fig:8e}} 
    \subfigure[]{\includegraphics[width=0.49\textwidth]{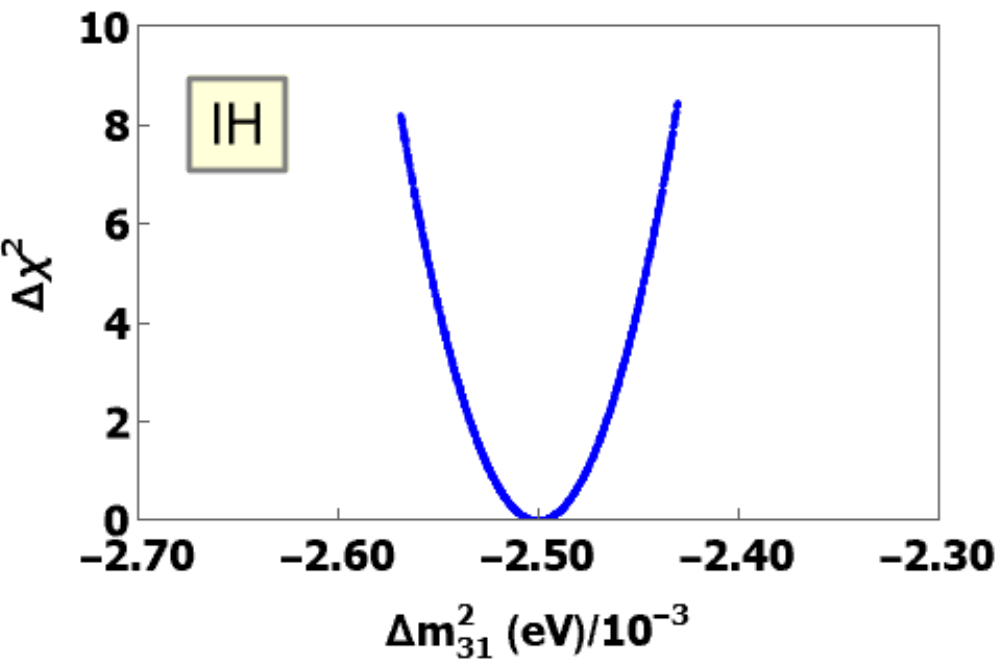}\label{fig:8f}}
   \caption{Variation of $\chi^2$ as a function of the studied parameters for IH. The reduced texture predictions for the physical parameters are compared with the corresponding experimental $1\sigma$ best-fit values within the allowed confidence interval. For $\theta_{23}$, the reduced texture predicts sharp bound $(45.08^\circ-45.68^\circ)$ which is far from the experimental $1\sigma$ best fit value $48.15^\circ$ for IH\,\cite{nufit}. Hence, we observe a linear fit.}
\label{fig:chiih}
\end{figure}

For graphical visualization, we generate the correlation plots for the sum of three neutrino mass eigenvalues for both NH and IH\,(see Fig.\,(\ref{fig:sum})). The Fig.\,(\ref{fig:texture parameters}) show possible parameters space for the texture parameters for both the hierarchies. The minimum and maximum values of the texture parameters are listed in Table\,(\ref{values of texture parameters}).
 
From the texture analysis, we find that the texture parameters share several common regions. In this regard, these possible additional constraints reduce the number of free parameters. These additional correlations are not only motivated from the texture point of view, rather, they are also relevant from a model-building perspective. The said correlations impose certain constraints on the model parameters. 

\begin{itemize}

  \item[\textbf{Case I}] 
  
  For both NH and IH, we encounter the possibility of $b_\nu \simeq c_\nu$\,(see Fig.\,\ref{fig:correlation1}). This agrees upto an accuracy of $\mathcal{O}(10^{-5})$. In this regard, we obtain the following constrain on the model parameters,

\begin{equation}
\frac{x_1^\nu}{z_1^\nu} \approx \frac{v_\kappa v_\sigma}{v_\eta \Lambda}.
\end{equation}

\item[\textbf{Case II}] For both NH and IH, we observe the possibility of $r_1^\nu \simeq r_2^\nu$\,(see Fig.\,\ref{fig:correlation2}). This correlation holds upto an accuracy of $\mathcal{O}(10^{-5})$. It put a constrain over the model parameter is shown below,

\begin{equation}
z_2^\nu \approx \frac{(v_\eta^2 \Lambda^2 ({x_2^{\nu}}^2+{x_3^{\nu}}^2)- v_\kappa^2 v_\sigma^2 {z_3^{\nu}}^2)^{\frac{1}{2}}}{v_\kappa v_\sigma}
\end{equation}

\item[\textbf{Case III}] For both NH and IH, there exists the possibility of $\theta_1^\nu \simeq \theta_2^\nu$\,(see Fig.\,\ref{fig:correlation3}). Since, $\theta_1^\nu$ and $\theta_2^\nu$ are angles, the order of equality is taken upto $\mathcal{O}(10^{-3})$. In this light, we use $\tan \theta_1^\nu \approx \tan \theta_2^\nu$ and obtain a constrain on the model parameters as shown in the following,

\begin{equation}
z_3^{\nu} \approx \frac{x_3^{\nu} z_2^{\nu}}{x_2^{\nu}}.
\end{equation}

\end{itemize}

Needless to mention, a general Hermitian matrix contains nine real parameters. The above mentioned three correlations coexist simultaneously and reduce the number of free parameters from nine to six. The reduced neutrino mass matrix texture ($M^R_\nu$) takes the following form,

\begin{equation}
M^R_{\nu} \approx \begin{bmatrix}
a^\nu & r_1^\nu e^{i \theta_1^\nu} & r_1^\nu e^{i\theta_1^\nu}\\
r_1^\nu e^{-i \theta_1^\nu} & b^\nu & r_3^\nu e^{i\theta_3^\nu}\\
r_1^\nu e^{-i \theta_1^\nu} & r_3^\nu e^{-i\theta_3^\nu} & b^\nu\\
\end{bmatrix}.
\label{Texture}
\end{equation}

\begin{table}
\centering
\begin{tabular*}{\textwidth}{@{\extracolsep{\fill}} ccc}
\hline
\hline
Parameters & Model best fit value &  $\chi^2_{min}$ \\
\hline
\hline
$\theta_{12} /^\circ$  & 33.76  &  $5.76 \times 10^{-6}$ \\
\hline
$\theta_{13} /^\circ$  & 8.61 & , $6.05 \times 10^{-8}$  \\
\hline
$\theta_{23} /^\circ$ & 43.27 & $4.49 \times 10^{-6}$ \\
\hline
$\delta /^\circ$ & 203.37 & 0.025 \\
\hline
$\Delta m_{21}^2$/eV$^2$ &  $7.54 \times 10^{-5}$  & $1.39 \times 10^{-5}$ \\
\hline
$\Delta m_{31}^2$/eV$^2$ & $2.52 \times 10^{-3}$ & $4.16 \times 10^{-5}$ \\
\hline
\end{tabular*}
\caption{Shows the best fit points from the reduced texture limits for NH.}
\label{chisquare1}
\end{table}

\begin{table}
\centering
\begin{tabular*}{\textwidth}{@{\extracolsep{\fill}} ccc}
\hline
\hline
Parameters & Model best fit value &  $ \chi^2_{min}$ \\
\hline
\hline
$\theta_{12} /^\circ$  & 33.76, & $1.93 \times 10^{-6}$ \\
\hline
$\theta_{13} /^\circ$  & 8.65  & $1.15 \times 10^{-5}$  \\
\hline
$\theta_{23} /^\circ$ & 45.68 & 7.22 \\
\hline
$\delta /^\circ$ & 283.20 & $6.99 \times 10^{-5}$ \\
\hline
$\Delta m_{21}^2$/eV$^2$ & $7.54 \times 10^{-5}$ & $5.01 \times 10^{-7}$ \\
\hline
$\Delta m_{31}^2$/eV$^2$ & -$2.53 \times 10^{-3}$  & $6.71 \times 10^{-9}$  \\
\hline
\end{tabular*}
\caption{Shows the best fit points from the reduced texture limits for IH.}
\label{chisquare2}
\end{table}

It is important to mention that the $M^R_\nu$ with six real parameters is now predictive. It significantly reduces the overall parameter space of the general Hermitian texture and addresses two key issues. For NH, we observe that the $M^R_\nu$ is consistent with the lower octant of $\theta_{23}$ and suggests a forbidden region for the Dirac CP phase $\delta$. For IH, the $M^R_\nu$ favours the higher octant of $\theta_{23}$. In this regard, the $ \chi^2$ analysis becomes relevant for $M^R_\nu$ for both NH and IH. We calculate $ \chi^2$ for the physical parameters with respect to their corresponding best-fit values using the following formula,

\begin{equation}
\chi^2 = \sum_i \left( \frac{O_i^{\mathrm{th}} - O_i^{\mathrm{exp}}}{\sigma_i} \right)^2,
\end{equation}

where, $O_i^{\mathrm{th}}$ represents the dataset corresponds to reduced texture limits, $O_i^{\mathrm{exp}}$ denotes the experimental best-fit value and $\sigma_i$ stands for the corresponding uncertainty at the $1\sigma$ confidence level. The dataset of $M^R_\nu$ exhibits a good fit for all observables except $\theta_{23}$ in the case of IH. The matrix $M^R_\nu$ predicts a narrow range $(45.08^\circ-45.68^\circ)$, which lies away from the experimental $1\sigma$ best-fit value $48.15^\circ$ for IH\,\cite{nufit}, resulting in an approximately linear $\chi^2$ profile. The graphical visualization can be found in Figs.\,(\ref{fig:chinh})-(\ref{fig:chiih}). For both NH and IH, we list the best-fit points obtained from $M_\nu^R$, corresponding to the minimum values of the $\chi^2$ function, ($\chi^2_{\min}$) in Table\,(\ref{chisquare1})-(\ref{chisquare2}).

In the next section, we address the naturalness of the Yukawa couplings.

\begin{figure}
  \centering
    \subfigure[]{\includegraphics[width=0.47\textwidth]{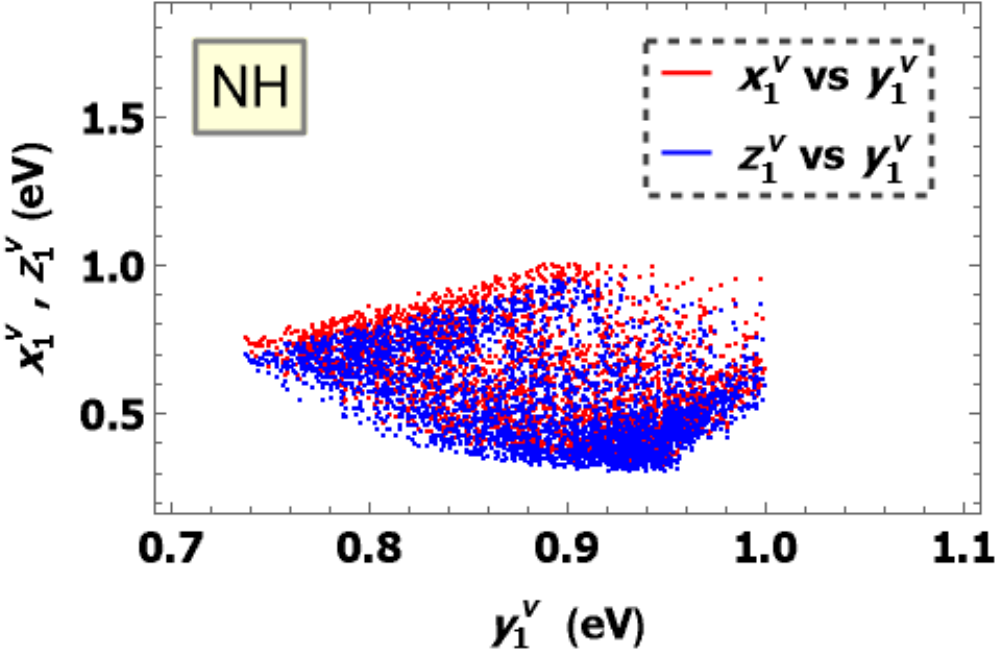}} 
    \subfigure[]{\includegraphics[width=0.49\textwidth]{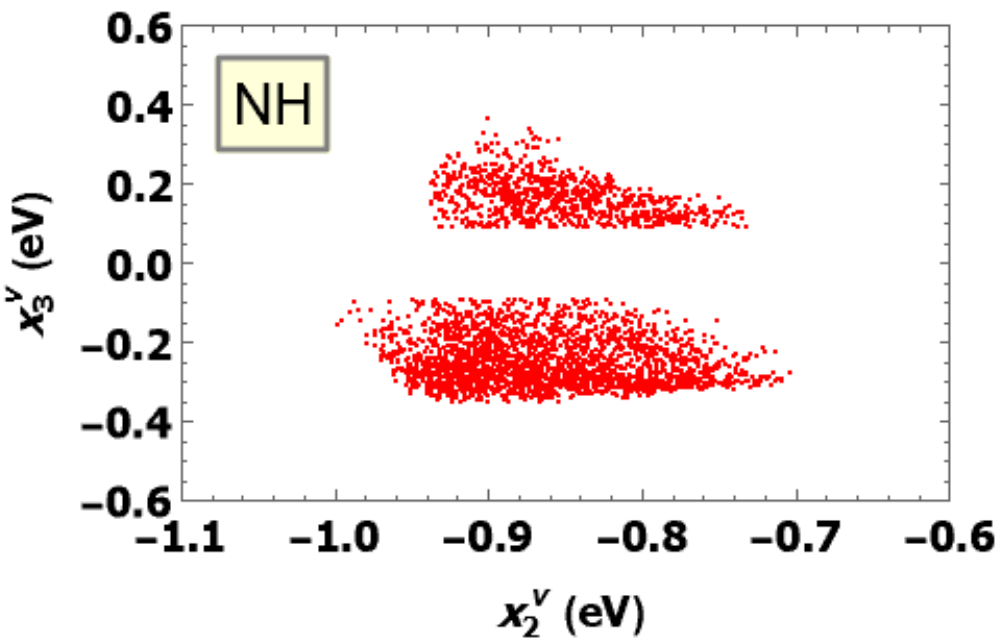}}\\
    \subfigure[]{\includegraphics[width=0.49\textwidth]{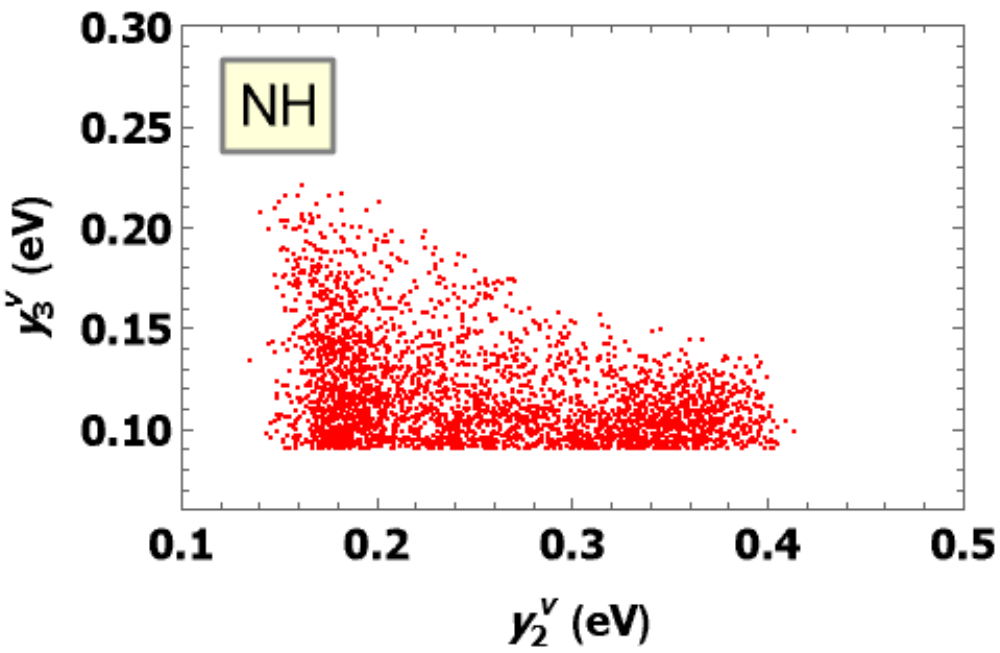}} 
    \subfigure[]{\includegraphics[width=0.49\textwidth]{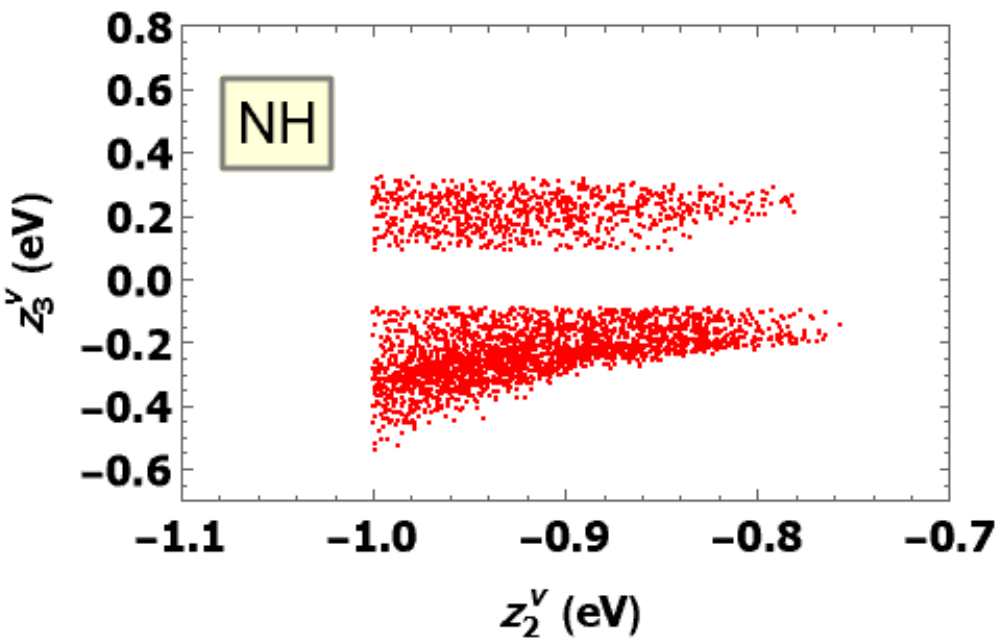}} 
   \caption{The correlation plots between the Yukawa couplings (a) $y_1^\nu$ vs $x_1^\nu$ and $z_1^\nu$, (b) $x_2^\nu$ vs $x_3^\nu$, (c) $y_2^\nu$ vs $y_3^\nu$, (d) $z_2^\nu$ vs $z_3^\nu$ for NH. It is seen that naturalness of the Yukawa coupling is maintained.}
\label{YC1}
\end{figure}

\begin{figure}
  \centering
    \subfigure[]{\includegraphics[width=0.47\textwidth]{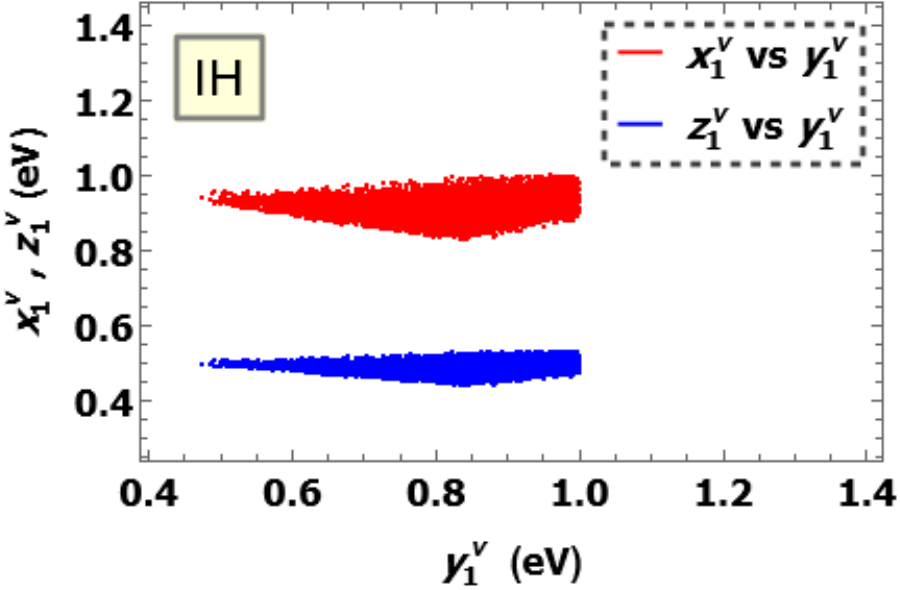}} 
    \subfigure[]{\includegraphics[width=0.49\textwidth]{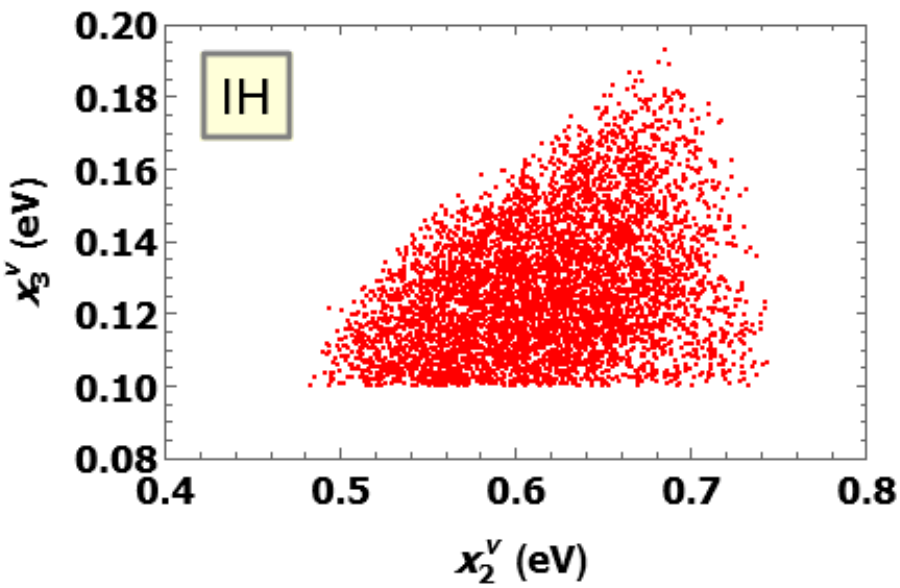}}\\
    \subfigure[]{\includegraphics[width=0.49\textwidth]{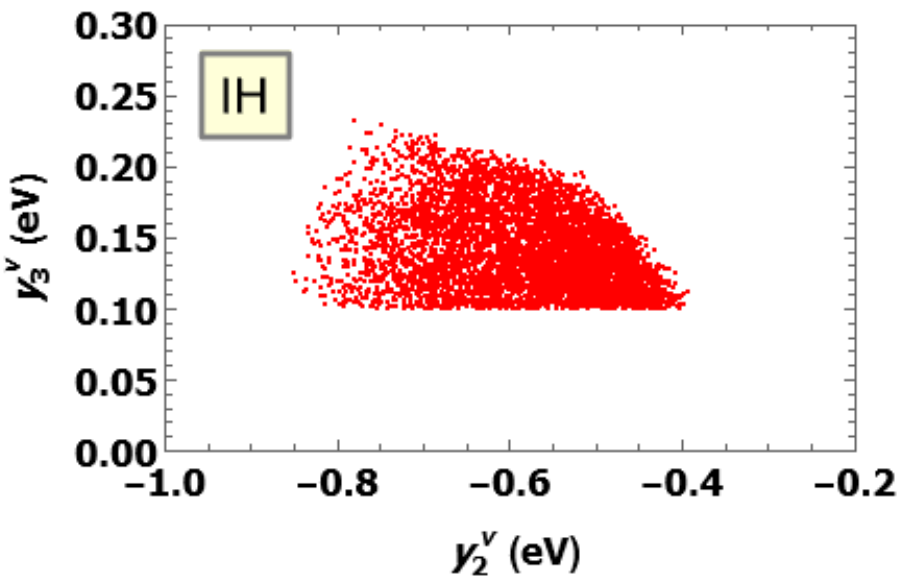}} 
    \subfigure[]{\includegraphics[width=0.49\textwidth]{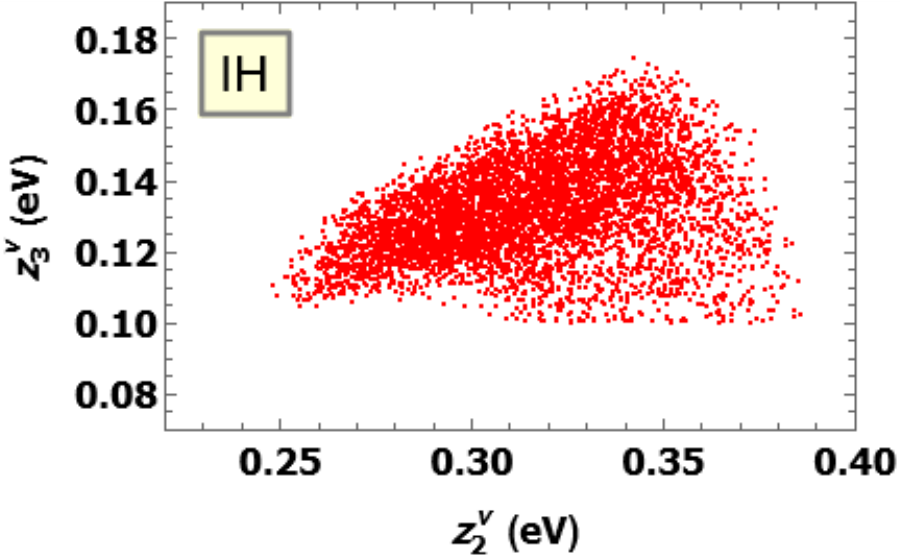}} 
   \caption{The correlation plots between the Yukawa couplings (a) $y_1^\nu$ vs $x_1^\nu$ and $z_1^\nu$, (b) $x_2^\nu$ vs $x_3^\nu$, (c) $y_2^\nu$ vs $y_3^\nu$, (d) $z_2^\nu$ vs $z_3^\nu$ for IH. It is seen that naturalness of the Yukawa coupling is maintained.}
\label{YC2}
\end{figure}

\subsection{Naturalness of the Yukawa Couplings \label{sec4}}

In the present framework, the Yukawa couplings in the neutrino sector are expected to be of $\sim \mathcal{O}(1)$. This is desirable from both theoretical and phenomenological considerations. From a theoretical point of view, assuming $\mathcal{O}(1)$ Yukawa couplings ensures naturalness by avoiding fine-tuning and unexplained hierarchies among fundamental parameters. Since the smallness of neutrino masses is already accounted for by the suppression due to a high seesaw scale, there is no compelling need to invoke tiny Yukawa couplings.

In this context, we obtain the ratio between the Yukawa couplings of the two sectors as shown below:

\begin{equation}
\frac{\Upsilon_i^\nu}{\Upsilon_i^d}=\frac{m^\nu_{ij}}{\gamma^\nu m^d_{ij}}, 
\end{equation}

where,
\begin{eqnarray}
m^\nu_{ij}&=& f(\theta_{ij}^\nu, m_i^\nu, \delta^\nu) \quad \text{and} \quad m^d_{ij}= f(\theta_{ij}^q, m_i^q, \delta^q).
\end{eqnarray}

The ratio $\gamma^\nu$ is a very small dimensionless parameter $\sim10^{-11}$. From the quark sector, we estimate $\Sigma_i$'s $\sim 10^9$ eV and $\Upsilon_i^d \sim \mathcal{O}(1)$. Given that the Yukawa couplings in the down-type quark sector are of $\mathcal{O}(1)$, it is quite natural to expect similar magnitudes in the neutrino sector to preserve the structural correlation. In this regard, we perform a numerical scan to identify the allowed parameter space for the neutrino Yukawa couplings that yield physical observables consistent with experimental data.\,(see Figs.\,(\ref{YC1})-(\ref{YC2})). The other model parameters align as: $\Lambda \sim 10^{14}$\,GeV, $v_{\eta} \sim 10^{12}$\,GeV, $v_{\kappa} \sim 10^{13}$\,GeV, $v_{\xi} \sim 10^{12}$\,GeV, $v_{\rho} \sim 10^{12}$\,GeV, $v_{\sigma} \sim 10^{12}$\,GeV, $v_{\phi} \sim 10^{11}$\,GeV, $v_{\chi} \sim 10^{13}$\,GeV and $v_H = 246$\,GeV respectively. 

In the next section, we shall study some phenomenological constrains of the proposed model.

\section{Phenomenological Constraints \label{sec6}}

In this section, we study some important phenomenological consequences of our model and compare them with current experimental limits. Our analysis includes charged lepton flavour violation\,(CLFV)\,\cite{Branco:2011iw} and  possible nonunitarity effects in the leptonic mixing matrix\,\cite{Antusch:2006vwa}. In BSM models, the mixing between the active neutrinos and heavy neutral fermions can induce flavour violating processes such as $\mu \rightarrow e\gamma$, $\tau \rightarrow e\gamma$ and $\tau \rightarrow \mu \gamma$ while simultaneously leading to deviations from the exact unitarity of the PMNS matrix. These effects are constrained by several experimental observations and therefore provide important complementary probes of the underlying model parameters. We therefore examine the implications of the model for CLFV observables and nonunitarity constraints in the following discussion.

\subsection{Charged Lepton Flavour Violation}
In this section, we study various possible CLFV processes allowed by the proposed model. One of the possible sources of possible CLFV are the gauge interaction Lagrangian ( $-\mathcal{L}_W=(g/\sqrt{2})\bar{l}_L \gamma^{\mu} \nu_L W_{\mu} + \text{h.c.}$) enabled by the neutrino sector of the Yukawa Lagrangian through Type-I Dirac seesaw mechanism. The CLFV processes originating from the interaction Lagrangian have the gauge boson, $W^\pm$ as the mediator. 

\begin{figure}
  \centering
    \subfigure[]{\includegraphics[width=0.49\textwidth]{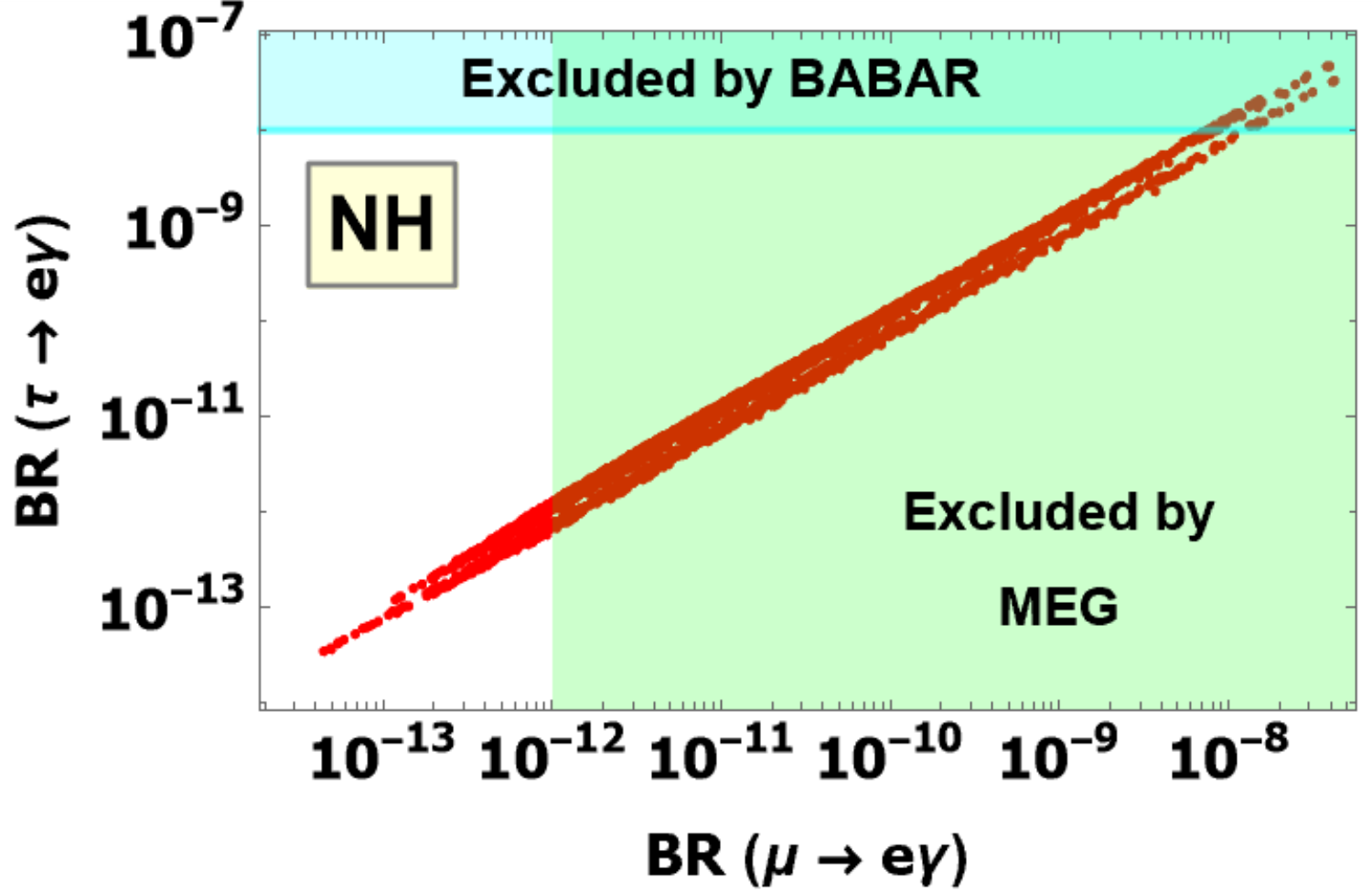}
   } 
    \subfigure[]{\includegraphics[width=0.49\textwidth]{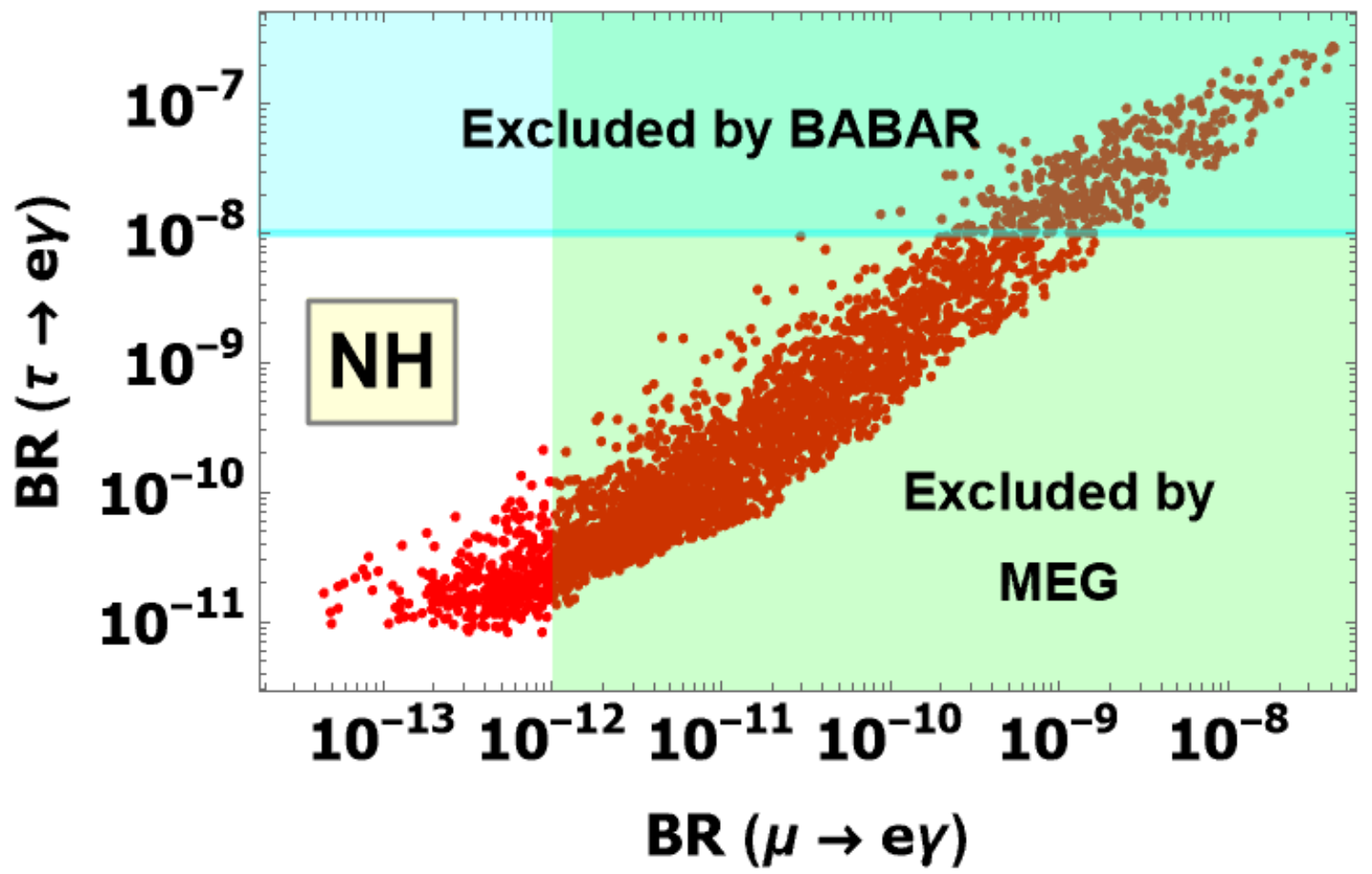}}\\
    \subfigure[]{\includegraphics[width=0.49\textwidth]{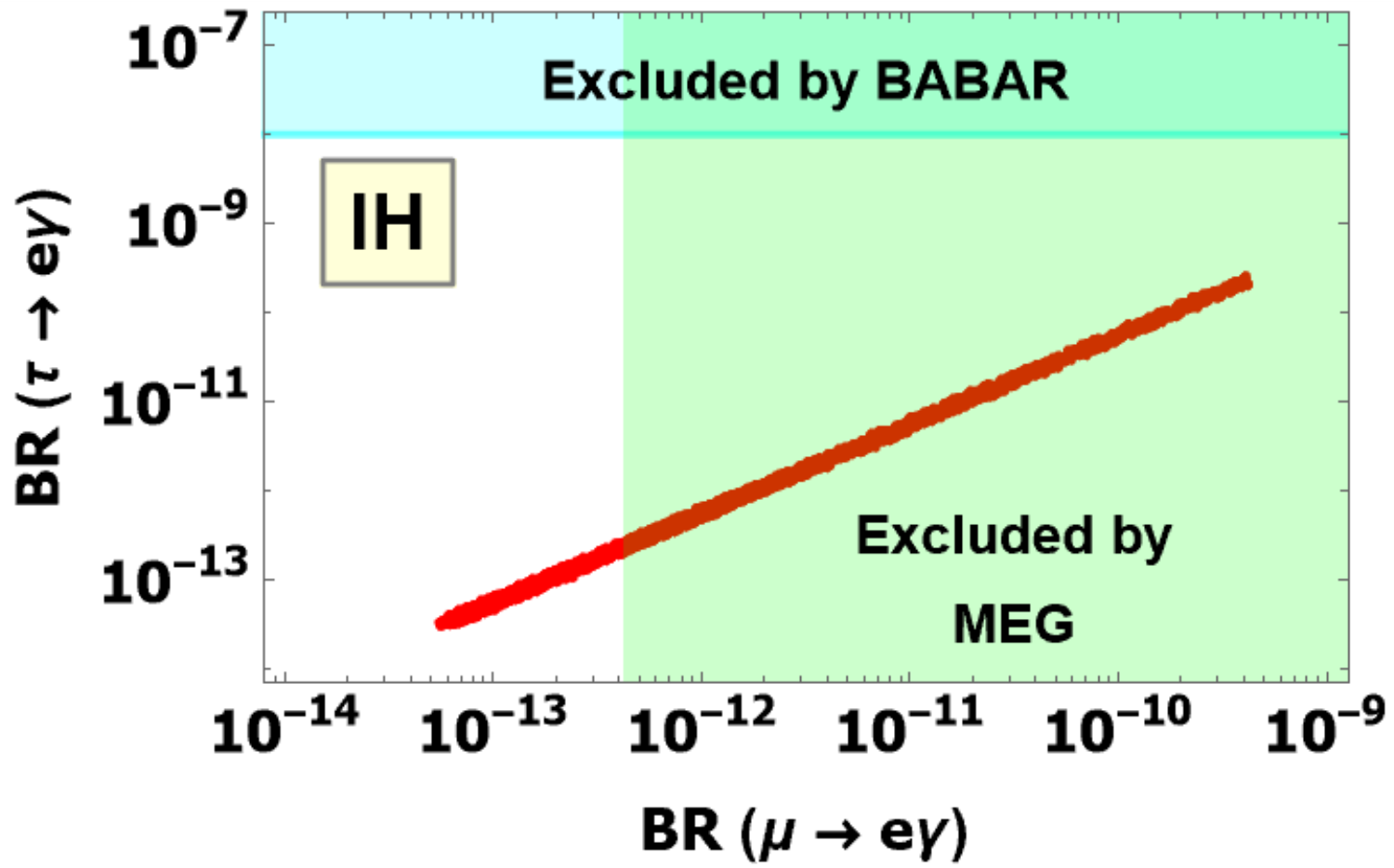}} 
    \subfigure[]{\includegraphics[width=0.49\textwidth]{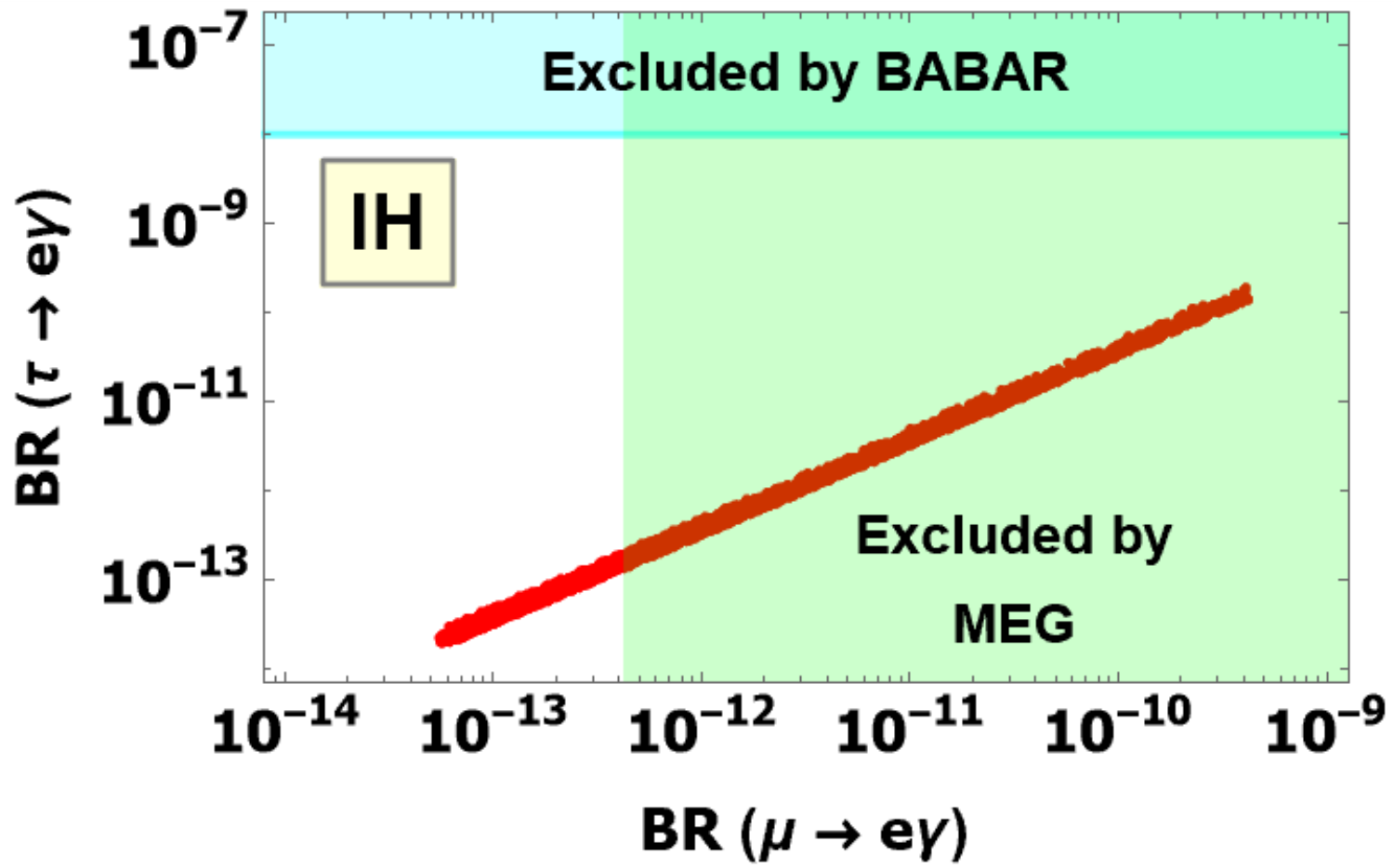}} 
   \caption{The study of CLFV branching ratios through $\mu \rightarrow e\gamma$, $\tau \rightarrow e\gamma$ and $\tau \rightarrow \mu \gamma$ channels for both NH and IH. It is seen that the prediction for BR\,($\mu \rightarrow e\gamma$) is partially excluded by MEG and can be tested in ongoing MEG\,II experiment.}
\label{clfv}
\end{figure}

In our model, the charged Higgs scalar field, $H^+$ is a would be Goldstone boson, ultimately `eaten' by the charged gauge boson, $W^+$ via the Higgs mechanism. The $W^+$ boson connects a left handed (LH) charged lepton to a LH neutrino via the interaction Lagrangian. The branching ratio of a CLFV process with a light neutrino as a mediator is very low (e.g., $BR(\mu \to e \gamma) \approx 10^{-54}$\,\cite{ParticleDataGroup:2024cfk}. However, in Type-I Dirac seesaw, we can connect $\nu_L$ to a heavy right handed neutrinos with a mixing, $\Theta \approx B M^{-1}$. Neglecting the highly suppressed light-neutrino contribution and retaining only the heavy-neutrino contribution, the approximate formula looks like the following,
\begin{align}
\label{eqlalbg}
\mathrm{BR}(\ell_\alpha \to \ell_\beta \gamma)
\approx
\frac{3\alpha}{32\pi}
\left|
\sum_i
\Theta_{\alpha i}\,
\Theta^*_{\beta i}\,
F\!\left(\frac{M_i^2}{M_W^2}\right)
\right|^2,
\end{align}
where $M_i$ is the mass of $i$th heavy neutrino and the function, $F(x) \approx4/3$ for $M_i >> M_W$.  With $B,\,M$ and $U_L$ in our model as shown in Eq.\,(\ref{BMD}) and (\ref{UL}), all the non-diagonal elements of $\Theta \Theta^\dagger$ survive in flavour basis. In this regard, all three $\mu \rightarrow e\gamma$, $\tau \rightarrow e\gamma$ and $\tau \rightarrow \mu \gamma$ channels contribute to CLFV. From our analysis, it is seen that for NH, the prediction of CLFV branching ratios from our model are partially excluded by MEG and BABAR experiments\,\cite{MEG:2016leq, BaBar:2009hkt}. On the other hand, the model predictions are partially discarded by MEG. The allowed predictions can be tested by the ongoing experiment MEG II\,\cite{MEGII:2023ltw}. The graphical analysis can be found in Fig.\,(\ref{clfv}).

In the next section, we shall investigate the effective nonunitary of the PMNS matrix.

\subsection{Nonunitary of the Lepton Mixing matrix}

In general, the extensions of the SM involving additional neutral fermionic states can lead to deviations from the exact unitarity of the leptonic mixing matrix. In the conventional three generation framework, the PMNS matrix remains unitary due to the presence of only three active light neutrinos. However, the BSM models contain heavy neutral fermions, the mixing between the active and heavy sectors modifies the effective low-energy leptonic mixing matrix, thereby inducing nonunitarity effects. 

In the present type-I Dirac seesaw framework, the nonunitarity effects originate from the mixing between the active neutrinos and the heavy singlet fermions. The active-heavy mixing matrix is then approximately $\Theta \simeq B M^{-1}$. As a consequence of this mixing, the effective leptonic mixing matrix associated with the light neutrinos deviates from exact unitarity and can be written as

\begin{table}
\centering
\begin{tabular*}{\textwidth}{@{\extracolsep{\fill}} ccccc}
\hline
\hline
Parameters/ & Min Value &  Max Value & Min Value & Max Value\\
Values & (NH) & (NH) & (IH) & (IH)\\
\hline
\hline
$|\eta_{11}|$ & $2.31 \times 10^{-7}$  & $2.91 \times 10^{-3}$ & $6.42 \times 10^{-5}$ & $5.95 \times 10^{-3}$\\
\hline
$|\eta_{22}|$ & $7.84 \times 10^{-5}$ & $1.46 \times 10^{-2}$ & $4.29 \times 10^{-5}$ & $4.21 \times 10^{-3}$ \\
\hline
$|\eta_{33}|$ & $6.01 \times 10^{-5}$ & $1.58 \times 10^{-2}$  & $2.36 \times 10^{-5}$ & $2.59 \times 10^{-3}$\\
\hline
$|\eta_{12}|$ & $4.67 \times 10^{-6}$ & $5.97 \times 10^{-3}$  & $5.38 \times 10^{-5}$ & $4.81 \times 10^{-3}$\\
\hline
$|\eta_{13}|$ & $4.14 \times 10^{-6}$ & $6.78 \times 10^{-3}$  & $4.01 \times 10^{-5}$ & $3.76 \times 10^{-3}$\\
\hline
$|\eta_{23}|$ & $7.12 \times 10^{-5}$ & $1.39 \times 10^{-2}$  & $3.29 \times 10^{-5}$ & $3.31 \times 10^{-3}$\\
\hline
\end{tabular*}
\caption{Shows the maximum and minimum values of the parameter $|\eta_{\alpha \beta}|$ predicted by the present model. It is seen that the predicted values align with the experimental observation.}
\label{nonunitarity}
\end{table}

\begin{equation}
N \simeq \left(I-\frac{1}{2}\Theta\Theta^\dagger\right)U_{\rm PMNS},
\end{equation}
where, $U_{\rm PMNS}$ denotes the unitary PMNS matrix in the absence of heavy-light mixing. The corresponding nonunitarity parameter is therefore defined as
\begin{equation}
\eta = \frac{1}{2}\Theta\Theta^\dagger.
\end{equation}

\begin{figure}
  \centering
    \subfigure[]{\includegraphics[width=0.46\textwidth]{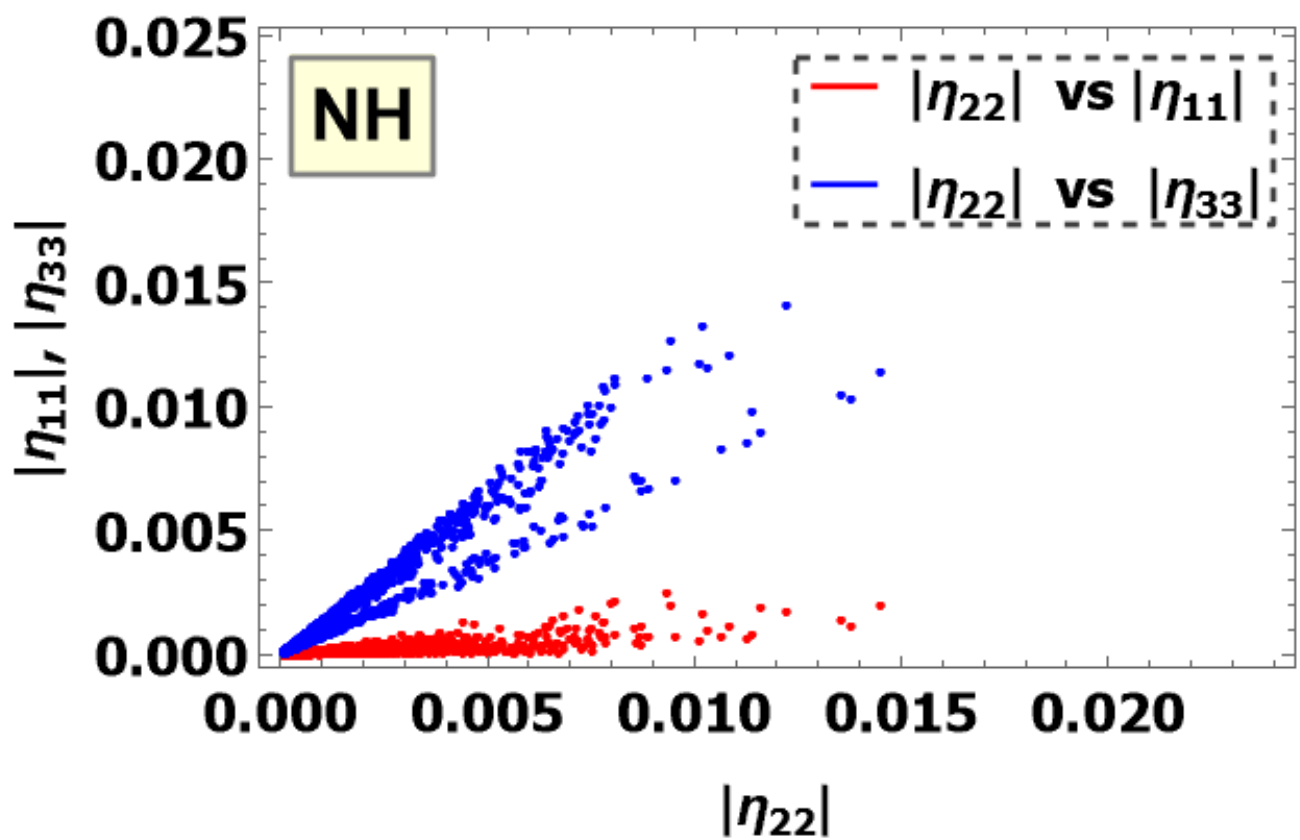}
   } 
    \subfigure[]{\includegraphics[width=0.49\textwidth]{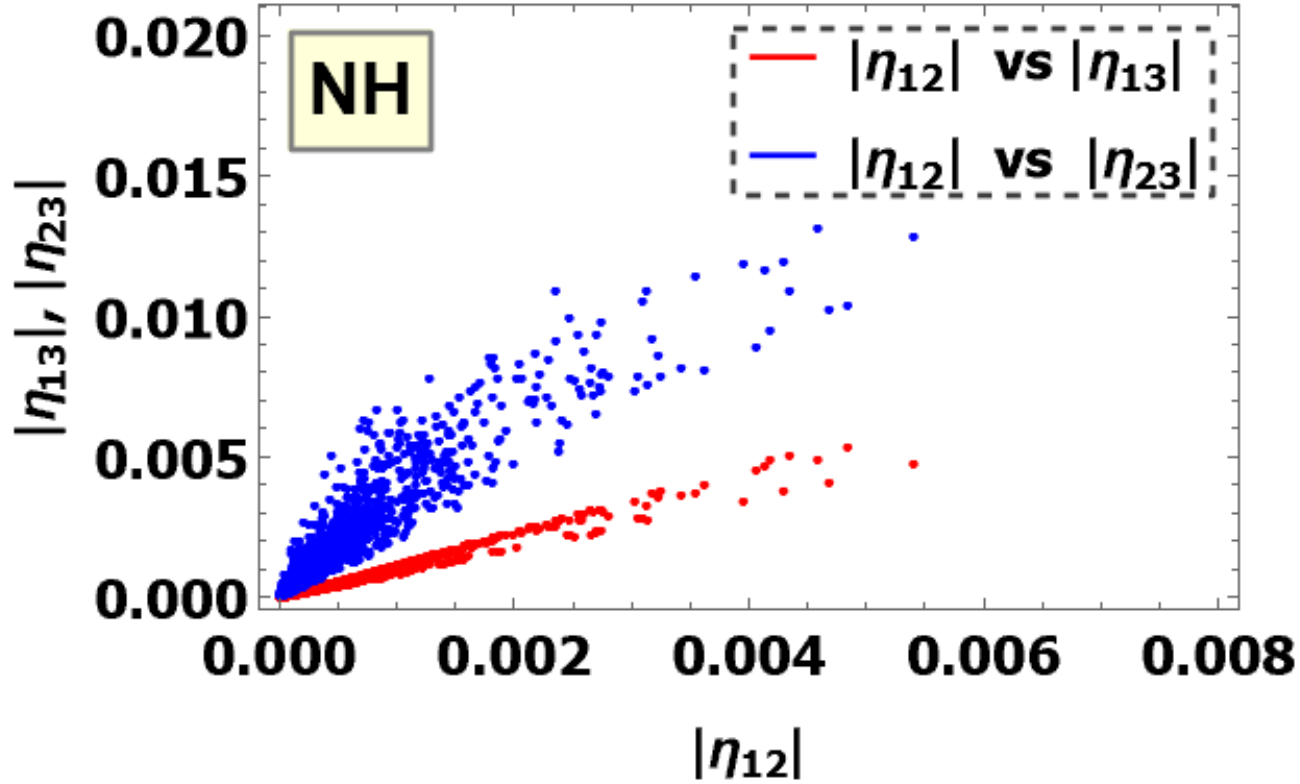}}\\
    \subfigure[]{\includegraphics[width=0.49\textwidth]{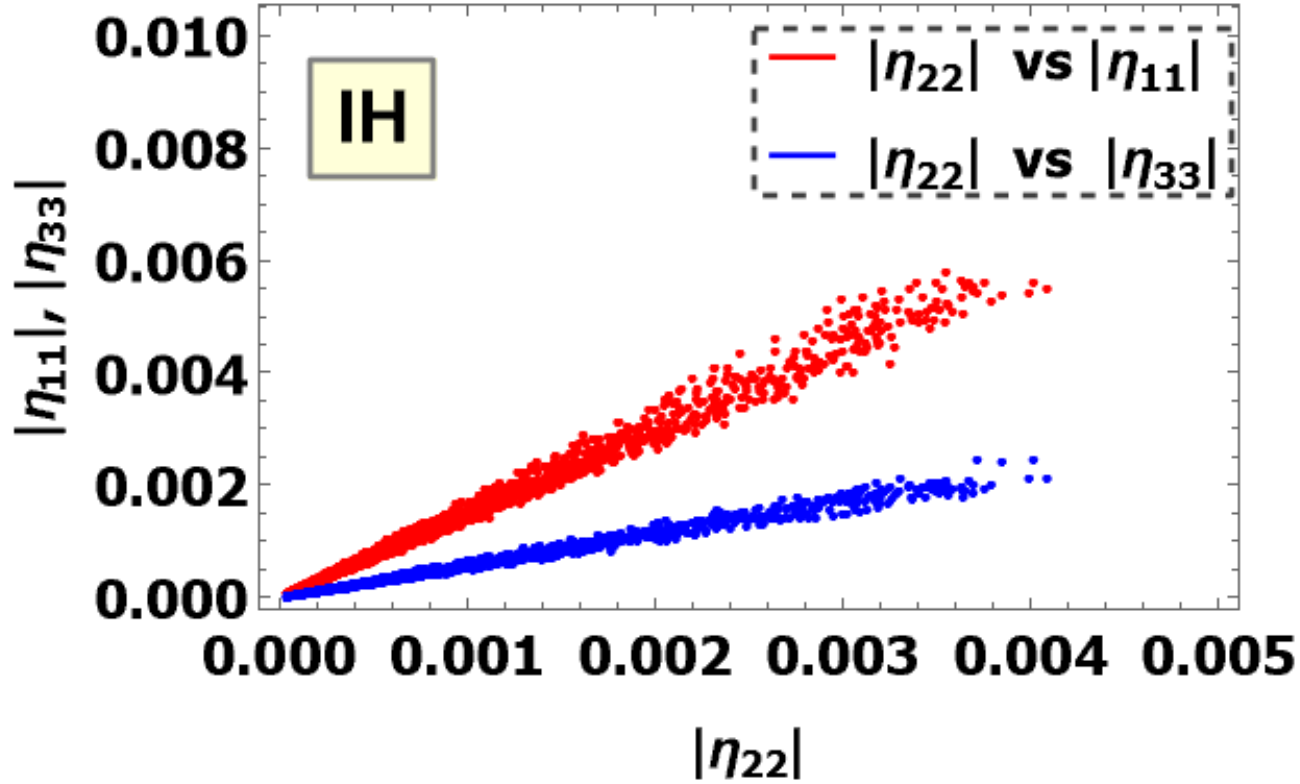}} 
    \subfigure[]{\includegraphics[width=0.49\textwidth]{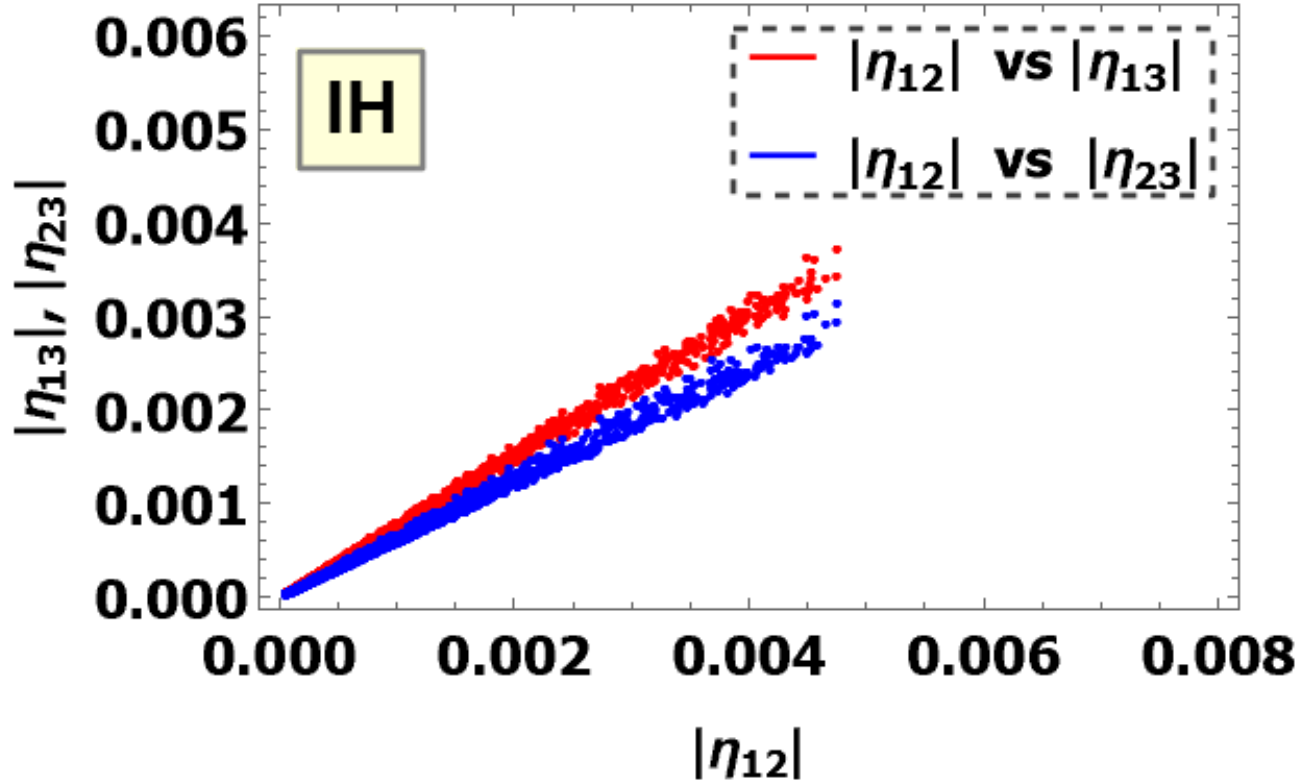}} 
   \caption{The prediction of $|\eta_{\alpha \beta}|$ from the present model for both NH and IH. It is seen that the predicted values of the parameters align with the experimental observation.}
\label{nuparameter}
\end{figure}

The deviations are tightly constrained by electroweak precision data, neutrino oscillation experiments and weak decay observables\,\cite{Fernandez-Martinez:2007iaa}: 

\begin{equation}
|\eta_{\alpha \beta}| <
\begin{pmatrix}
|\eta_{ee}| < 5.5 \times 10^{-3} &
|\eta_{e\mu}| < 3.5 \times 10^{-5} &
|\eta_{e\tau}| < 8.0 \times 10^{-3} \\

|\eta_{\mu e}| < 3.5 \times 10^{-5} &
|\eta_{\mu\mu}| < 5.0 \times 10^{-3} &
|\eta_{\mu\tau}| < 5.1 \times 10^{-3} \\

|\eta_{\tau e}| < 8.0 \times 10^{-3} &
|\eta_{\tau\mu}| < 5.1 \times 10^{-3} &
|\eta_{\tau\tau}| < 5.0 \times 10^{-3}
\end{pmatrix}.
\end{equation}

The off-diagonal entries of $\eta_{\alpha\beta}$ are particularly important since they can induce flavour violating effects, while the diagonal entries modify the normalization of weak interaction couplings. Using the model predicted forms of $B$ and $M$, we explicitly evaluate the active-heavy mixing matrix $\Theta$ and subsequently determine the corresponding nonunitarity parameters.

From our model, we investigate the variation of the parameter $\eta_{\alpha \beta}$ and it is found to be consistent with the experimental bound. The graphical analysis is shown in Fig.\,(\ref{nuparameter}). The maximum and minimum values of the said parameter is listed in Table\,(\ref{nonunitarity}). 

In the next section, we study the stability of the obtained textures under renormalization group evolution\,(RGE).

\section{RGE Stability \label{sec7}}

The RG running equations govern the scale dependence of the parameters appearing in the effective Lagrangian and provide the framework to connect physics defined at a high scale to experimentally measured low-energy observables\,\cite{Antusch:2005gp, Tanimoto:1995bf, Casas:1999tp, Casas:1999ac, King:2000hk}. In flavour models based on discrete symmetries, including the present construction, the Yukawa matrices are generated at the flavour symmetry breaking scale through non renormalizable operators suppressed by the cut off scale $\Lambda$. In this work the flavour textures are defined at the electroweak scale $\mu_0=v\simeq 246~\mathrm{GeV}$ and we test their stability under renormalization group evolution up to the flavour cut-off scale $\Lambda_F \simeq 10^{14}$\,GeV, corresponding to $t_{\max}=\ln(\Lambda_F/\mu_0)\simeq 26.73$. We assume that the heavy states responsible for generating the effective Dirac neutrino sector lie above the cut-off scale $\Lambda_F$. Therefore, between $\mu_0$ and $\Lambda_F$ the running is described by the one-loop SM RGEs supplemented by the Dirac neutrino Yukawa matrix $Y_\nu$, without intermediate threshold matching. At $\mu_0$, we start with the measured values of fermion masses and mixing parameters measured at low energy. For neutrinos, we stick to both NH and IH and consider the best fit values for the parameters\,\cite{nufit}.

We define $t=\ln(\mu/\mu_0)$. The gauge couplings $g_{1,2,3}$ satisfy the following relations: 

\begin{figure}
  \centering
    \subfigure[]{\includegraphics[width=0.3\textwidth]{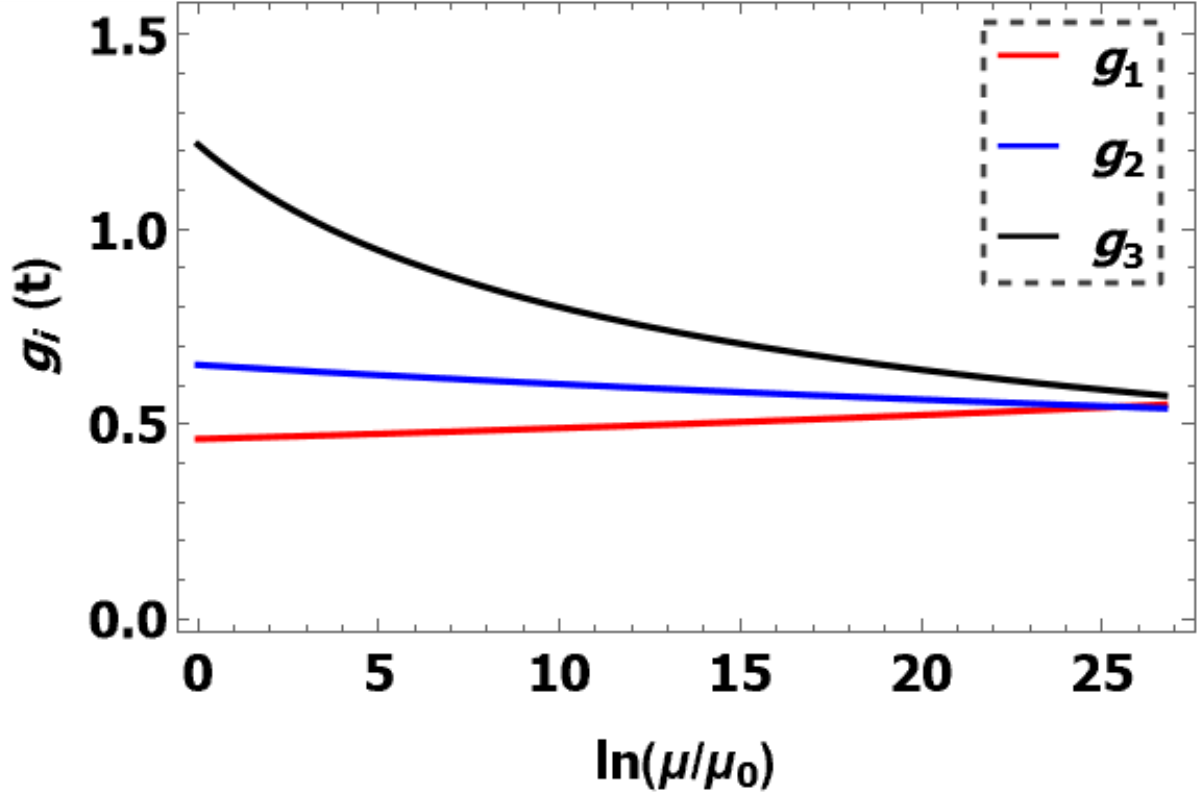}
   \label{g1g2g3}} 
    \subfigure[]{\includegraphics[width=0.33\textwidth]{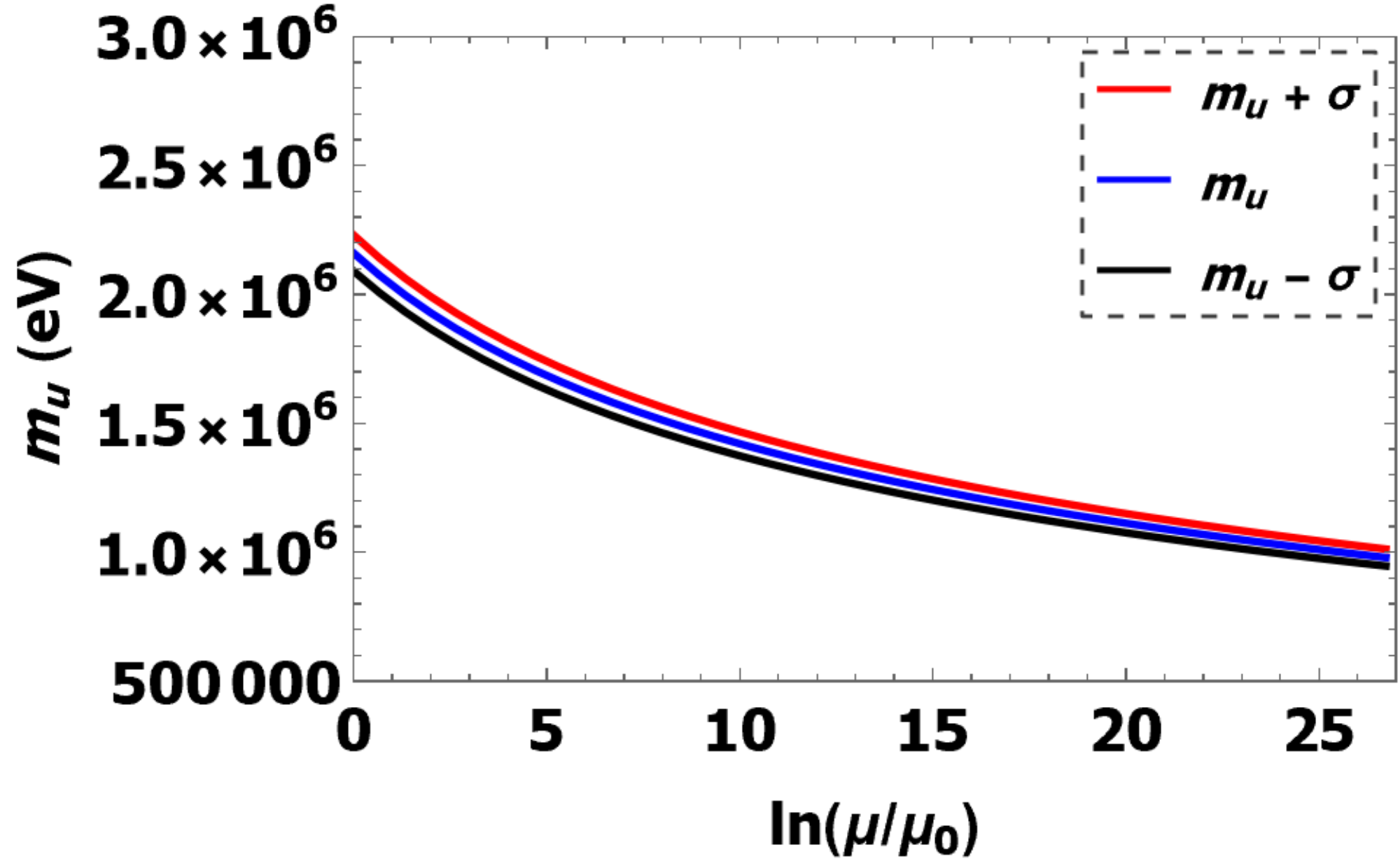}\label{up}}
    \subfigure[]{\includegraphics[width=0.32\textwidth]{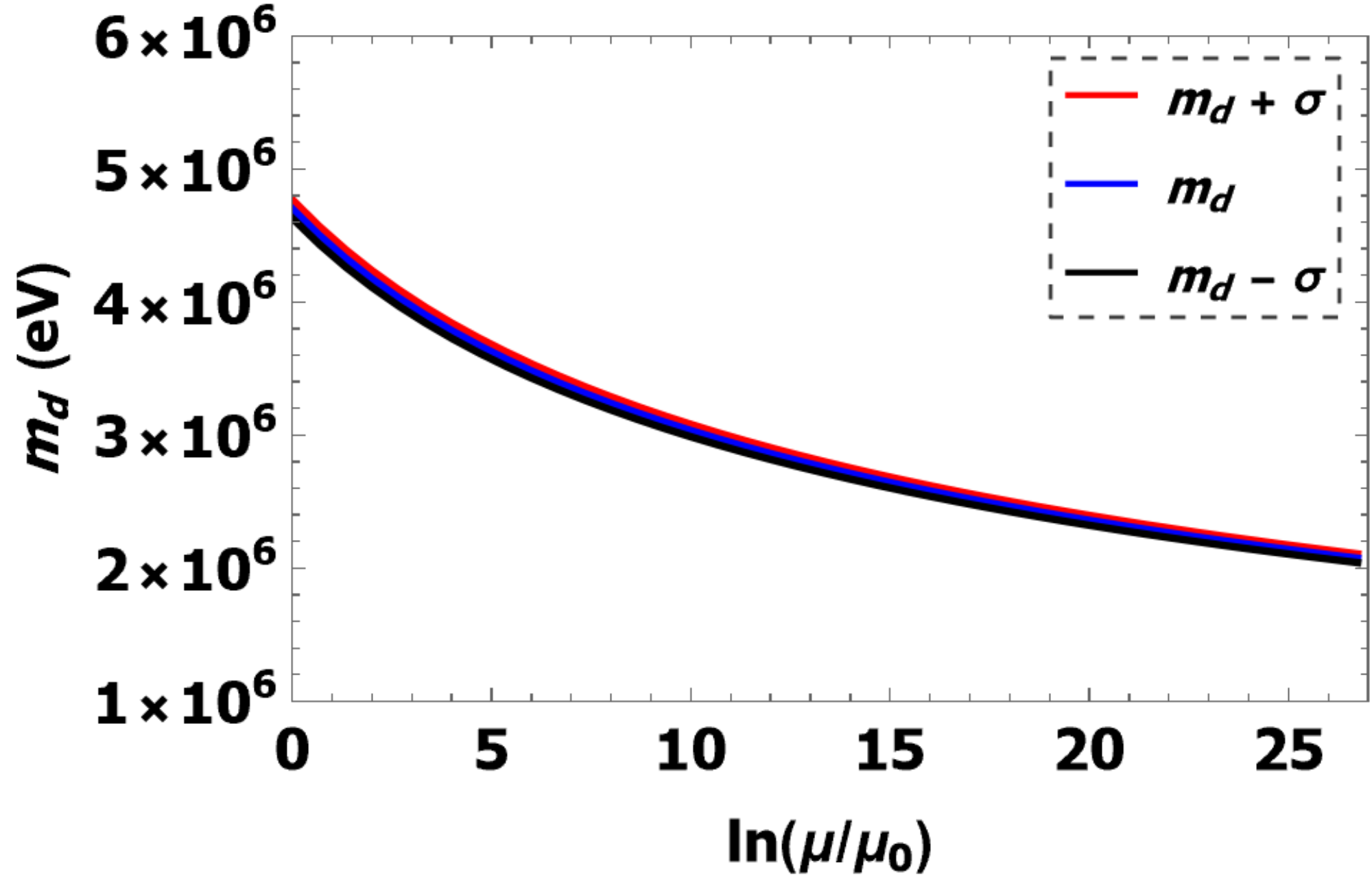}\label{down}} 
    \subfigure[]{\includegraphics[width=0.32\textwidth]{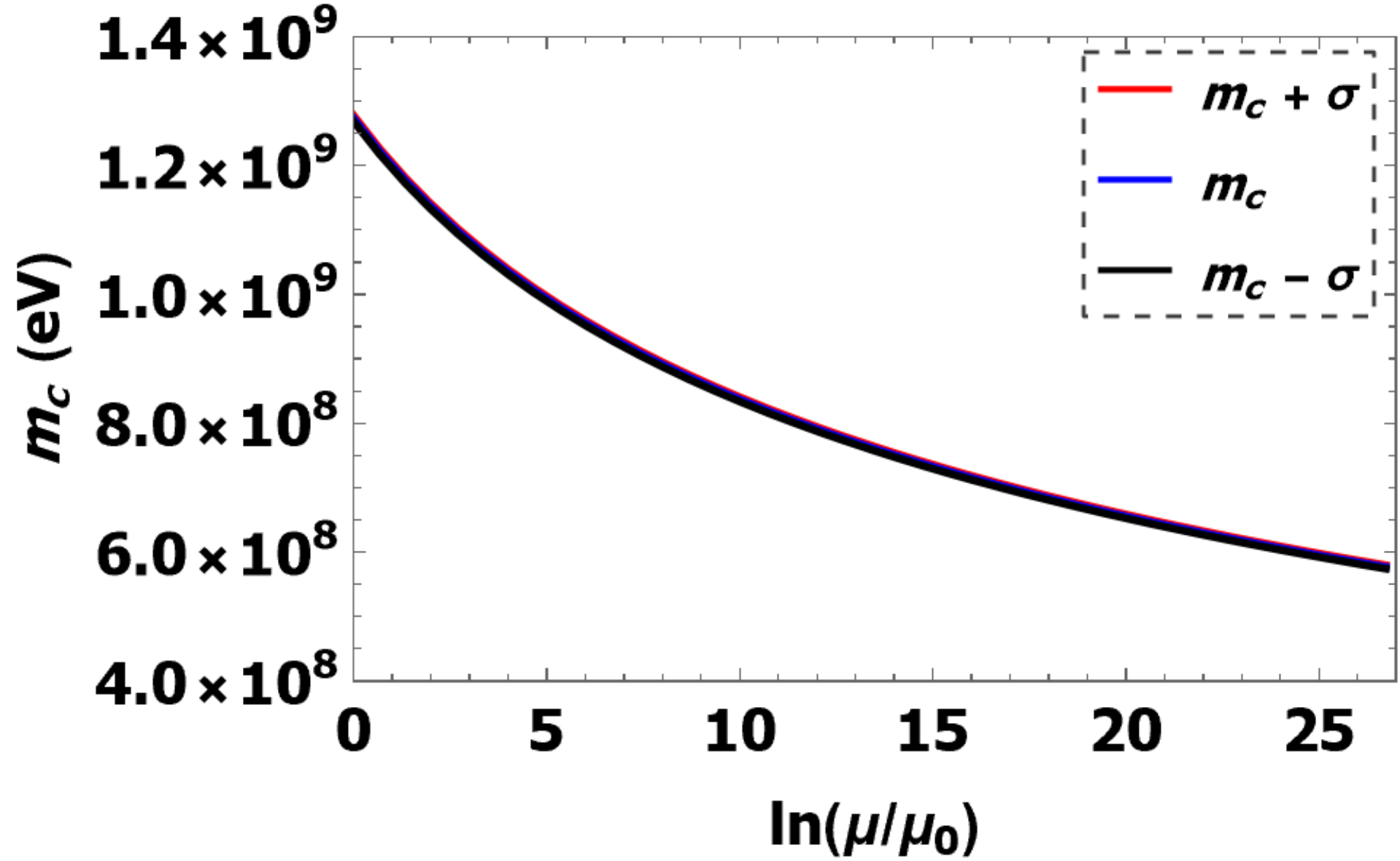}\label{cl}} 
     \subfigure[]{\includegraphics[width=0.32\textwidth]{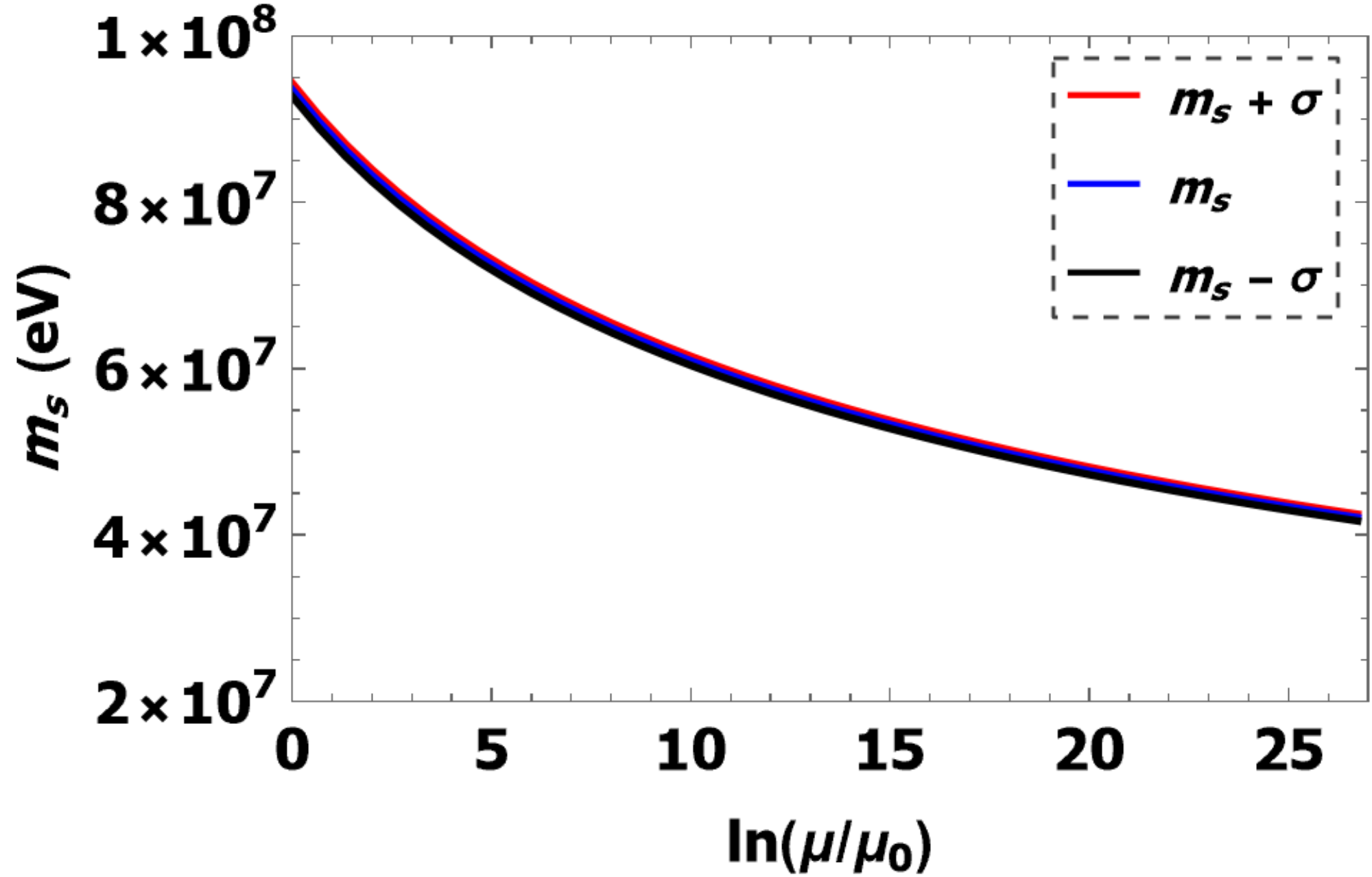}
   \label{g1g2g3}} 
    \subfigure[]{\includegraphics[width=0.32\textwidth]{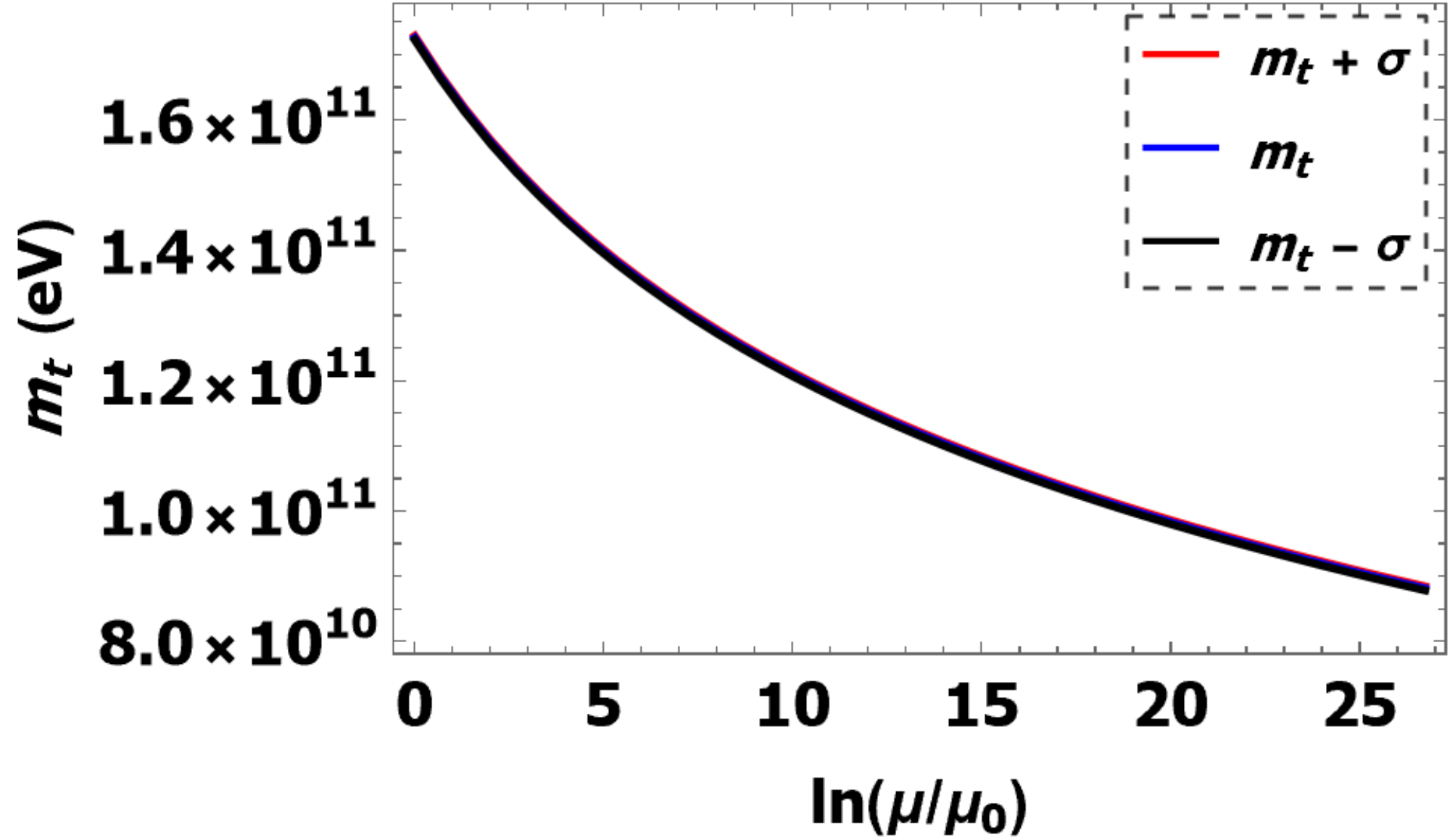}\label{up}}
    \subfigure[]{\includegraphics[width=0.32\textwidth]{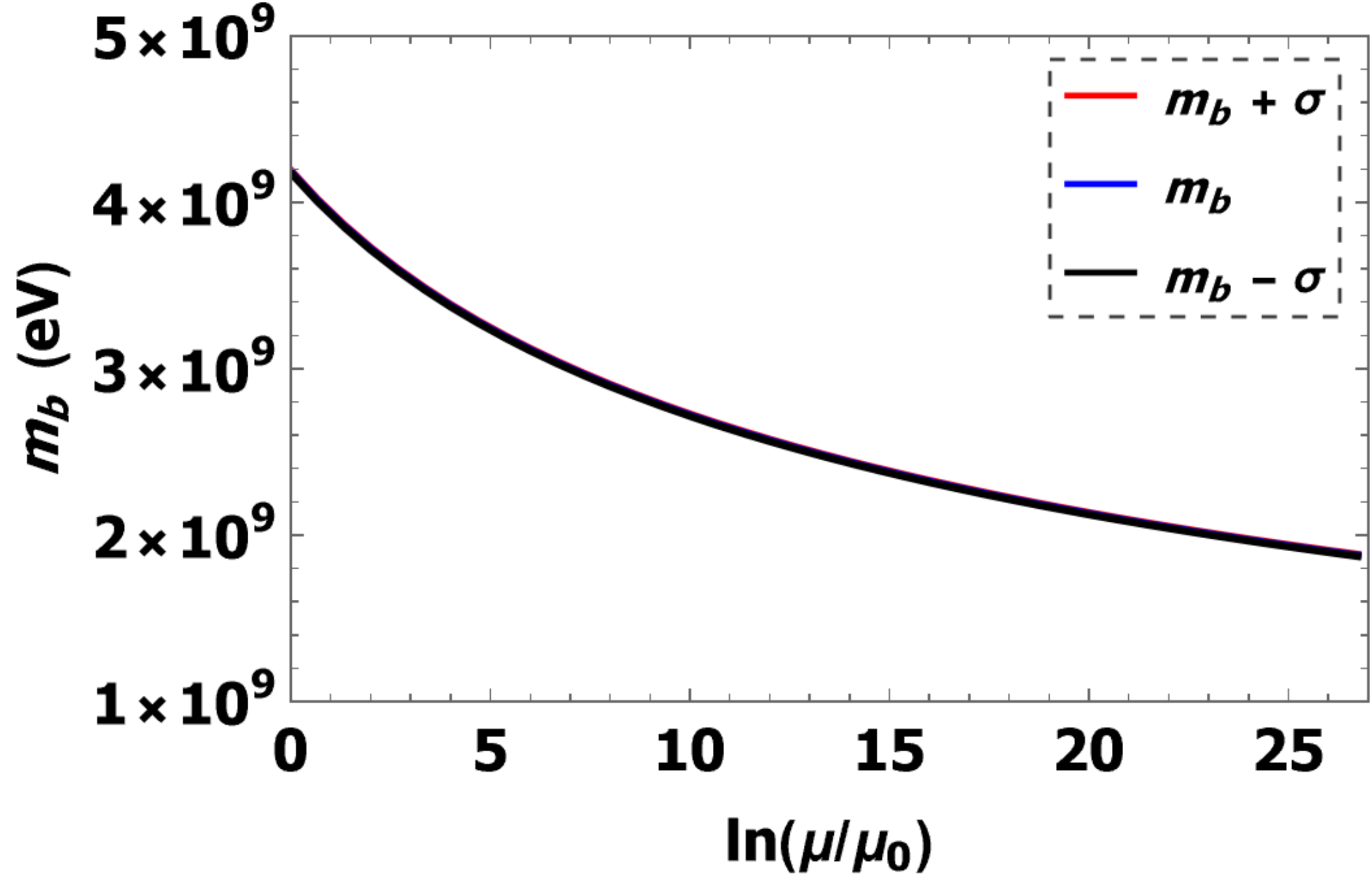}\label{down}} 
    \subfigure[]{\includegraphics[width=0.32\textwidth]{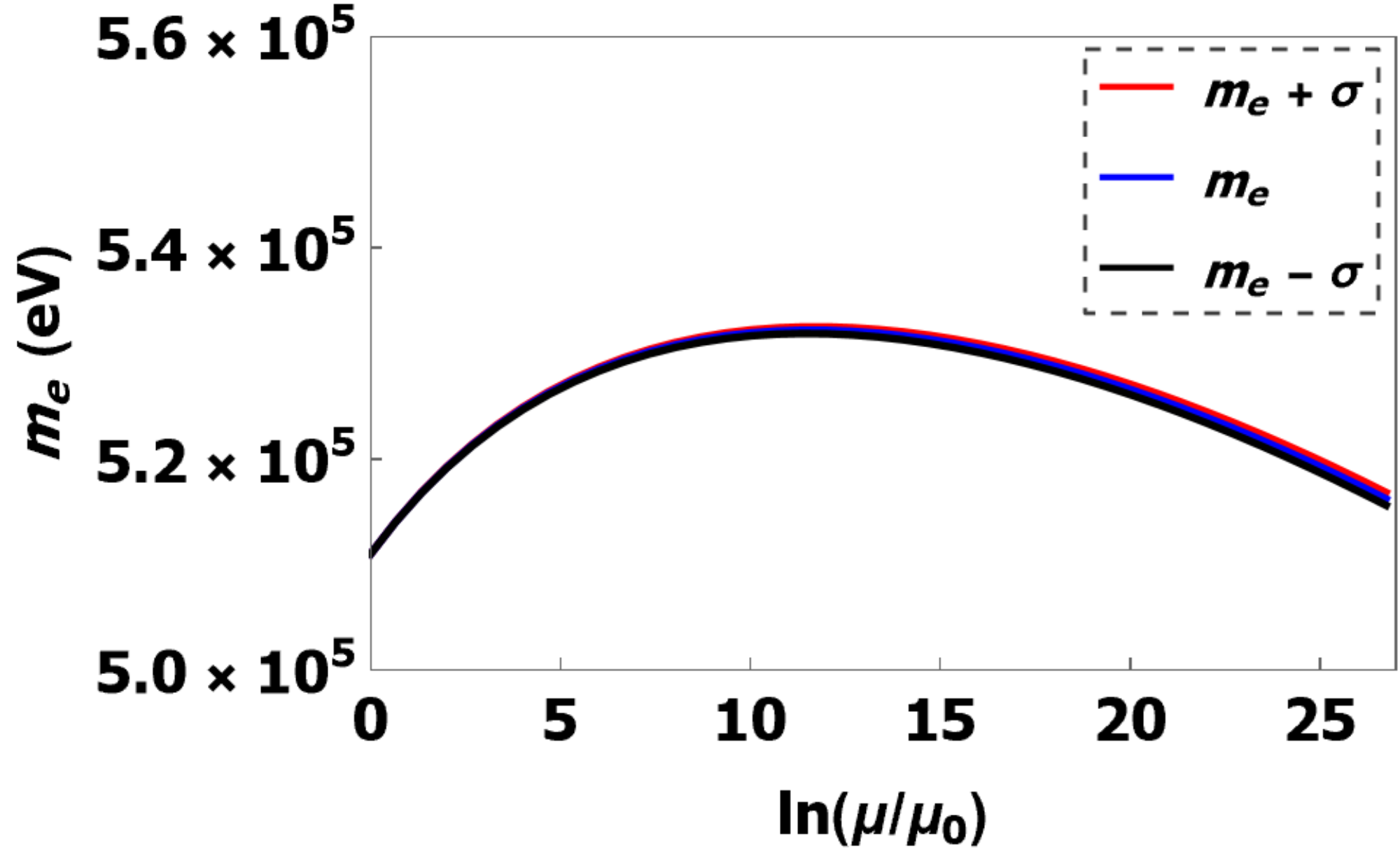}\label{cl}} 
     \subfigure[]{\includegraphics[width=0.32\textwidth]{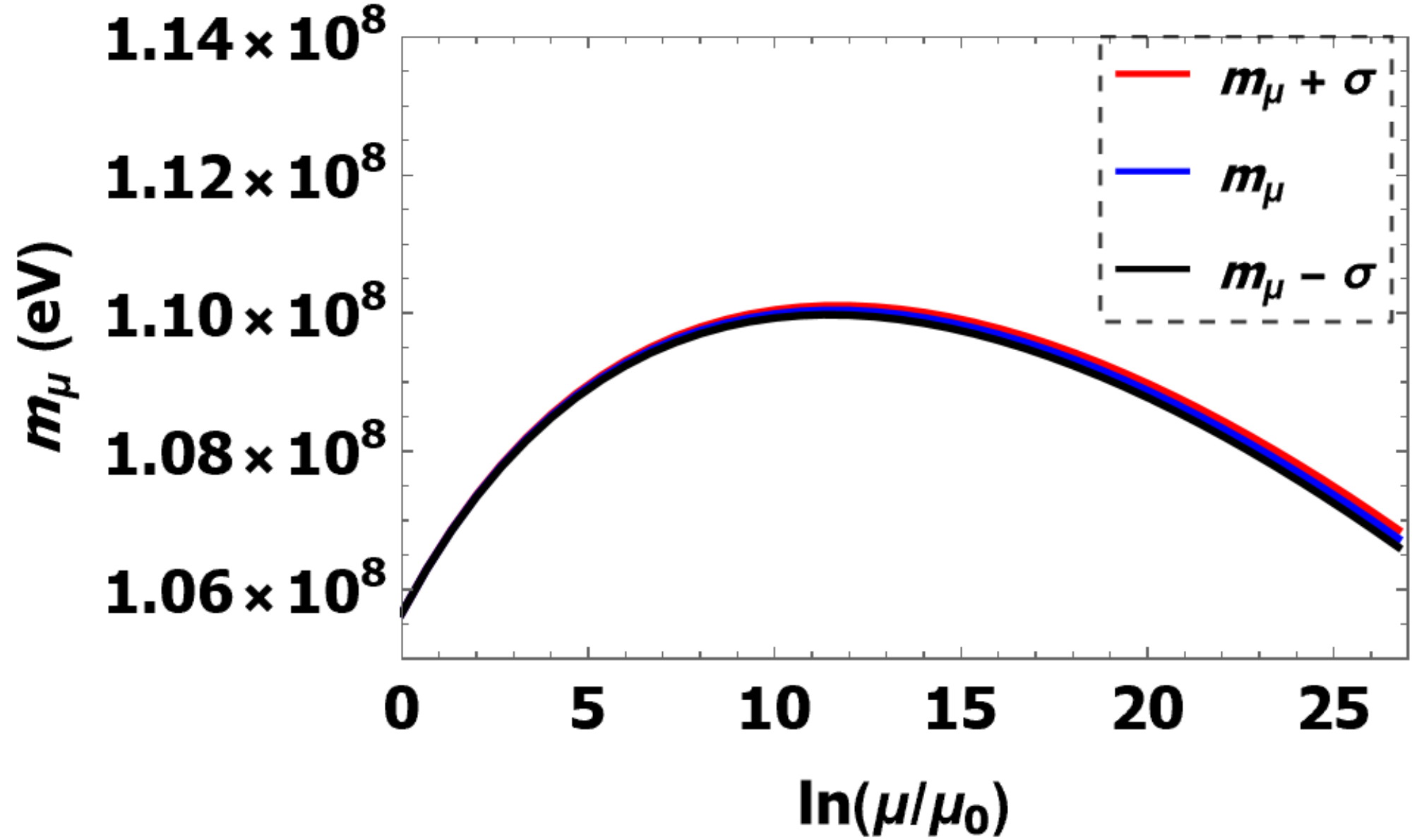}
   \label{g1g2g3}} 
    \subfigure[]{\includegraphics[width=0.32\textwidth]{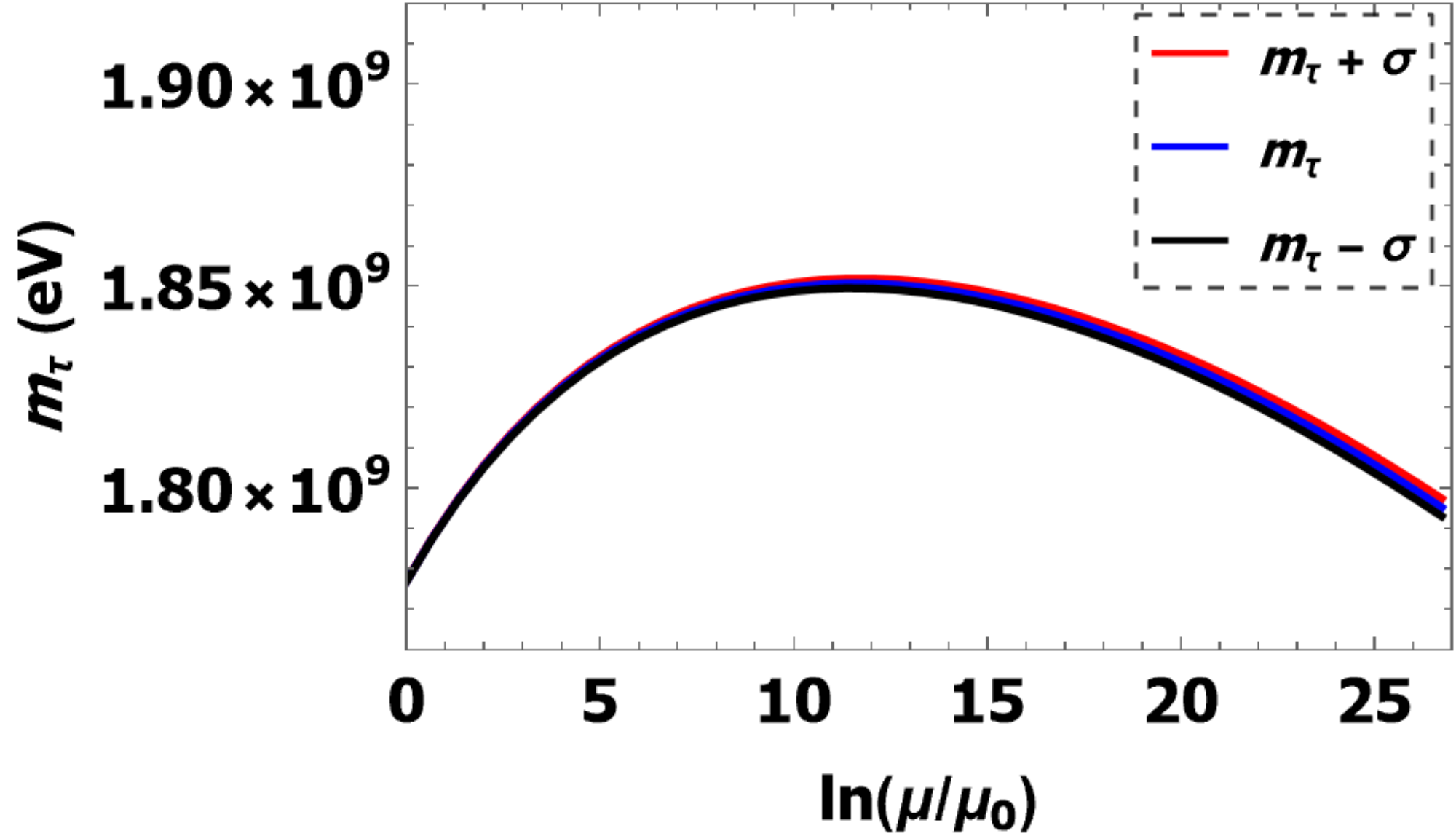}\label{up}}
    \subfigure[]{\includegraphics[width=0.32\textwidth]{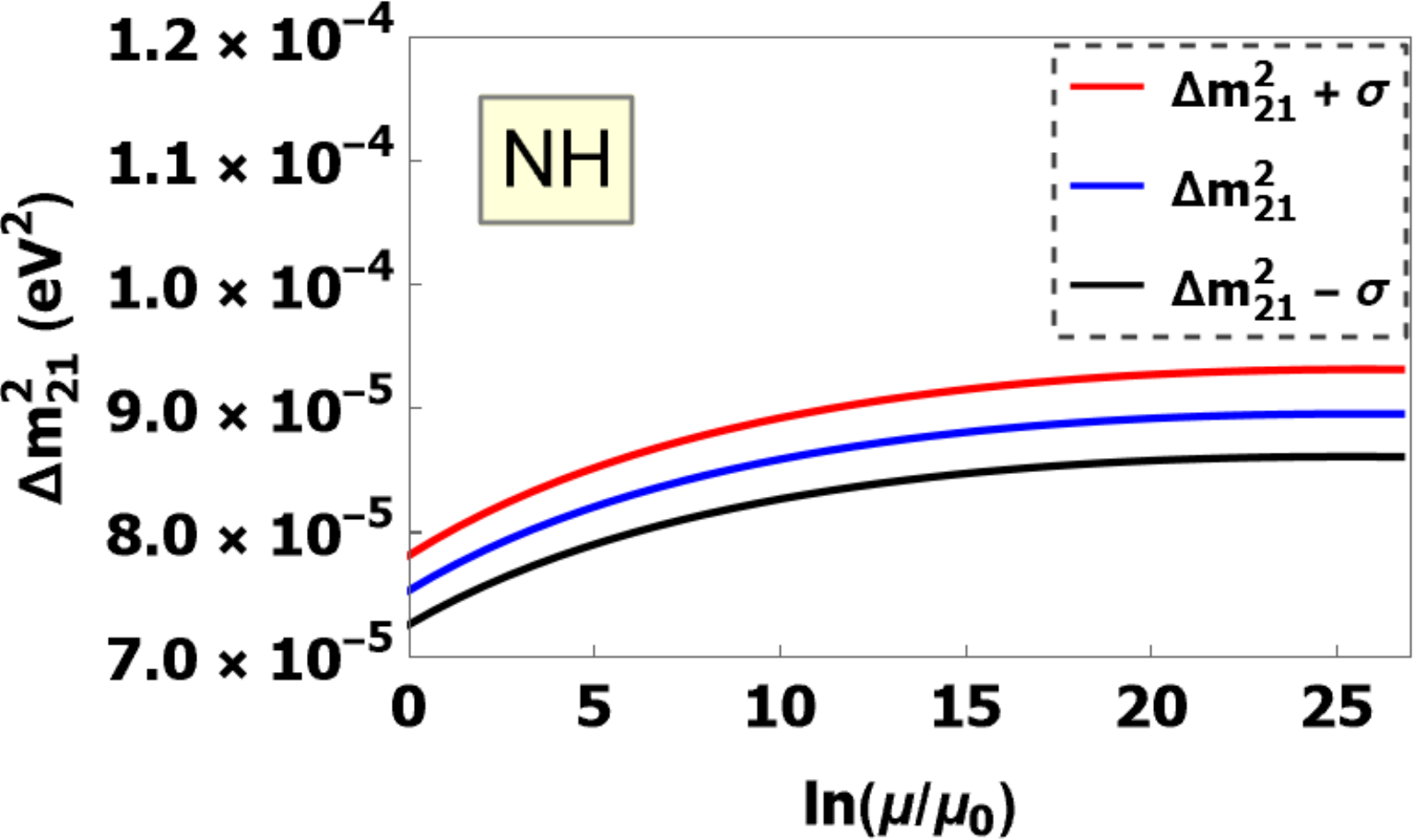}\label{down}} 
    \subfigure[]{\includegraphics[width=0.32\textwidth]{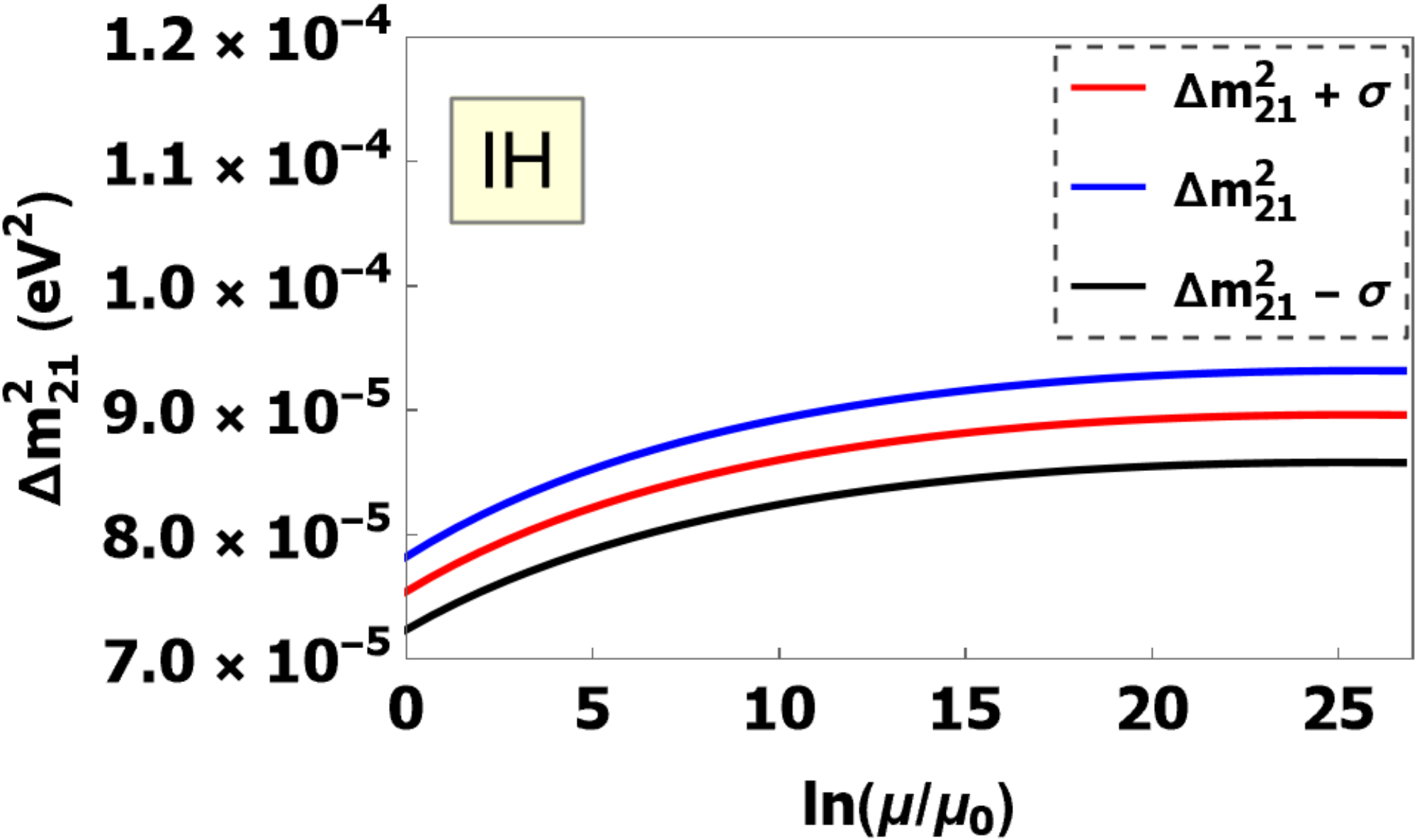}\label{cl}} 
     \subfigure[]{\includegraphics[width=0.32\textwidth]{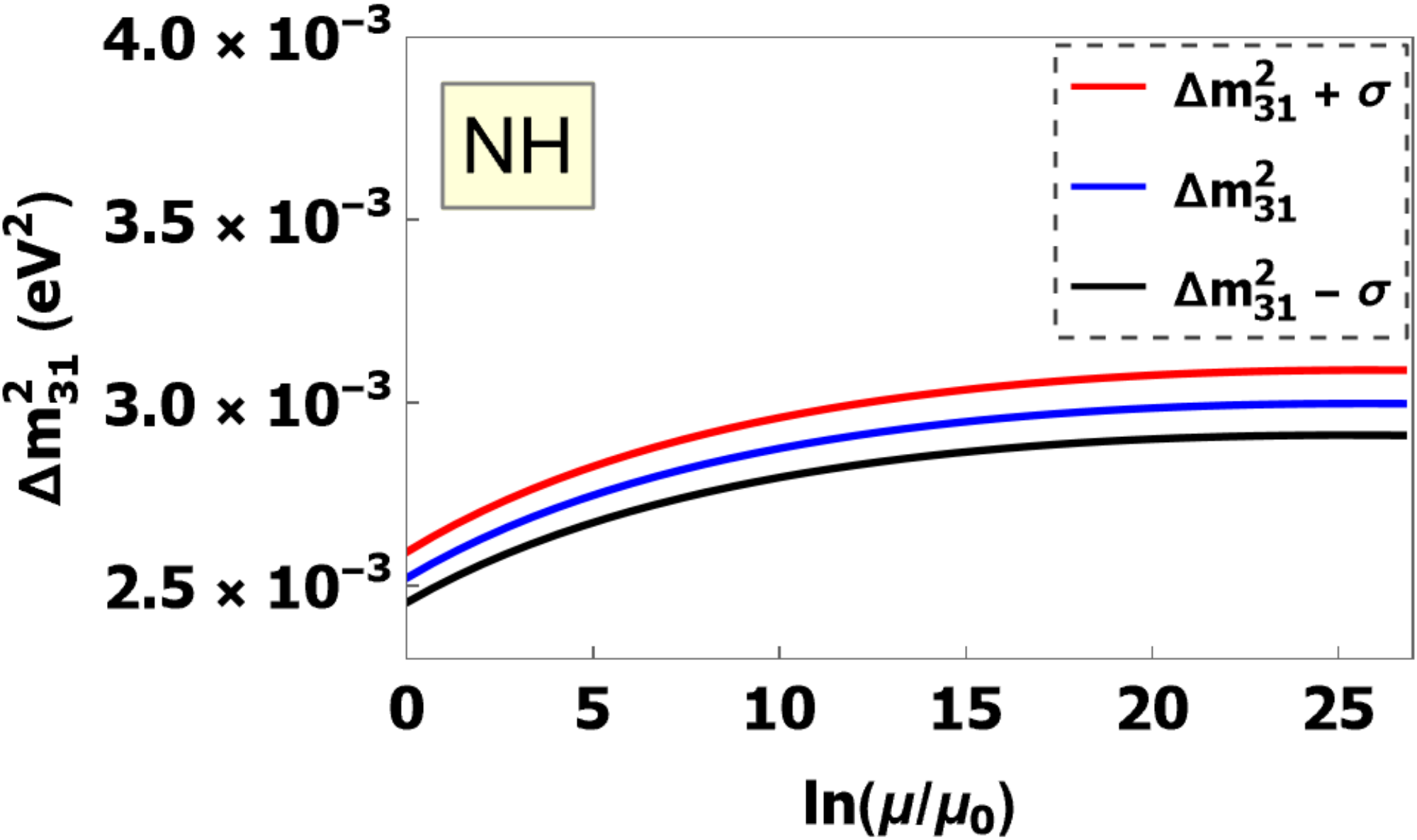}
   \label{g1g2g3}} 
    \subfigure[]{\includegraphics[width=0.32\textwidth]{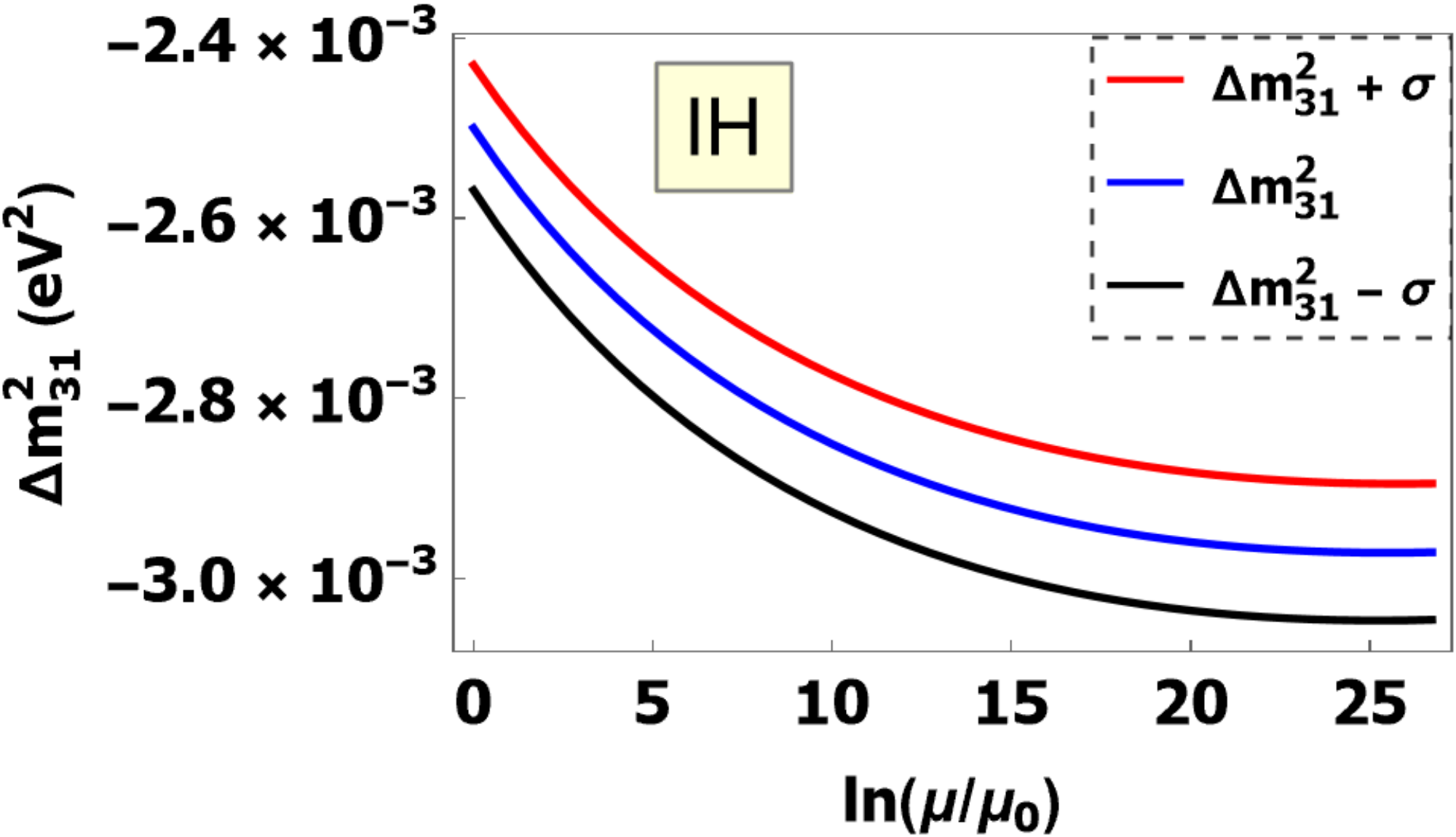}\label{up}}
   \caption{The running of the gauge coupling constants, charged fermion masses and neutrino mass squared differences from electroweak scale to flavour symmetry breaking scale.}
\label{RGE1}
\end{figure}

\begin{figure}
  \centering
  \subfigure[]{\includegraphics[width=0.32\textwidth]{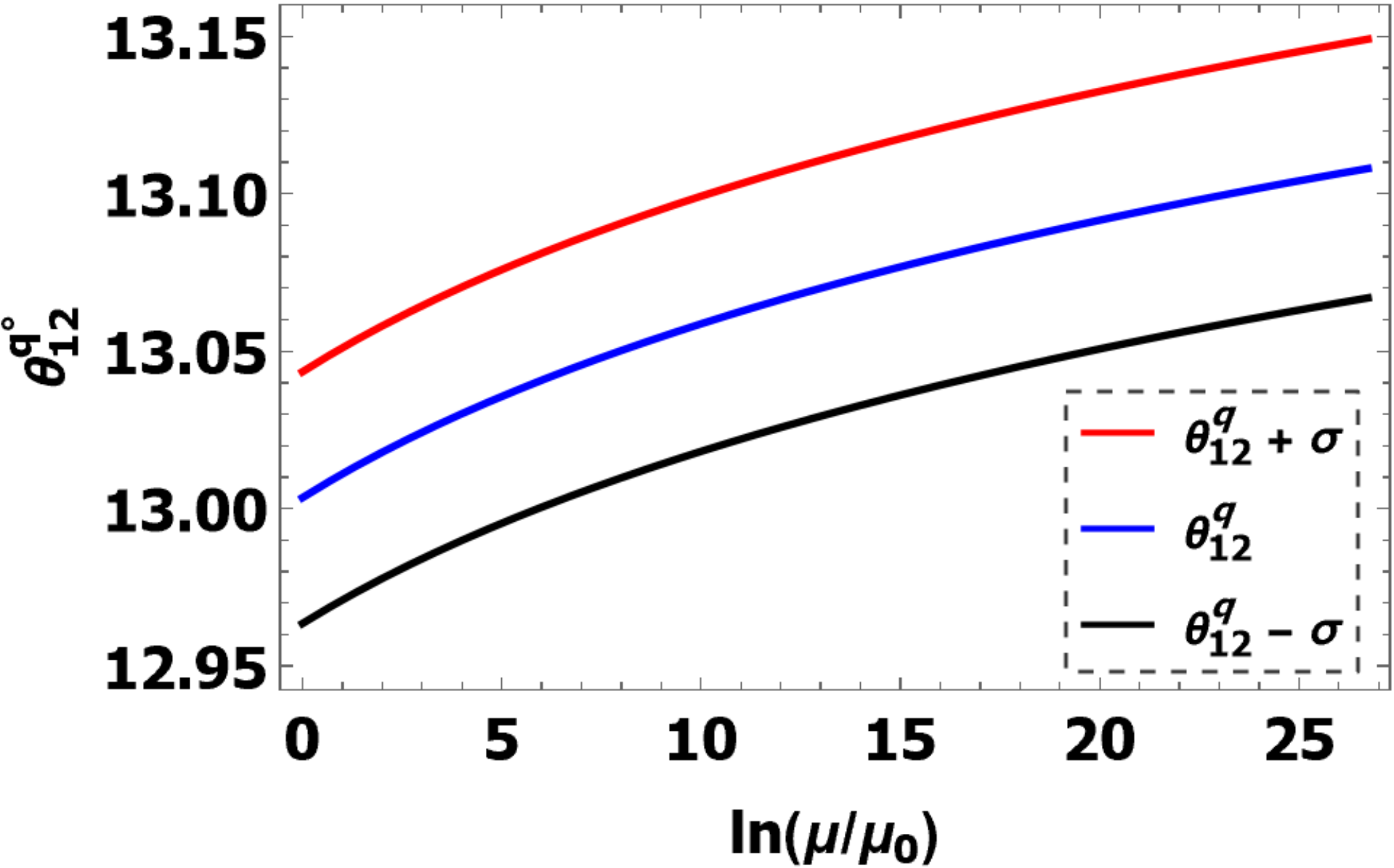}\label{down}}
    \subfigure[]{\includegraphics[width=0.32\textwidth]{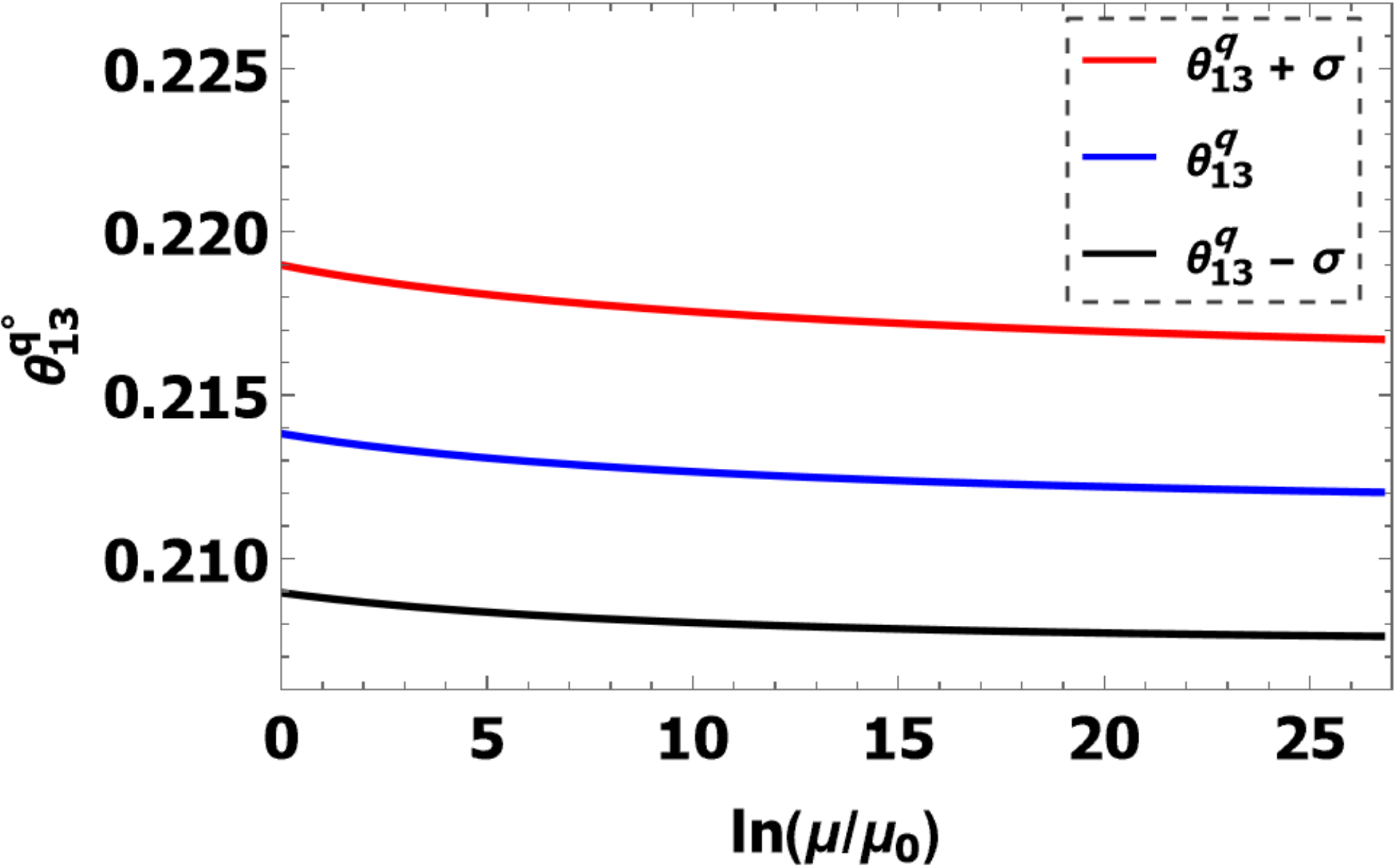}\label{cl}} 
      \subfigure[]{\includegraphics[width=0.32\textwidth]{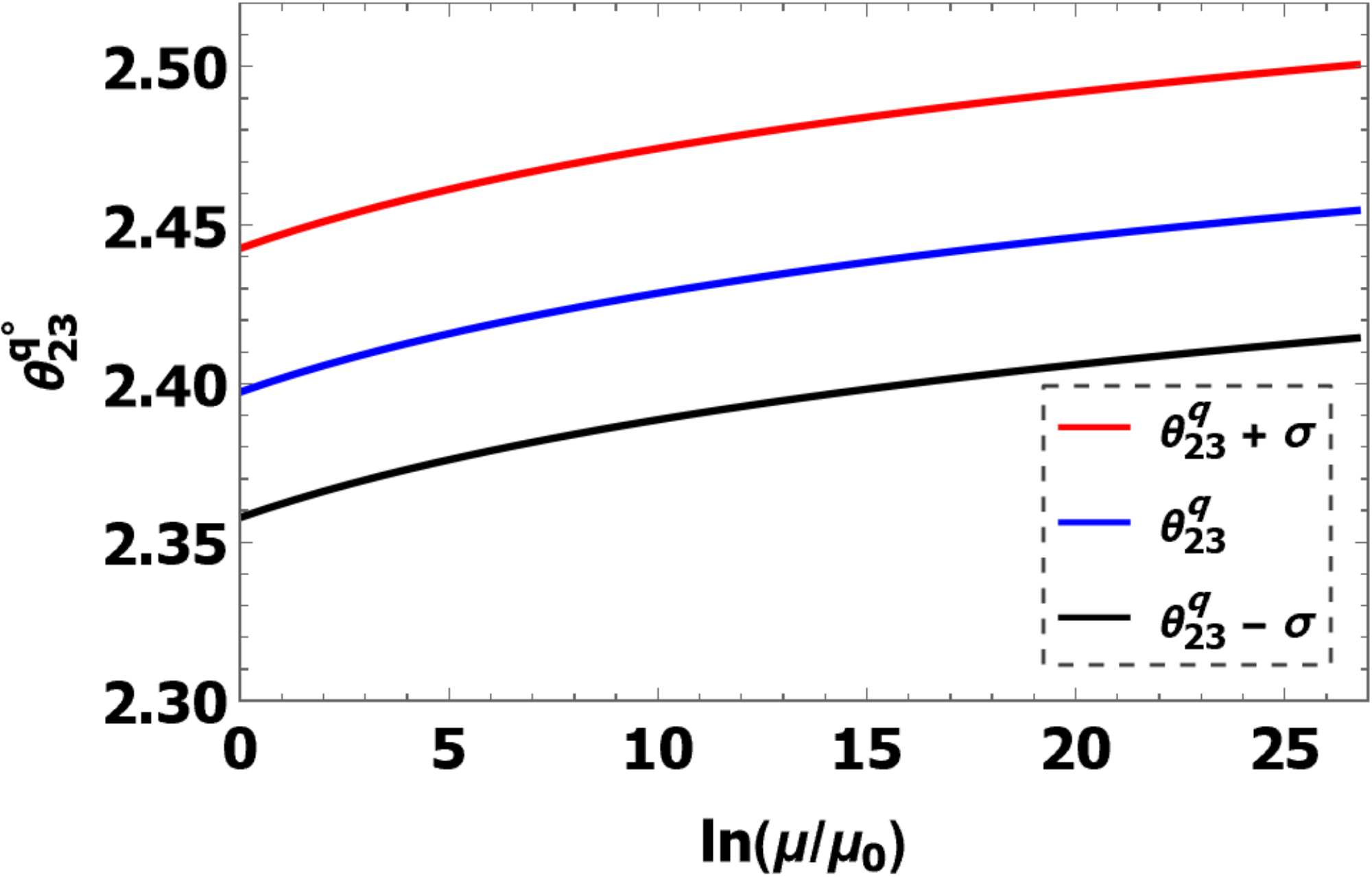}
   \label{g1g2g3}} 
    \subfigure[]{\includegraphics[width=0.33\textwidth]{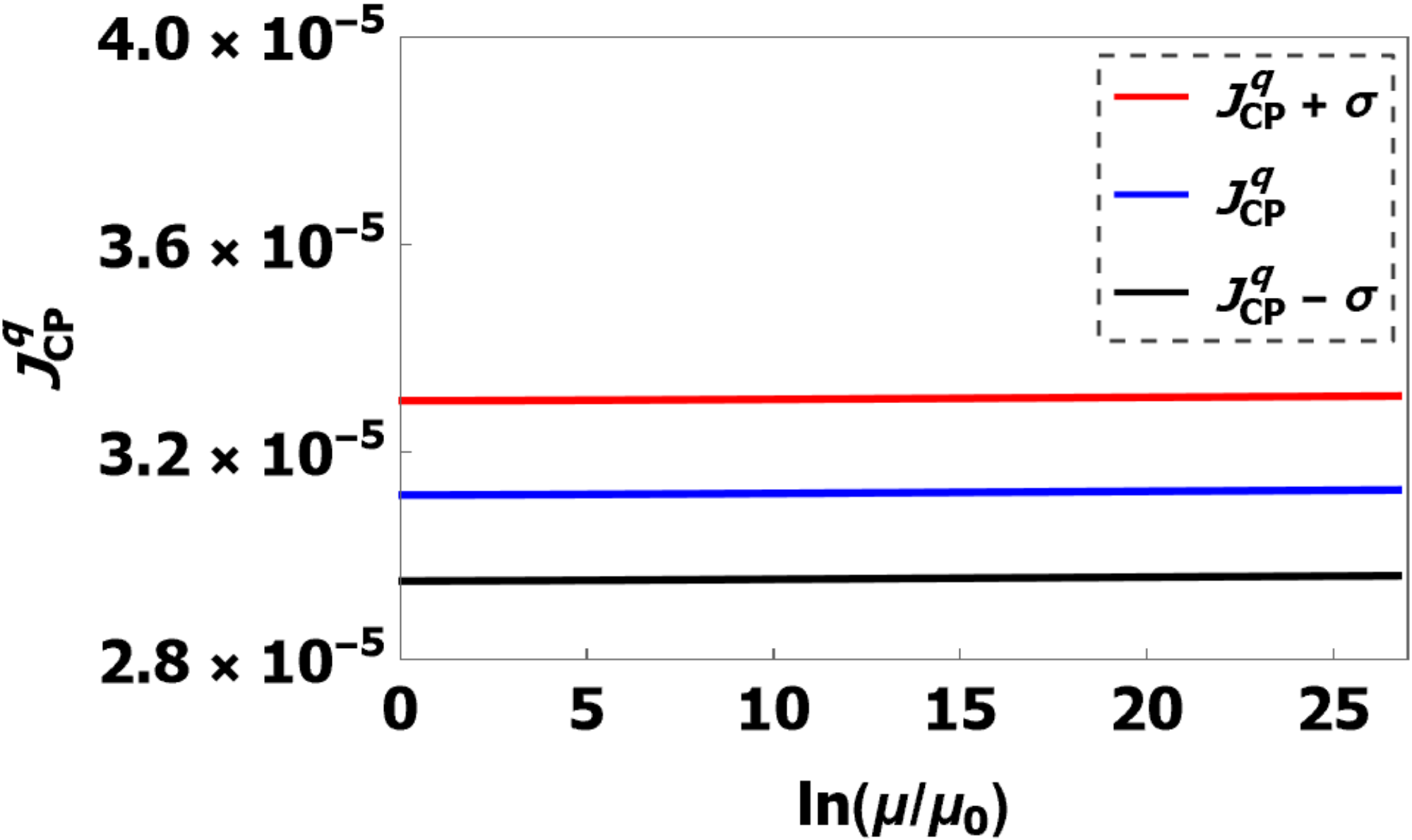}\label{up}}
    \subfigure[]{\includegraphics[width=0.32\textwidth]{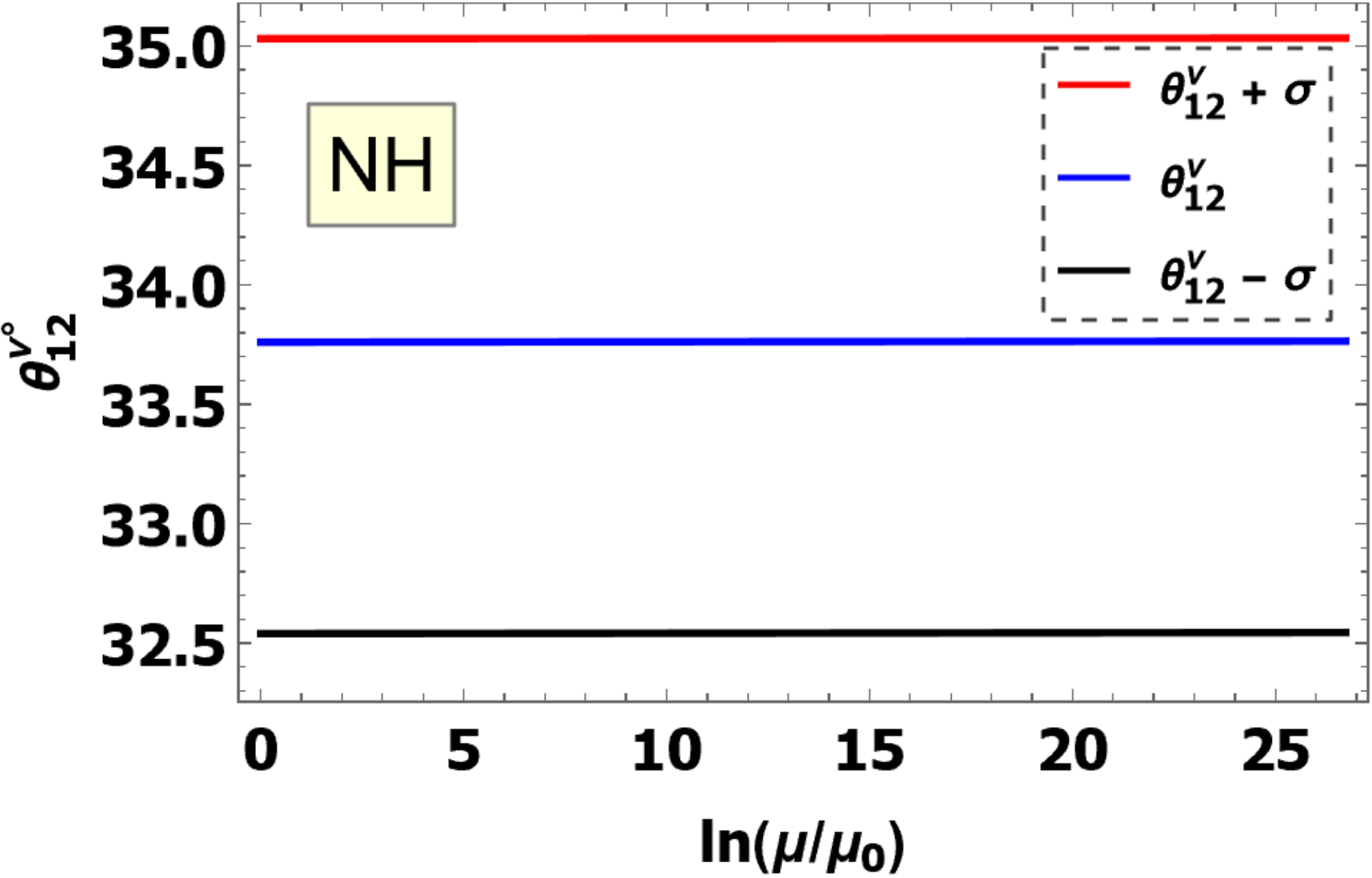}\label{down}} 
    \subfigure[]{\includegraphics[width=0.32\textwidth]{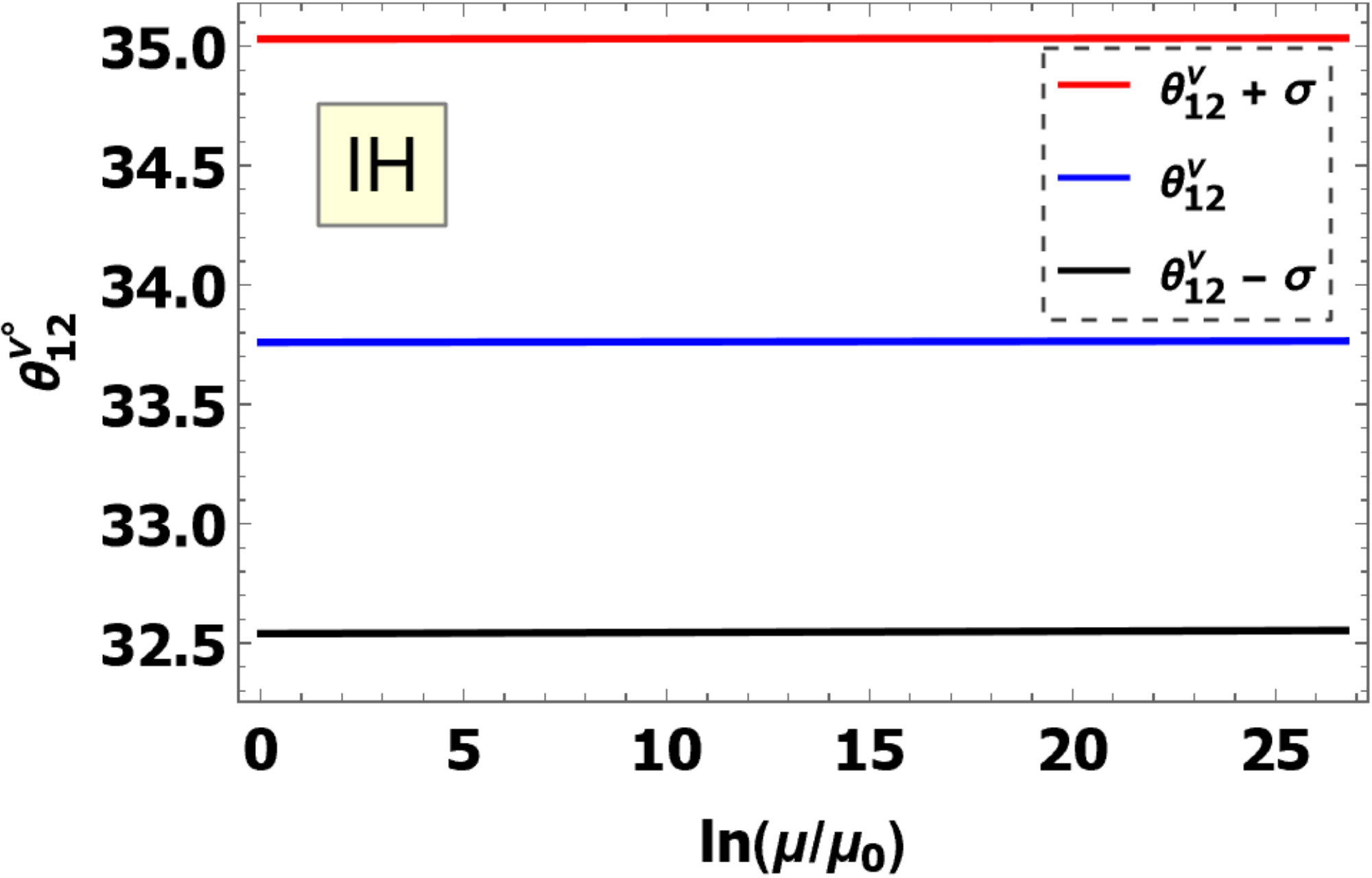}\label{cl}} 
     \subfigure[]{\includegraphics[width=0.32\textwidth]{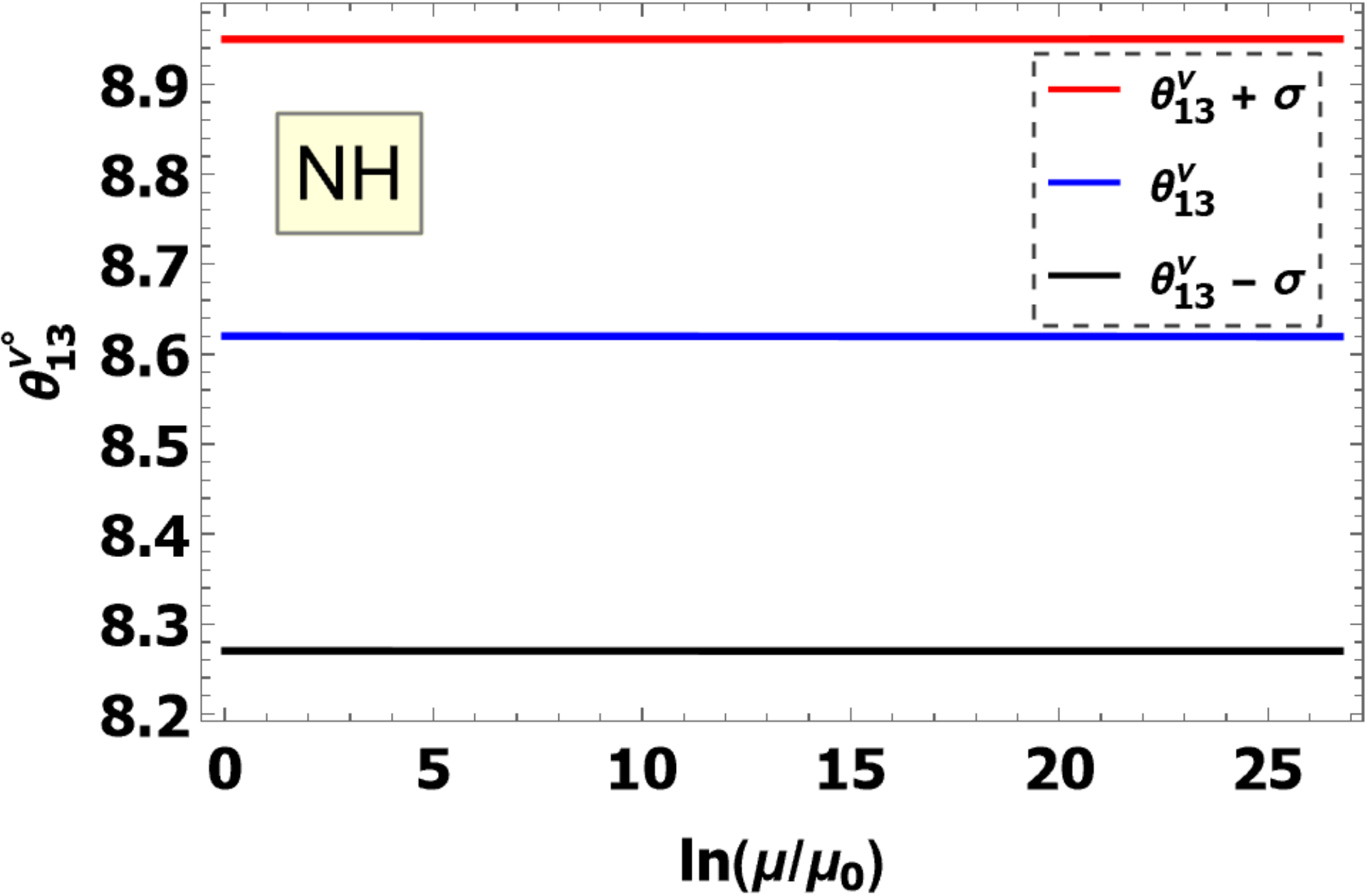}
   \label{g1g2g3}} 
      \subfigure[]{\includegraphics[width=0.32\textwidth]{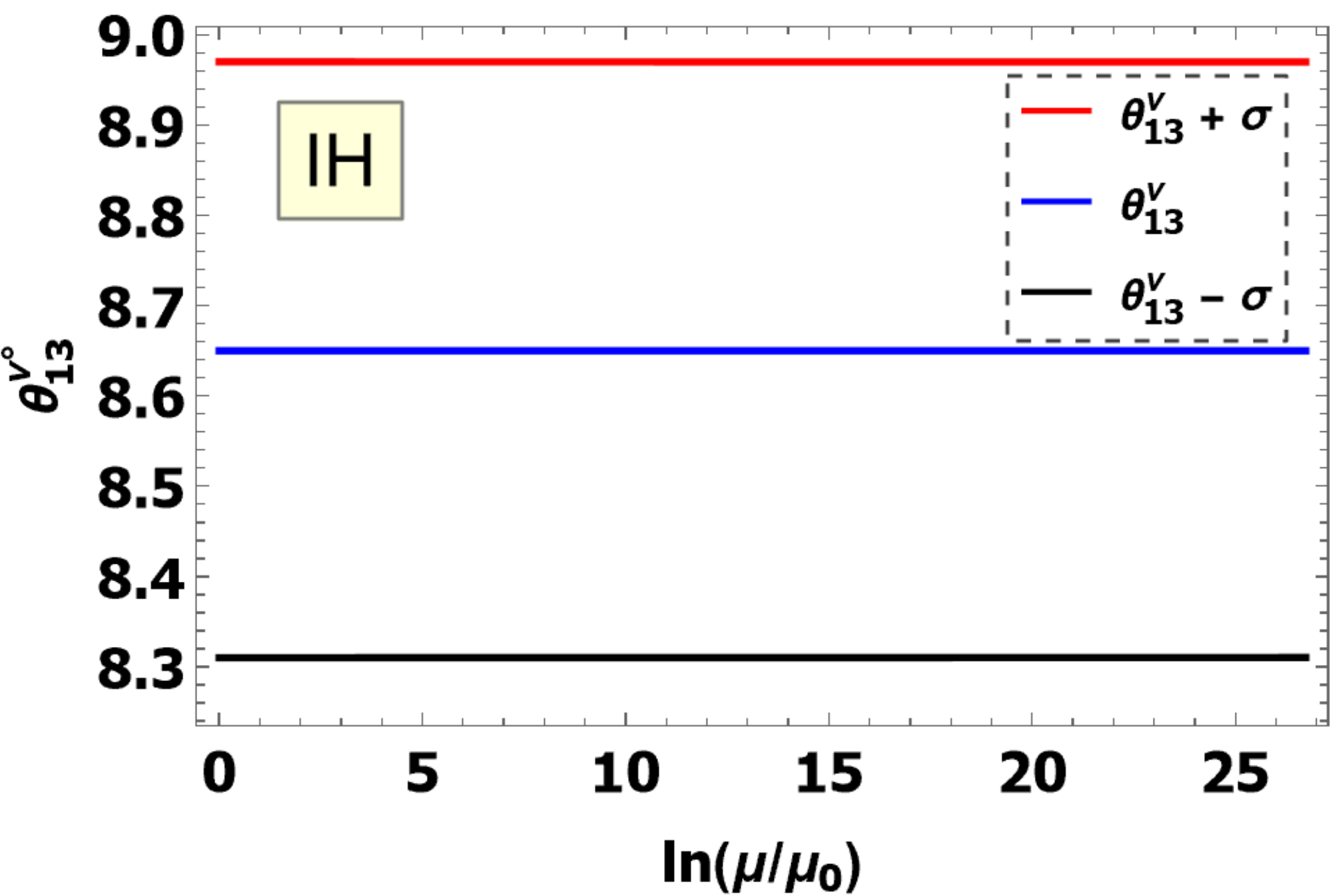}
   \label{g1g2g3}}
    \subfigure[]{\includegraphics[width=0.3\textwidth]{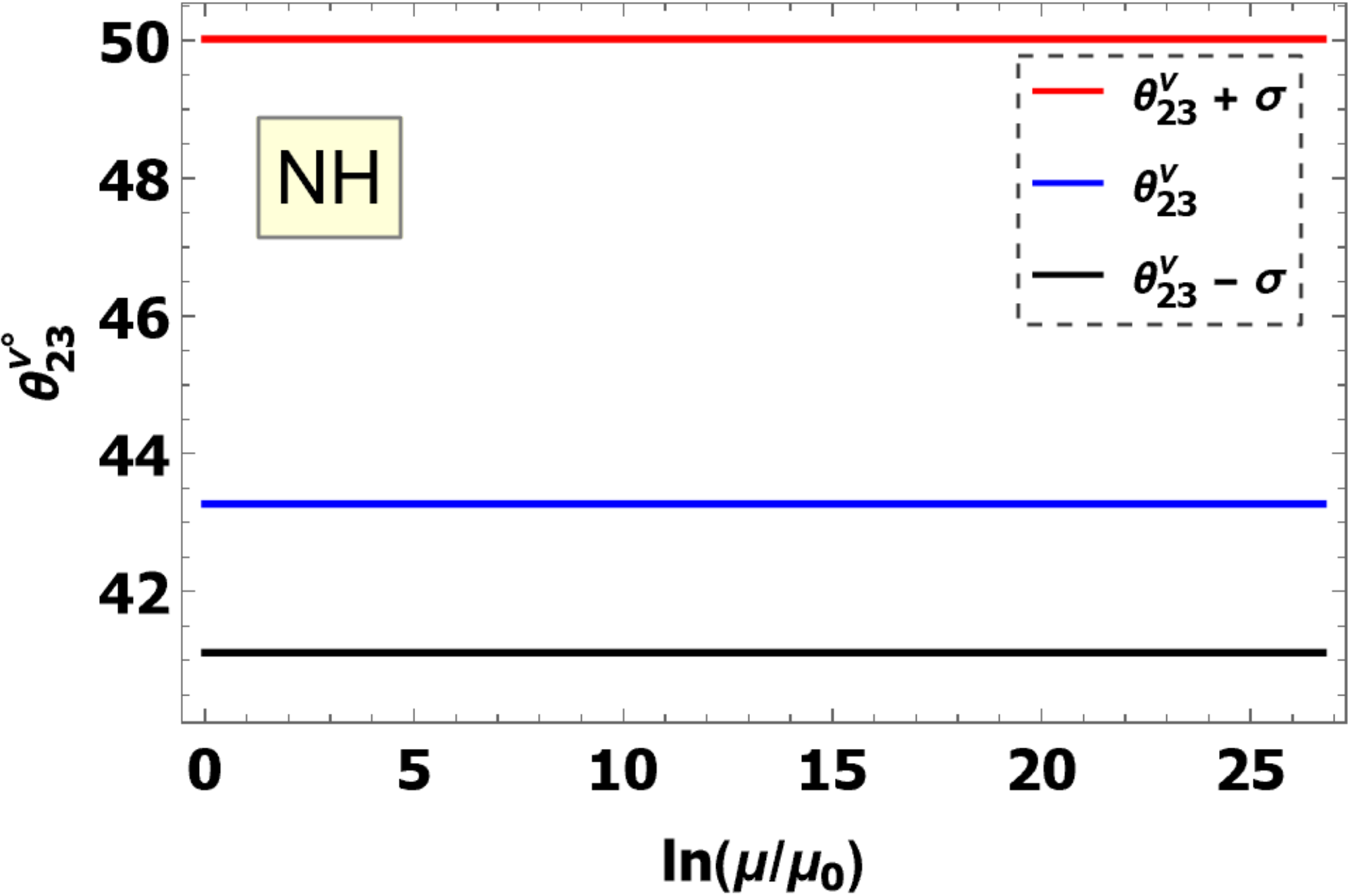}
   \label{g1g2g3}}
    \subfigure[]{\includegraphics[width=0.32\textwidth]{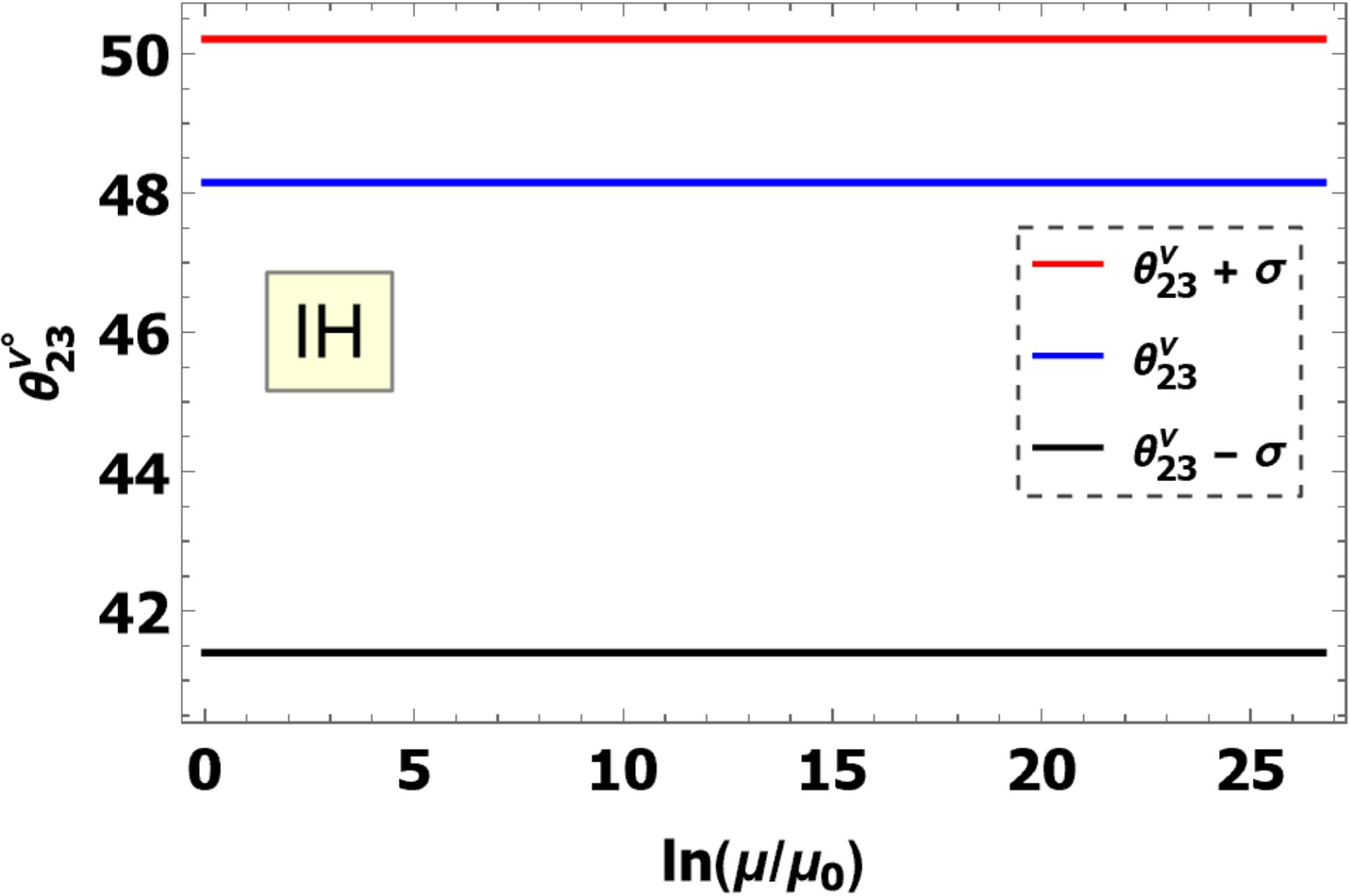}
   \label{g1g2g3}}
    \subfigure[]{\includegraphics[width=0.32\textwidth]{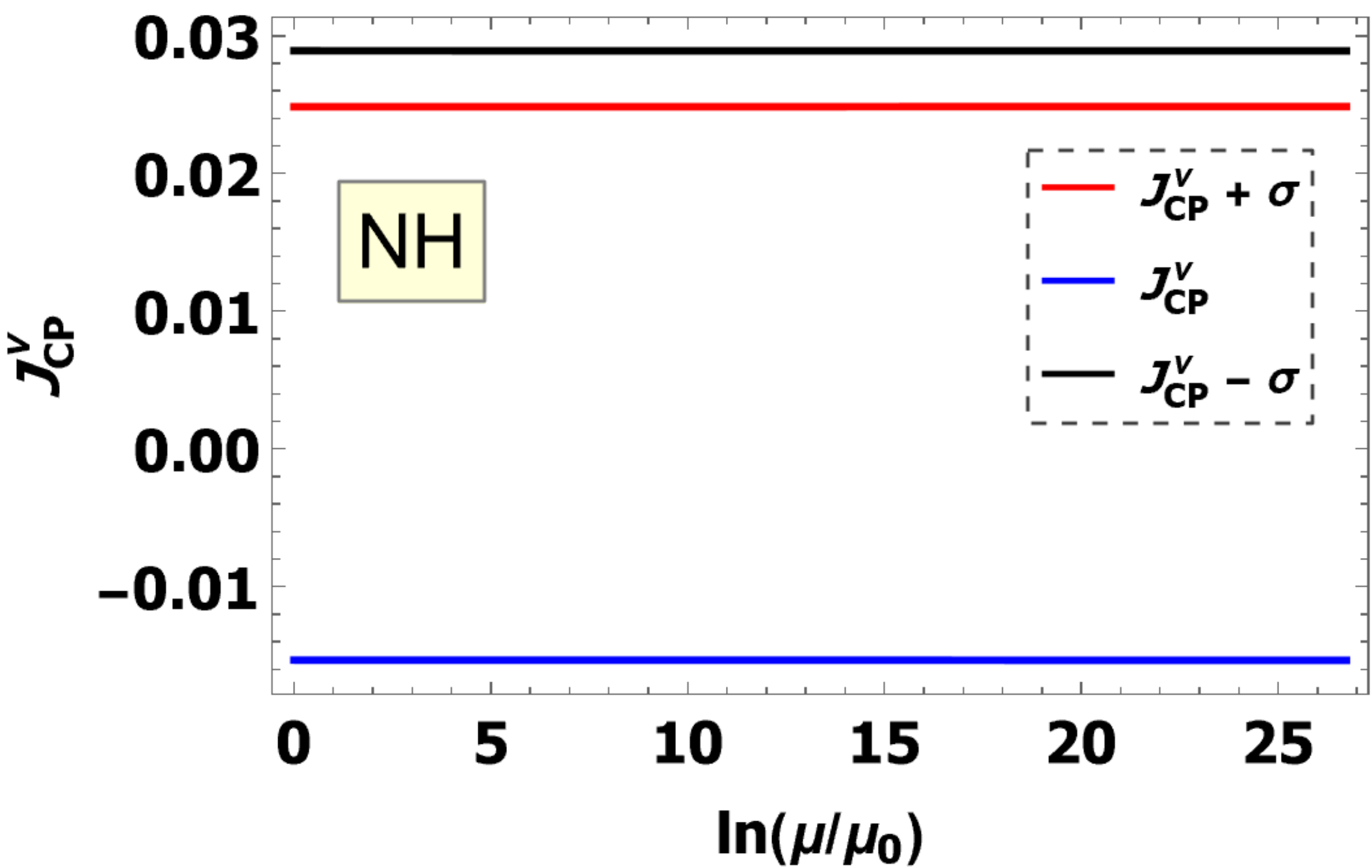}\label{up}}
     \subfigure[]{\includegraphics[width=0.32\textwidth]{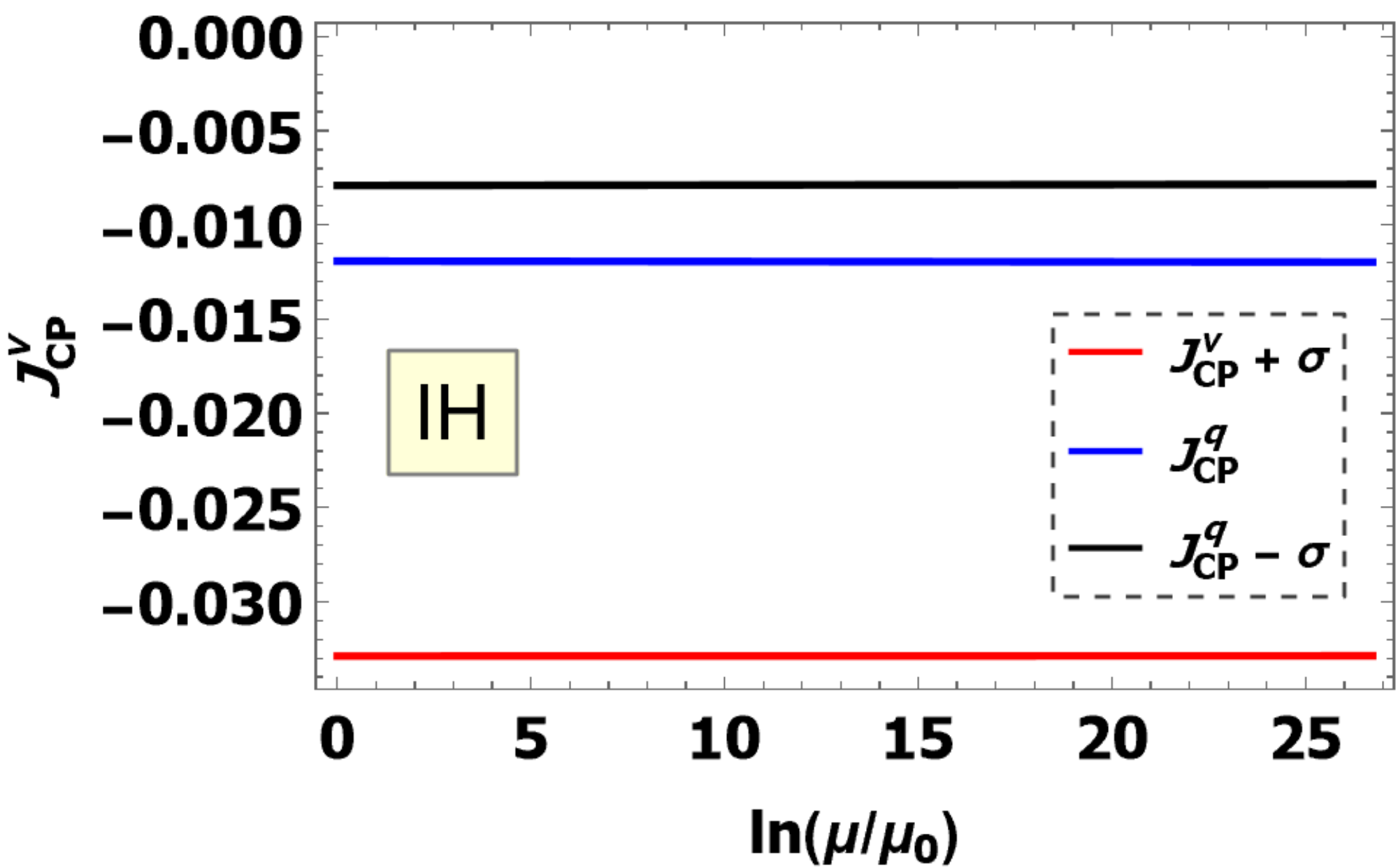}\label{up}}
   \caption{The running of the quark and neutrino mixing parameters from electroweak scale to flavour symmetry breaking scale. The CKM mixing angles show mild running, whereas the PMNS mixing angles remain almost
constant. The $J_{CP}$ for both sectors almost remain unchanged.}
\label{RGE2}
\end{figure}

\begin{align}
16\pi^2\,\frac{dg_1}{dt}=\frac{41}{10}g_1^3,\qquad
16\pi^2\,\frac{dg_2}{dt}=-\frac{19}{6}g_2^3,\qquad
16\pi^2\,\frac{dg_3}{dt}=-7g_3^3.
\end{align}

\vspace{1cm}

For the Yukawa matrices, we solve:
\begin{align}
16\pi^2\,\frac{dY_u}{dt} &=
Y_u\Bigg[\frac{3}{2}Y_u^\dagger Y_u-\frac{3}{2}Y_d^\dagger Y_d
+T\,\mathbf{I}
-\Big(\frac{17}{20}g_1^2+\frac{9}{4}g_2^2+8g_3^2\Big)\mathbf{I}\Bigg],\\
16\pi^2\,\frac{dY_d}{dt} &=
Y_d\Bigg[\frac{3}{2}Y_d^\dagger Y_d-\frac{3}{2}Y_u^\dagger Y_u
+T\,\mathbf{I}
-\Big(\frac{1}{4}g_1^2+\frac{9}{4}g_2^2+8g_3^2\Big)\mathbf{I}\Bigg],
\end{align}

\begin{align}
16\pi^2\,\frac{dY_e}{dt} &=
Y_e\Bigg[\frac{3}{2}Y_e^\dagger Y_e+\frac{1}{2}Y_\nu^\dagger Y_\nu
+T\,\mathbf{I}
-\Big(\frac{9}{4}g_1^2+\frac{9}{4}g_2^2\Big)\mathbf{I}\Bigg],\\
16\pi^2\,\frac{dY_\nu}{dt} &=
Y_\nu\Bigg[\frac{3}{2}Y_\nu^\dagger Y_\nu+\frac{1}{2}Y_e^\dagger Y_e
+T\,\mathbf{I}
-\Big(\frac{9}{20}g_1^2+\frac{9}{4}g_2^2\Big)\mathbf{I}\Bigg]
\end{align}
where the trace combination is
\begin{align}
T = 3\,\mathrm{Tr}(Y_u^\dagger Y_u)+3\,\mathrm{Tr}(Y_d^\dagger Y_d)
+\mathrm{Tr}(Y_e^\dagger Y_e)+\mathrm{Tr}(Y_\nu^\dagger Y_\nu).
\end{align}

At each scale, we diagonalise the Dirac mass matrices
$M_f(\mu)=\frac{v}{\sqrt{2}}\,Y_f(\mu)$. We extract the mixing angles and Dirac CP phase. The gauge couplings evolve smoothly up to $\Lambda_F$. We find $g_1$ increases while $g_2$ and $g_3$ decrease over the full interval. More importantly, the fermion mixing parameters show only mild running in our set-up. While the mixing angles from the quark sector vary a little, the PMNS mixing angles remain almost invariant under RGE evolution. Thus, the Jarlskog invariant $J_{CP}$ does not undergo any abrupt change from both sectors. Although the mixing angles run only
mildly, the mass parameters from both lepton and quark sectors show visible running. We summarise the running graphically in Fig.\,(\ref{RGE1})-(\ref{RGE2}).

The apparent difference between the running of masses and mixing angles follows from the fact that the dominant one-loop RGE terms are flavour-universal. These terms mainly rescale the mass matrix and hence modify the mass eigenvalues, while leaving the diagonalising matrix almost unchanged. The mixing angles run only through flavour non-universal terms such as $Y_\nu^\dagger Y_\nu$ and $Y_e^\dagger Y_e$ which are small in the effective Dirac-neutrino framework considered here. Consequently, the masses and mass-squared differences may show visible scale dependence, whereas the PMNS angles remain nearly stable. Thus, even if the numerical variations induced by the one-loop RG evolution are small compared to the present experimental uncertainties, the analysis serves as a consistency check of the radiative stability of the proposed texture structures.

In the next section, we shall summarize our work.

\section{Summary and Discussion \label{sec8}}

In the present work, we investigate whether both quarks and lepton sectors can be studied starting from common mass matrix textures. While the quark and charged lepton mass matrices are obtained through the standard Higgs mechanism, we adopt the type-I Dirac seesaw mechanism to realize the neutrino mass matrix, assuming neutrinos to be purely Dirac type. As an important feature of our model, we observe that both the quark and neutrino sectors are connected through a common Hermitian texture with three common parameters $\Sigma_1$, $\Sigma_2$, and $\Sigma_3$. In addition, to preserve naturalness and avoid fine-tuning, we ensure the Yukawa couplings of both down-type and neutrino sectors to be $\sim \mathcal{O}(1)$ coefficients.  We observe that this texture is consistent with experimental data, successfully explaining the observed fermion masses and mixing patterns. In addition, we discuss some special cases that may arise in the neutrino sector and decrease the number of free parameters. The reduced texture $M_\nu^R$ with six free parameters becomes predictive. Specifically, for NH, $M_\nu^R$ predicts $\theta_{23}$ in the lower octant: $(42.14^\circ-44.50^\circ)$. On the other hand, for IH, $M_\nu^R$ predicts $\theta_{23}$ in the upper octant: $(45.08^\circ-45.68^\circ)$. It is important to note that the predicted value of $\theta_{23}$ for IH lies far from the experimental $1\sigma$ best-fit value of $48.15^\circ$ and is therefore testable in future experiments.

We would like to mention a key feature of the proposed model. We address the hierarchy of charged leptons and up-type quarks starting from the FN mechanism\,\cite{Froggatt:1978nt}, where both cut-off scales and the VEVs of the scalar fields are involved. At first glance, it may seem misleading that the Yukawa terms of the down-type quarks and neutrino sectors are suppressed by $\sim 1/\Lambda, 1/\Lambda^2$, whereas for charged leptons and up-type quarks the suppressions are of the order $1/\Lambda^{10}$ and $1/\Lambda^{7}$, respectively. Here, it is important to mention that the involvement of the VEVs of different scalar fields plays a crucial role in addressing this anomaly. In both lepton and quark sectors, the powers of the scalar fields appear differently and thus naturally explain the desired results. Hence, the cut-off scale $\Lambda$ is not the sole suppression factor in our proposed model. We also highlight that the involvement of auxiliary groups restricts other possibilities of $1/\Lambda^n$, with $n > 10$ for the charged lepton sector and $n > 7$ for the quark sector, from appearing in the theory. 

We study several phenomenological aspects of the model, such as lepton flavour violation and nonunitarity of the lepton mixing matrix, in the light of present experimental constraints. Furthermore, we investigate the stability of the proposed textures under renormalization group evolution from the electroweak scale to the flavour symmetry breaking scale. In this regard, we adopt a model-independent one-loop running and assume that no intermediate thresholds are crossed between the electroweak and cut-off scales. A complete threshold analysis in the present multi-scalar type-I Dirac seesaw framework would require the determination of the physical scalar spectrum and sequential EFT matching across multiple heavy thresholds. The threshold corrections, if present, are expected to enter at loop level, typically of order $1/(16\pi^2)$ multiplied by combinations of heavy-sector couplings and logarithmic factors involving the matching scales, and may therefore modify the numerical running quantitatively. Such an analysis is beyond the scope of the present model-independent one-loop study and is left for future work.

In the present work, the choice of the discrete flavour group $\Delta(27)$\,\cite{Vien:2016tmh, CarcamoHernandez:2016piw, Vien:2022sxh,Vien:2020hzy, Branco:1983tn, deMedeirosVarzielas:2006fc, Ma:2007wu, deMedeirosVarzielas:2012ylr, Bhattacharyya:2012pi, Ferreira:2012ri, Ma:2013xqa, Nishi:2013jqa, deMedeirosVarzielas:2013xas, Harrison:2014jqa, CarcamoHernandez:2018djj, Bjorkeroth:2019csz, Bazzocchi:2009qg, Varzielas:2012nn, Aranda:2013gga, Abbas:2014ewa, Abbas:2015zna, Bjorkeroth:2015uou} is motivated by its unique group theoretical properties, which are essential for the present construction. Unlike commonly used non-Abelian groups such as $A_4$\,\cite{Vien:2023xyq, Vien:2021eog, King:2013hj, Chen:2023mwt, Borah:2017dmk, Singh:2024imk, Goswami:2025jde,Chakraborty:2024hhq, He:2006dk, Feruglio:2008ht, deMedeirosVarzielas:2010ppv, Altarelli:2012bn, Ahn:2012tv, Memenga:2013vc, GonzalezFelipe:2013yhh, deMedeirosVarzielas:2012cet, Ishimori:2012fg, CarcamoHernandez:2013yiy, Babu:2002dz, Altarelli:2005yx, Gupta:2011ct, Morisi:2013eca, Altarelli:2005yp, Kadosh:2010rm, Kadosh:2013nra, delAguila:2010vg, Campos:2014lla, Vien:2014pta, Karmakar:2016cvb, Chattopadhyay:2017zvs, Srivastava:2017sno, Srivastava:2018ser, Borah:2018nvu, Pramanick:2019qpg} or $S_4$\,\cite{Vien:2019zhs, Vien:2022euq, Vien:2020aya, Vien:2016jkz, Dong:2010zu, Pakvasa:1978tx, Derman:1979nf, Yamanaka:1981pa, Brown:1984mq, Brown:1984dk, Lee:1994qx, Mohapatra:2003tw, Ma:2005pd, Cai:2006mf, Caravaglios:2006aq, Zhang:2006fv}, the $\Delta(27)$ possesses inequivalent complex triplet representations, $\mathbf{3}$ and $\mathbf{3}^*$, allowing for intrinsic complex Clebsch-Gordan coefficients. Furthermore, the tensor products $\mathbf{3}\times\mathbf{3}^*$ in $\Delta(27)$ contain multiple independent symmetric and antisymmetric triplet contractions, which provide sufficient freedom to realize the targeted fermion mass textures. For example, the structure of $\mathbf{(3\times3)_{3s_{1}^*}}$ under $\Delta(27)$ is essential for the present construction. In addition, the singlet representation structures provide greater flexibility in constructing hierarchical Yukawa textures which is more difficult to achieve within minimal $A_4$ or $S_4$ frameworks due to their limited number of singlet representations. Thus, the larger multiplicity of singlet representations in $\Delta(27)$, together with its non trivial triplet and anti triplet structure, enables the construction of predictive fermion mass matrices with reduced parameter redundancy. As a result, realistic fermion mass hierarchies and mixing patterns consistent with experimental data can be obtained without resorting to excessive fine-tuning of the model parameters. These features are not simultaneously achievable in $A_4$ or $S_4$  making $\Delta(27)$ a well motivated choice for the present model.

Although the discrete group $\Delta(27)$ has been extensively explored in the literature, the present work differs from existing frameworks in several essential aspects. In Ref.\,\cite{Vien:2016tmh, CarcamoHernandez:2016piw, Chakraborty:2023msb, Vien:2020hzy, CarcamoHernandez:2018iel, Branco:1983tn, deMedeirosVarzielas:2006fc, Ma:2007wu, deMedeirosVarzielas:2012ylr, Ferreira:2012ri, Ma:2013xqa, Nishi:2013jqa, Harrison:2014jqa, Bazzocchi:2009qg, Varzielas:2012nn, Aranda:2013gga, Abbas:2014ewa, Abbas:2015zna, Bjorkeroth:2015uou}, the several viable neutrino models are explored in the lepton sector for both Dirac and Majorana scenarios based on $\Delta(27)$ group. From the unification perspective, most previous $\Delta(27)$ based studies either employ Fritzsch-like or more general textures in the quark sector while consistently assuming a Majorana nature for neutrinos\,\cite{CarcamoHernandez:2018iel, Vien:2022sxh,Bhattacharyya:2012pi, deMedeirosVarzielas:2013xas, CarcamoHernandez:2018djj, Bjorkeroth:2019csz, Bazzocchi:2009qg, Varzielas:2012nn, Aranda:2013gga, Abbas:2014ewa, Abbas:2015zna, Bjorkeroth:2015uou}. In addition, there exist unified frameworks based on Majorana nature of neutrinos from several discrete flavour symmetries\,\cite{Barranco:2010we, GonzalezCanales:2009zz, deMedeirosVarzielas:2017sdv, Koide:2003rx, King:2006np, King:2013hoa, Vien:2023xyq, Vien:2023zid, Vien:2022sxh, Vien:2021eog, Hernandez:2021mxo, Vien:2021xfp, Vien:2021ciw, Vien:2020uzf, Vien:2019eju, Vien:2019zhs, Vien:2019lso, Vien:2016qbb, Gupta:2015iku, Gupta:2011zzg, Randhawa:1999hi, Gupta:2012fsl, Berezhiani:2024fsw, Garces:2018nar}. In contrast, the present analysis constitutes the first unified framework in which type-I Dirac seesaw mechanism is employed in order to explain tiny neutrino masses. While general Hermitian texture for neutrinos have been considered earlier\,\cite{Vien:2025cak}, a common Hermitian texture simultaneously governing the down-type quark and neutrino sectors has not been previously investigated. This unified texture structure is controlled by three parameters, $\Sigma_1$, $\Sigma_2$, and $\Sigma_3$, which encode the seed of quark-lepton unification within the model. Furthermore, since the suppression of neutrino masses originates dynamically from the Dirac seesaw mechanism, no additional fine-tuning of Yukawa couplings is required. As a result, all fundamental Yukawa couplings remain naturally of $\mathcal{O}(1)$ throughout the construction, ensuring the preservation of naturalness. These features collectively distinguish the present work from earlier $\Delta(27)$ models and unified frameworks based on other discrete symmetries.

\section{Acknowledgement}

The research work of PC is supported by Innovation in Science Pursuit for Inspired Research (INSPIRE), Department of Science and Technology, Government of India, New Delhi vide grant No. IF190651. PC thanks S.T. Goswami, Gauhati University for valuable discussions on CLFV.

\appendix

\section{The Role of Cyclic Symmetries. \label{appendix c}}
 
 The auxiliary cyclic symmetries $Z_5$, $Z_4$, $Z_3$ and $Z_7$ play important roles in eliminating several unwanted operators allowed by the $SU(2)_L \times \Delta(27)$ symmetry. In particular, the $Z_5$ symmetry forbids the lower dimensional operators
\begin{align}
&\frac{1}{\Lambda}(\overline{Q}_{l_L}\psi)\widetilde H c_R\xi,\quad
\frac{1}{\Lambda}(\overline{D}_{l_L}\psi)H e_R\xi,\quad
\frac{1}{\Lambda}(\overline{D}_{l_L}\psi)H \tau_R\xi,\quad
\frac{1}{\Lambda}(\overline{Q}_{l_L}\chi)H d_R,\nonumber\\
&\frac{1}{\Lambda}(\overline{D}_{l_L}\chi)\widetilde H N_R,\quad
\frac{1}{\Lambda}(\overline{N}_L\phi)N_R,
\end{align}
as well as the dimension-six operators
\begin{align}
&\frac{1}{\Lambda^2}(\overline{Q}_{l_L}\psi)\widetilde H u_R\xi^2,\quad
\frac{1}{\Lambda^2}(\overline{Q}_{l_L}\psi)\widetilde H c_R\xi^2,\quad
\frac{1}{\Lambda^2}(\overline{D}_{l_L}\psi)H e_R\xi^2,\nonumber\\
&\frac{1}{\Lambda^2}(\overline{D}_{l_L}\psi)H \mu_R\xi^2,\quad
\frac{1}{\Lambda^2}(\overline{D}_{l_L}\psi)H \tau_R\xi^2.
\end{align}

Similarly, the $Z_4$ symmetry forbids the operators
\begin{align}
&\frac{1}{\Lambda}(\overline{Q}_{l_L}\kappa)H d_R,\quad
\frac{1}{\Lambda}(\overline{D}_{l_L}\kappa)\widetilde H N_R,\quad
\frac{1}{\Lambda}(\overline{N}_L\phi)N_R,\quad
\frac{1}{\Lambda}(\overline{N}_L\kappa)\rho N_R,
\end{align}
along with the dimension-six operators
\begin{align}
&\frac{1}{\Lambda^2}(\overline{Q}_{l_L}\kappa)H\xi d_R,\quad
\frac{1}{\Lambda^2}(\overline{D}_{l_L}\kappa)\widetilde H\xi N_R.
\end{align}

The $Z_3$ symmetry forbids the operators
\begin{align}
&\frac{1}{\Lambda}(\overline{Q}_{l_L}\psi)\widetilde H u_R\xi,\quad
\frac{1}{\Lambda}(\overline{D}_{l_L}\psi)H \mu_R\xi,\quad
\frac{1}{\Lambda}(\overline{Q}_{l_L}\chi)H d_R,\quad
\frac{1}{\Lambda}(\overline{D}_{l_L}\chi)\widetilde H N_R,\nonumber\\
&\frac{1}{\Lambda}(\overline{Q}_{l_L}\kappa)H d_R,\quad
\frac{1}{\Lambda}(\overline{D}_{l_L}\kappa)\widetilde H N_R,\quad
\frac{1}{\Lambda}(\overline{N}_L\phi)\rho N_R,
\end{align}
and the dimension-six operators
\begin{align}
&\frac{1}{\Lambda^2}(\overline{Q}_{l_L}\chi)H\sigma d_R,\quad
\frac{1}{\Lambda^2}(\overline{D}_{l_L}\chi)\widetilde H\sigma N_R,\quad
\frac{1}{\Lambda^2}(\overline{Q}_{l_L}\kappa)H\xi d_R,\nonumber\\
&\frac{1}{\Lambda^2}(\overline{D}_{l_L}\kappa)\widetilde H\xi N_R.
\end{align}

Furthermore, the $Z_7$ symmetry forbids the bare Majorana mass terms
\begin{align}
\overline{N_R^c}N_R,\qquad
\overline{\nu_R^c}\nu_R,\qquad
\overline{N_L^c}N_L.
\end{align}
Moreover, no scalar singlet combination constructed from the scalar sector carries the appropriate compensating $Z_7$ charge required to generate the corresponding higher-dimensional operators of the form
\begin{align}
&\frac{1}{\Lambda}\overline{N_R^c}N_R\Phi,\qquad
\frac{1}{\Lambda}\overline{\nu_R^c}\nu_R\Phi,\qquad
\frac{1}{\Lambda}\overline{N_L^c}N_L\Phi,
\end{align}
or
\begin{align}
&\frac{1}{\Lambda^2}\overline{N_R^c}N_R\Phi^2,\qquad
\frac{1}{\Lambda^2}\overline{\nu_R^c}\nu_R\Phi^2,\qquad
\frac{1}{\Lambda^2}\overline{N_L^c}N_L\Phi^2,
\end{align}
where $\Phi$ denotes any scalar singlet combination.

\section{Product Rules of $\Delta(27)$ \label{appendix a}}

The non-Abelian discrete group $\Delta\,(27)$ is a subgroup of $SU(3)$ has 27 elements which can be divided into 11 equivalence classes. The group has two three dimensional representations and nine one-dimensional representations.

The group multiplication rules of two triplets $(a_1, a_2, a_3)^T$ and $(b_1, b_2, b_3)^T$ are shown below,

\begin{eqnarray}
3 \times 3&=&3^*_{S_1}+3^*_{S_2}+3^*_{A},\\
3 \times 3^* &=& \sum_{r=0}^{2} 1_{r,0}+\sum_{r=0}^{2} 1_{r,1} +\sum_{r=0}^{2} 1_{r,2},\\
1_{r,p}\times 1_{r', p'}&=&1_{(r+r') \,\text{mod}\,3,\,\, (p+p') \,\text{mod}\,3},
\end{eqnarray}

where,

\begin{align}
(3\times3)_{3^*_{S_1}}&=\begin{bmatrix}
a_1 b_1\\
a_2 b_2\\
a_3 b_3\\
\end{bmatrix},\,\,\, (3\times3)_{3^*_{S_2}}=\frac{1}{2}\begin{bmatrix}
a_2 b_3+a_3 b_2\\
a_3 b_1+a_1 b_3\\
a_1 b_2+a_2 b_1\\
\end{bmatrix},\,\,\, (3\times3)_{3^*_{A}}=\frac{1}{2}\begin{bmatrix}
a_2 b_3-a_3 b_2\\
a_3 b_1-a_1 b_3\\
a_1 b_2-a_2 b_1\\
\end{bmatrix},
\end{align}

and,

\begin{align}
(3\times3^*)_{1_{00}}&=(a_1\bar{b_1}+a_2\bar{b_2}+a_3\bar{b_3}),\quad
(3\times3^*)_{1_{10}}=(a_1\bar{b_1}+\omega^2 a_2\bar{b_2}+\omega a_3\bar{b_3}),\nonumber\\
(3\times3^*)_{1_{20}}&=(a_1\bar{b_1}+\omega a_2\bar{b_2}+\omega^2 a_3\bar{b_3}),\quad
(3\times3^*)_{1_{01}}=(a_1\bar{b_2}+ a_2\bar{b_3}+a_3\bar{b_1}),\nonumber\\
(3\times3^*)_{1_{11}}&=(a_1\bar{b_2}+\omega^2 a_2\bar{b_3}+\omega a_3\bar{b_1}),\quad
(3\times3^*)_{1_{21}}=(a_1\bar{b_2}+\omega a_2\bar{b_3}+\omega^2 a_3\bar{b_1})\nonumber\\
(3\times3^*)_{1_{02}}&=(a_1\bar{b_3}+ a_2\bar{b_1}+a_3\bar{b_2}),\quad
(3\times3^*)_{1_{12}}=(a_1\bar{b_3}+ \omega^2 a_2\bar{b_1}+\omega a_3\bar{b_2}),\nonumber\\
(3\times3^*)_{1_{22}}&=(a_1\bar{b_3}+ \omega a_2\bar{b_1}+\omega^2 a_3\bar{b_2}).
\end{align}

\section{The Scalar Potential \label{appendix b}}

To justify the vacuum alignments of the scalar fields, we construct the $SU(2)_L \times \Delta (27) \times Z_{7} \times Z_{5} \times  Z_4 \times Z_3$ invariant scalar potential as shown below,

\begin{align}
V \,&= V(H)+V(\psi)+V(\eta)+V(\chi)+V(\kappa)+v(\phi)+V(\sigma)+V(\xi)+v(\rho)\nonumber\\&+V(H,\psi)+\,V(H, \eta)+V(H, \chi)+\,V(H, \kappa)+V(H,\phi)+V(H, \sigma)\nonumber\\&+V(H, \xi)+V(H,\rho)+V(\psi, \eta)+V(\psi, \chi)+V(\psi, \kappa)+V(\psi, \phi)+\nonumber\\&V(\psi, \sigma)+V(\psi, \xi)+V(\psi, \rho)+V(\eta, \chi)+V(\eta, \kappa)+V(\eta, \phi)+V(\eta, \sigma)\nonumber\\&+V(\eta, \xi)+V(\eta, \rho)+V(\chi, \kappa)+V(\chi, \phi)+V(\chi, \sigma)+V(\chi, \xi)+V(\chi, \rho)\nonumber\\&+V(\kappa, \sigma)+V(\kappa, \phi)+V(\kappa, \xi)+V(\kappa, \rho)+V(\phi, \sigma)+V(\phi, \xi)+V(\phi, \rho)\nonumber\\&+V(\sigma, \xi)+V(\sigma, \rho)+V(\xi, \rho)+h.c.
\label{potential}
\end{align}

The explicit forms of the terms appearing in the scalar potential are shown in the following,

\begin{align}
V(H)&= -\mu^2_{H}(H^{\dagger}H)+ \lambda^{H}(H^{\dagger}H)(H^{\dagger}H),\\
V(\psi)&=-\mu^2_{\psi}(\psi^{\dagger}\psi)+\lambda^{\psi}_1(\psi^{\dagger}\psi)(\psi^{\dagger}\psi)+\lambda^{\psi}_2(\psi^{\dagger}\psi)_{1_{10}}(\psi^{\dagger}\psi)_{1_{20}}+\lambda^{\psi}_3(\psi^{\dagger}\psi)_{1_{01}}\nonumber\\&(\psi^{\dagger}\psi)_{1_{02}}+\lambda^{\psi}_4(\psi^{\dagger}\psi)_{1_{11}}(\psi^{\dagger}\psi)_{1_{22}}+\lambda^{\psi}_5(\psi^{\dagger}\psi)_{1_{21}}(\psi^{\dagger}\psi)_{1_{12}},\\
V(\eta)&=-\mu^2_{\eta}(\eta^{\dagger}\eta)+\lambda^{\eta}_1(\eta^{\dagger}\eta)(\eta^{\dagger}\eta)+\lambda^{\eta}_2(\eta^{\dagger}\eta)_{1_{10}}(\eta^{\dagger}\eta)_{1_{20}}+\lambda^{\eta}_3(\eta^{\dagger}\eta)_{1_{01}}(\eta^{\dagger}\eta)_{1_{02}}\nonumber\\&+\lambda^{\eta}_4(\eta^{\dagger}\eta)_{1_{11}}(\eta^{\dagger}\eta)_{1_{22}}+\lambda^{\eta}_5(\eta^{\dagger}\eta)_{1_{21}}(\eta^{\dagger}\eta)_{1_{12}},\\
V(\chi)&=-\mu^2_{\chi}(\chi^{\dagger}\chi)+\lambda^{\chi}_1(\chi^{\dagger}\chi)(\chi^{\dagger}\chi)+\lambda^{\chi}_2(\chi^{\dagger}\chi)_{1_{10}}(\chi^{\dagger}\chi)_{1_{20}}+\lambda^{\chi}_3(\chi^{\dagger}\chi)_{1_{01}}(\chi^{\dagger}\nonumber\\&\chi)_{1_{02}}+\lambda^{\chi}_4(\chi^{\dagger}\chi)_{1_{11}}(\chi^{\dagger}\chi)_{1_{22}}+\lambda^{\chi}_5(\chi^{\dagger}\chi)_{1_{21}}(\chi^{\dagger}\chi)_{1_{12}},\\
V(\kappa)&=-\mu^2_{\kappa}(\kappa^{\dagger}\kappa)+\lambda^{\kappa}_1(\kappa^{\dagger}\kappa)(\kappa^{\dagger}\kappa)+\lambda^{\kappa}_2(\kappa^{\dagger}\kappa)_{1_{10}}(\kappa^{\dagger}\kappa)_{1_{20}}+\lambda^{\kappa}_3(\kappa^{\dagger}\kappa)_{1_{01}}(\kappa^{\dagger}\nonumber\\&\kappa)_{1_{02}}+\lambda^{\kappa}_4(\kappa^{\dagger}\kappa)_{1_{11}}(\kappa^{\dagger}\kappa)_{1_{22}}+\lambda^{\kappa}_5(\kappa^{\dagger}\kappa)_{1_{21}}(\kappa^{\dagger}\kappa)_{1_{12}},\\
V(\phi)&=-\mu^2_{\phi}(\phi^{\dagger}\phi)+\lambda^{\phi}_1(\phi^{\dagger}\phi)(\phi^{\dagger}\phi)+\lambda^{\phi}_2(\phi^{\dagger}\phi)_{1_{10}}(\phi^{\dagger}\phi)_{1_{20}}+\lambda^{\phi}_3(\phi^{\dagger}\phi)_{1_{01}}(\phi^{\dagger}\nonumber\\&\phi)_{1_{02}}+\lambda^{\phi}_4(\phi^{\dagger}\phi)_{1_{11}}(\phi^{\dagger}\phi)_{1_{22}}+\lambda^{\phi}_5(\phi^{\dagger}\phi)_{1_{21}}(\phi^{\dagger}\phi)_{1_{12}},\\
V(\xi)&=  -\mu^2_{\xi}(\xi^{\dagger}\xi)+ \lambda^{\xi}(\xi^{\dagger}\xi)^2,\quad
V(\sigma)= -\mu^2_{\sigma}(\sigma^{\dagger}\sigma)+ \lambda^{\sigma}(\sigma^{\dagger}\sigma)^2,\\
V(\rho)&= -\mu^2_{\rho}(\rho^{\dagger}\rho)+ \lambda^{\rho}(\rho^{\dagger}\rho)^2,\quad
V(H,\phi)=\lambda^{H \phi}(H^{\dagger}H)(\phi^{\dagger}\phi),\\
V(H,\psi)&=\lambda^{H\psi}(H^{\dagger}H)(\psi^{\dagger}\psi),\quad
V(H,\eta)=\lambda^{H\eta}(H^{\dagger}H)(\eta^{\dagger}\eta),\\
V(H,\eta)&=\lambda^{H\eta}(H^{\dagger}H)(\eta^{\dagger}\eta),\quad
V(H,\chi)=\lambda^{H \chi}(H^{\dagger}H)(\chi^{\dagger}\chi),\\
V(H,\kappa)&=\lambda^{H \kappa}(H^{\dagger}H)(\kappa^{\dagger}\kappa),\quad
V(H,\xi)=\lambda^{H \xi}(H^{\dagger}H)(\xi^{\dagger}\xi),\\
V(H,\sigma)&=\lambda^{H \sigma}(H^{\dagger}H)(\sigma^{\dagger}\sigma),\quad
V(H,\rho)=\lambda^{H \rho}(H^{\dagger}H)(\rho^{\dagger}\rho),\\
V(\psi,\xi)&=\lambda^{\psi\xi}(\psi^{\dagger}\psi)\xi^{\dagger}\xi,\quad
V(\psi,\sigma)=\lambda^{\psi\sigma}(\psi^{\dagger}\psi)\sigma^{\dagger}\sigma,\quad
V(\sigma,\rho)=\lambda^{\sigma\rho}(\sigma^{\dagger}\sigma)\rho^{\dagger}\rho,\\
V(\psi,\rho)&=\lambda^{\psi\rho}(\psi^{\dagger}\psi)\rho^{\dagger}\rho,\quad
V(\phi,\xi)=\lambda^{\phi\xi}(\phi^{\dagger}\phi)\xi^{\dagger}\xi,\quad
V(\phi,\sigma)=\lambda^{\phi\sigma}(\phi^{\dagger}\phi)\sigma^{\dagger}\sigma,\\
V(\phi,\rho)&=\lambda^{\phi\rho}(\phi^{\dagger}\phi)\rho^{\dagger}\rho,\quad
V(\eta,\xi)=\lambda^{\eta\xi}(\eta^{\dagger}\eta)\xi^{\dagger}\xi,\quad
V(\eta,\sigma)=\lambda^{\eta\sigma}(\eta^{\dagger}\eta)\sigma^{\dagger}\sigma,\\
V(\eta,\rho)&=\lambda^{\eta\rho}(\eta^{\dagger}\eta)\rho^{\dagger}\rho,\quad
V(\chi,\xi)=\lambda^{\chi\xi}(\chi^{\dagger}\chi)\xi^{\dagger}\xi,\quad
V(\chi,\sigma)=\lambda^{\chi\sigma}(\chi^{\dagger}\chi)\sigma^{\dagger}\sigma,\\
V(\chi,\rho)&=\lambda^{\chi\rho}(\chi^{\dagger}\chi)\rho^{\dagger}\rho,\quad
V(\kappa,\xi)=\lambda^{\kappa\xi}(\kappa^{\dagger}\kappa)\xi^{\dagger}\xi,\quad
V(\kappa,\sigma)=\lambda^{\kappa\sigma}(\kappa^{\dagger}\kappa)\sigma^{\dagger}\sigma,\\
V(\kappa,\rho)&=\lambda^{\kappa\rho}(\kappa^{\dagger}\kappa)\rho^{\dagger}\rho,\quad
V(\xi,\sigma)=\lambda^{\xi\sigma}(\xi^{\dagger}\xi)\sigma^{\dagger}\sigma,\quad
V(\xi,\rho)=\lambda^{\xi\rho}(\xi^{\dagger}\xi)\rho^{\dagger}\rho,\\
V(\psi,\eta)&= \lambda^{\psi \eta}_1(\psi^{\dagger}\psi)(\eta^{\dagger}\eta)+\lambda^{\psi \eta}_2(\psi^{\dagger}\psi)_{1_{10}}(\eta^{\dagger}\eta)_{1_{20}}+\lambda^{\psi \eta}_3(\psi^{\dagger}\psi)_{1_{01}}(\eta^{\dagger}\eta)_{1_{02}}+\lambda^{\psi \eta}_4\nonumber\\&(\psi^{\dagger}\psi)_{1_{11}}(\eta^{\dagger}\eta)_{1_{22}}+\lambda^{\psi \eta}_5(\psi^{\dagger}\psi)_{1_{21}}(\eta^{\dagger}\eta)_{1_{12}}+\lambda^{\psi \eta}_6(\psi^{\dagger}\psi)_{1_{20}}(\eta^{\dagger}\eta)_{1_{10}}-\nonumber\\&\lambda^{\psi \eta}_3(\psi^{\dagger}\psi)_{1_{02}}(\eta^{\dagger}\eta)_{1_{01}}+\lambda^{\psi \eta}_8(\psi^{\dagger}\psi)_{1_{22}}(\eta^{\dagger}\eta)_{1_{11}}+\lambda^{\psi \eta}_9(\psi^{\dagger}\psi)_{1_{12}}(\eta^{\dagger}\eta)_{1_{21}},
\end{align}

\begin{align}
V(\psi,\chi)&= \lambda^{\psi \chi}_1(\psi^{\dagger}\psi)(\chi^{\dagger}\chi)+\lambda^{\psi \chi}_2(\psi^{\dagger}\psi)_{1_{10}}(\chi^{\dagger}\chi)_{1_{20}}+\lambda^{\psi \chi}_3(\psi^{\dagger}\psi)_{1_{01}}(\chi^{\dagger}\chi)_{1_{02}}+\nonumber\\&\lambda^{\psi \chi}_4(\psi^{\dagger}\psi)_{1_{11}}(\chi^{\dagger}\chi)_{1_{22}}+\lambda^{\psi \chi}_5(\psi^{\dagger}\psi)_{1_{21}}(\chi^{\dagger}\chi)_{1_{12}}+\lambda^{\psi \chi}_6(\psi^{\dagger}\psi)_{1_{20}}(\chi^{\dagger}\chi)_{1_{10}}\nonumber\\&-\lambda^{\psi \chi}_3(\psi^{\dagger}\psi)_{1_{02}}(\chi^{\dagger}\chi)_{1_{01}}+\lambda^{\psi \chi}_8(\psi^{\dagger}\psi)_{1_{22}}(\chi^{\dagger}\chi)_{1_{11}}+\lambda^{\psi \chi}_9(\psi^{\dagger}\psi)_{1_{12}}(\chi^{\dagger}\chi)_{1_{21}},\\
V(\psi,\kappa)&= \lambda^{\psi \kappa}_1(\psi^{\dagger}\psi)(\kappa^{\dagger}\kappa)+\lambda^{\psi \kappa}_2(\psi^{\dagger}\psi)_{1_{10}}(\kappa^{\dagger}\kappa)_{1_{20}}+\lambda^{\psi \kappa}_3(\psi^{\dagger}\psi)_{1_{01}}(\kappa^{\dagger}\kappa)_{1_{02}}+\lambda^{\psi \kappa}_4\nonumber\\&(\psi^{\dagger}\psi)_{1_{11}}(\kappa^{\dagger}\kappa)_{1_{22}}+\lambda^{\psi \kappa}_5(\psi^{\dagger}\psi)_{1_{21}}(\kappa^{\dagger}\kappa)_{1_{12}}+\lambda^{\psi \kappa}_6(\psi^{\dagger}\psi)_{1_{20}}(\kappa^{\dagger}\kappa)_{1_{10}}-\lambda^{\psi \kappa}_3\nonumber\\&(\psi^{\dagger}\psi)_{1_{02}}(\kappa^{\dagger}\kappa)_{1_{01}}+\lambda^{\psi \kappa}_8(\psi^{\dagger}\psi)_{1_{22}}(\kappa^{\dagger}\kappa)_{1_{11}}+\lambda^{\psi \kappa}_9(\psi^{\dagger}\psi)_{1_{12}}(\kappa^{\dagger}\kappa)_{1_{21}},\\
V(\psi,\phi)&= \lambda^{\psi \phi}_1(\psi^{\dagger}\psi)(\phi^{\dagger}\phi)+\lambda^{\psi \phi}_2(\psi^{\dagger}\psi)_{1_{10}}(\phi^{\dagger}\phi)_{1_{20}}+\lambda^{\psi \phi}_3(\psi^{\dagger}\psi)_{1_{01}}(\phi^{\dagger}\phi)_{1_{02}}+\lambda^{\psi \phi}_4\nonumber\\&(\psi^{\dagger}\psi)_{1_{11}}(\phi^{\dagger}\phi)_{1_{22}}+\lambda^{\psi \phi}_5(\psi^{\dagger}\psi)_{1_{21}}(\phi^{\dagger}\phi)_{1_{12}}+\lambda^{\psi \phi}_6(\psi^{\dagger}\psi)_{1_{20}}(\phi^{\dagger}\phi)_{1_{10}}-\nonumber\\&\lambda^{\psi \phi}_3(\psi^{\dagger}\psi)_{1_{02}}(\phi^{\dagger}\phi)_{1_{01}}+\lambda^{\psi \phi}_8(\psi^{\dagger}\psi)_{1_{22}}(\phi^{\dagger}\phi)_{1_{11}}+\lambda^{\psi \phi}_9(\psi^{\dagger}\psi)_{1_{12}}(\phi^{\dagger}\phi)_{1_{21}},\\
V(\eta,\chi)&= \lambda^{\eta \chi}_1(\eta^{\dagger}\eta)(\chi^{\dagger}\chi)+\lambda^{\eta \chi}_2(\eta^{\dagger}\eta)_{1_{10}}(\chi^{\dagger}\chi)_{1_{20}}+\lambda^{\eta \chi}_3(\eta^{\dagger}\eta)_{1_{01}}(\chi^{\dagger}\chi)_{1_{02}}+\lambda^{\eta \chi}_4\nonumber\\&(\eta^{\dagger}\eta)_{1_{11}}(\chi^{\dagger}\chi)_{1_{22}}+\lambda^{\eta \chi}_5(\eta^{\dagger}\eta)_{1_{21}}(\chi^{\dagger}\chi)_{1_{12}}+\lambda^{\eta \chi}_6(\eta^{\dagger}\eta)_{1_{20}}(\chi^{\dagger}\chi)_{1_{10}}+\nonumber\\&\lambda^{\eta \chi}_7(\eta^{\dagger}\eta)_{1_{02}}(\chi^{\dagger}\chi)_{1_{01}}+\lambda^{\eta \chi}_8(\eta^{\dagger}\eta)_{1_{22}}(\chi^{\dagger}\chi)_{1_{11}}+\lambda^{\eta \chi}_9(\eta^{\dagger}\eta)_{1_{12}}(\chi^{\dagger}\chi)_{1_{21}},\\
V(\eta,\kappa)&= \lambda^{\eta \kappa}_1(\eta^{\dagger}\eta)(\kappa^{\dagger}\kappa)+\lambda^{\eta \kappa}_2(\eta^{\dagger}\eta)_{1_{10}}(\kappa^{\dagger}\kappa)_{1_{20}}+\lambda^{\eta \kappa}_3(\eta^{\dagger}\eta)_{1_{01}}(\kappa^{\dagger}\kappa)_{1_{02}}+\lambda^{\eta \kappa}_4\nonumber\\&(\eta^{\dagger}\eta)_{1_{11}}(\kappa^{\dagger}\kappa)_{1_{22}}+\lambda^{\eta \kappa}_5(\eta^{\dagger}\eta)_{1_{21}}(\kappa^{\dagger}\kappa)_{1_{12}}+\lambda^{\eta \kappa}_6(\eta^{\dagger}\eta)_{1_{20}}(\kappa^{\dagger}\kappa)_{1_{10}}+\nonumber\\&\lambda^{\eta \kappa}_7(\eta^{\dagger}\eta)_{1_{02}}(\kappa^{\dagger}\kappa)_{1_{01}}+\lambda^{\eta \kappa}_8(\eta^{\dagger}\eta)_{1_{22}}(\kappa^{\dagger}\kappa)_{1_{11}}+\lambda^{\eta \kappa}_9(\eta^{\dagger}\eta)_{1_{12}}(\kappa^{\dagger}\kappa)_{1_{21}},\\
V(\eta,\phi)&= \lambda^{\eta \phi}_1(\eta^{\dagger}\eta)(\phi^{\dagger}\phi)+\lambda^{\eta \phi}_2(\eta^{\dagger}\eta)_{1_{10}}(\phi^{\dagger}\phi)_{1_{20}}+\lambda^{\eta \phi}_3(\eta^{\dagger}\eta)_{1_{01}}(\phi^{\dagger}\phi)_{1_{02}}+\lambda^{\eta \phi}_4\nonumber\\&(\eta^{\dagger}\eta)_{1_{11}}(\phi^{\dagger}\phi)_{1_{22}}+\lambda^{\eta \phi}_5(\eta^{\dagger}\eta)_{1_{21}}(\phi^{\dagger}\phi)_{1_{12}}+\lambda^{\eta \phi}_6(\eta^{\dagger}\eta)_{1_{20}}(\phi^{\dagger}\phi)_{1_{10}}+\nonumber\\&\lambda^{\eta \phi}_7(\eta^{\dagger}\eta)_{1_{02}}(\phi^{\dagger}\phi)_{1_{01}}+\lambda^{\eta \phi}_8(\eta^{\dagger}\eta)_{1_{22}}(\phi^{\dagger}\phi)_{1_{11}}+\lambda^{\eta \phi}_9(\eta^{\dagger}\eta)_{1_{12}}(\phi^{\dagger}\phi)_{1_{21}},\\
V(\chi,\kappa)&= \lambda^{\chi \kappa}_1(\chi^{\dagger}\chi)(\kappa^{\dagger}\kappa)+\lambda^{\chi \kappa}_2(\chi^{\dagger}\chi)_{1_{10}}(\kappa^{\dagger}\kappa)_{1_{20}}+\lambda^{\chi \kappa}_3(\chi^{\dagger}\chi)_{1_{01}}(\kappa^{\dagger}\kappa)_{1_{02}}+\lambda^{\chi \kappa}_4\nonumber\\&(\chi^{\dagger}\chi)_{1_{11}}(\kappa^{\dagger}\kappa)_{1_{22}}+\lambda^{\chi \kappa}_5(\chi^{\dagger}\chi)_{1_{21}}(\kappa^{\dagger}\kappa)_{1_{12}}+\lambda^{\chi \kappa}_6(\chi^{\dagger}\chi)_{1_{20}}(\kappa^{\dagger}\kappa)_{1_{10}}+\nonumber\\&\lambda^{\chi \kappa}_7(\chi^{\dagger}\chi)_{1_{02}}(\kappa^{\dagger}\kappa)_{1_{01}}+\lambda^{\chi \kappa}_8(\chi^{\dagger}\chi)_{1_{22}}(\kappa^{\dagger}\kappa)_{1_{11}}+\lambda^{\chi \kappa}_9(\chi^{\dagger}\chi)_{1_{12}}(\kappa^{\dagger}\kappa)_{1_{21}},\\
V(\chi,\phi)&= \lambda^{\chi \phi}_1(\chi^{\dagger}\chi)(\phi^{\dagger}\phi)+\lambda^{\chi \phi}_2(\chi^{\dagger}\chi)_{1_{10}}(\phi^{\dagger}\phi)_{1_{20}}+\lambda^{\chi \phi}_3(\chi^{\dagger}\chi)_{1_{01}}(\phi^{\dagger}\phi)_{1_{02}}+\lambda^{\chi \phi}_4\nonumber\\&(\chi^{\dagger}\chi)_{1_{11}}(\phi^{\dagger}\phi)_{1_{22}}+\lambda^{\chi \phi}_5(\chi^{\dagger}\chi)_{1_{21}}(\phi^{\dagger}\phi)_{1_{12}}+\lambda^{\chi \phi}_6(\chi^{\dagger}\chi)_{1_{20}}(\phi^{\dagger}\phi)_{1_{10}}+\nonumber\\&\lambda^{\chi \phi}_7(\chi^{\dagger}\chi)_{1_{02}}(\phi^{\dagger}\phi)_{1_{01}}+\lambda^{\chi \phi}_8(\chi^{\dagger}\chi)_{1_{22}}(\phi^{\dagger}\phi)_{1_{11}}+\lambda^{\chi \phi}_9(\chi^{\dagger}\chi)_{1_{12}}(\phi^{\dagger}\phi)_{1_{21}},\\
V(\phi,\kappa)&= \lambda^{\phi \kappa}_1(\phi^{\dagger}\phi)(\kappa^{\dagger}\kappa)+\lambda^{\phi \kappa}_2(\phi^{\dagger}\phi)_{1_{10}}(\kappa^{\dagger}\kappa)_{1_{20}}+\lambda^{\phi \kappa}_3(\phi^{\dagger}\phi)_{1_{01}}(\kappa^{\dagger}\kappa)_{1_{02}}+\lambda^{\phi \kappa}_4\nonumber\\&(\phi^{\dagger}\phi)_{1_{11}}(\kappa^{\dagger}\kappa)_{1_{22}}+\lambda^{\phi \kappa}_5(\phi^{\dagger}\phi)_{1_{21}}(\kappa^{\dagger}\kappa)_{1_{12}}+\lambda^{\phi \kappa}_6(\phi^{\dagger}\phi)_{1_{20}}(\kappa^{\dagger}\kappa)_{1_{10}}+\nonumber\\&\lambda^{\phi \kappa}_7(\phi^{\dagger}\phi)_{1_{02}}(\kappa^{\dagger}\kappa)_{1_{01}}+\lambda^{\phi \kappa}_8(\phi^{\dagger}\phi)_{1_{22}}(\kappa^{\dagger}\kappa)_{1_{11}}+\lambda^{\phi \kappa}_9(\phi^{\dagger}\phi)_{1_{12}}(\kappa^{\dagger}\kappa)_{1_{21}},
\end{align}

Without loss of generality, for the chosen vevs for the scalar fields, we obtain the following minimisation conditions of the scalar potential,

\begin{align}
&v_H (2 v_H^2 \lambda ^H-2 \mu _H^2+\lambda ^{\text{H$\eta $}} v_{\eta }^2+\lambda ^{\text{H$\kappa $}} v_{\kappa }^2+\lambda ^{\text{H$\xi $}} v_{\xi }^2+\lambda ^{\text{H$\rho $}} v_{\rho }^2+\lambda ^{\text{H$\sigma $}} v_{\sigma }^2+\lambda ^{\text{H$\chi $}} v_{\chi }^2+3 \lambda ^{\text{H$\psi $}} v_{\psi }^2+\nonumber\\ &\lambda ^{\text{H$\phi $}} v_{\phi }^2)= 0,
\end{align}

\begin{align}
&\frac{1}{2} v_{\psi } ( 
2 (v_H^2 \lambda^{H\psi} - 2 \mu_{\psi}^2 
+ 6 (\lambda_1^{\psi} + \lambda_3^{\psi}) v_{\psi}^2 
+ v_{\chi}^2 (\lambda_1^{\psi\chi} + \lambda_2^{\psi \chi} + \lambda_6^{\psi \chi}) 
+ \lambda^{\psi \xi} v_{\xi}^2 
+ \lambda^{\psi \rho} v_{\rho}^2 
+ \lambda^{\psi \sigma} \nonumber\\ &v_{\sigma}^2 
) 
 + v_{\eta }^2 (2 \lambda_1^{\psi \eta} - \lambda_2^{\psi \eta} - \lambda_6^{\psi \eta}) 
+ v_{\kappa }^2 (2 \lambda_1^{\psi \kappa} - \lambda_2^{\psi \kappa} - \lambda_6^{\psi \kappa}) 
+ v_{\phi }^2 (2 \lambda_1^{\psi \phi} - \lambda_2^{\psi \phi} - \lambda_6^{\psi \phi})
)=0,
\end{align}

\begin{align}
&\frac{1}{2} v_{\psi } (2 (v_H^2 \lambda^{ H\psi} - 2 \mu_\psi^2 + 6 (\lambda_1^\psi + \lambda_3^\psi) v_{\psi }^2 + \lambda^{ \psi \xi} v_\xi^2 + \lambda ^{\psi \rho} v_\rho^2 + \lambda_{ \psi \sigma} v_\sigma^2 ) 
+ 2 v_{\eta }^2 (\lambda_1^{\psi \eta} + \lambda_2^{\psi \eta} +\nonumber\\ & \lambda_6^{\psi \eta}) 
+ v_{\kappa }^2 (2 \lambda_1^{\psi \kappa} - \lambda_2^{\psi \kappa} - \lambda_6^{\psi \kappa}) 
+ v_{\chi }^2 (2 \lambda_1^{\psi \chi} - \lambda_2^{\psi \chi} - \lambda_6^{\psi \chi}) 
+ v_{\phi }^2 (2 \lambda_1^{\psi \phi} - \lambda_2^{\psi \phi} - \lambda_6^{\psi \phi}) )=0,
\end{align}

\begin{align}
&\frac{1}{2} v_{\psi } (2 (v_H^2 \lambda^{H\psi} - 2 \mu_\psi^2 
+ 6 (\lambda_1^\psi + \lambda_3^\psi) v_{\psi }^2 
+ \lambda^{\psi \xi} v_\xi^2 
+ \lambda^{\psi \rho} v_\rho^2 
+ \lambda^{\psi \sigma} v_\sigma^2) 
+ v_{\eta }^2 (2 \lambda_1^{\psi \eta} - \lambda_2^{\psi \eta} - \nonumber\\ & \lambda_6^{\psi \eta}) 
+ 2 v_{\kappa }^2 (\lambda_1^{\psi \kappa}  \lambda_2^{\psi \kappa} + \lambda_6^{\psi \kappa}) 
+ v_{\chi }^2 (2 \lambda_1^{\psi \chi} - \lambda_2^{\psi \chi} - \lambda_6^{\psi \chi}) 
+ 2 v_{\phi }^2 (\lambda_1^{\psi \phi} + \lambda_2^{\psi \phi} + \lambda_6^{\psi \phi})
)=0,
\end{align}

\begin{align}
&\frac{1}{2} v_{\eta } ( 
     2 (v_H^2 \lambda^{H\eta} - 2 \mu_\eta^2  + 3 \lambda_1^{\psi \eta} v_{\psi }^2 
    + \lambda^{\eta \xi} v_\xi^2 + \lambda^{ \eta \rho} v_\rho^2 + \lambda^{\eta \sigma} v_\sigma^2 ) 
     + 4 (\lambda_1^{\eta} + \lambda_2^\eta) v_{\eta }^2 
     + v_{\kappa }^2 (2 \lambda_1^{\eta \kappa} -\nonumber\\ & \lambda_2^{\eta \kappa} - \lambda_6^{\eta \kappa}) +
     v_{\chi }^2 (2 \lambda_1^{\eta \chi} - \lambda_2^{\eta \chi} - \lambda_6^{\eta \chi}) 
     + v_{\phi }^2 (2\lambda_1^{\eta \phi} - \lambda_2^{\eta \phi} - \lambda_6^{\eta \phi})
)=0,
\end{align}

\begin{align}
&\frac{1}{2} v_{\chi } ( 2 (v_H^2 \lambda^{H\chi} - 2 \mu_\chi^2+ 2 (\lambda_1^{\chi} + \lambda_2^{\chi}) v_{\chi }^2 + 3 \lambda_1^{\psi\chi} v_{\psi }^2 + \lambda^{\chi\xi} v_{\xi }^2 + \lambda^{\chi\rho} v_{\rho }^2 + \lambda^{\chi\sigma} v_{\sigma }^2 )  + v_{\eta }^2 (2 \lambda_1^{\eta\chi} \nonumber\\ &- \lambda_2^{\eta\chi} - \lambda_6^{\eta\chi}) + v_{\kappa }^2 (2 \lambda_1^{\chi\kappa} - \lambda_2^{\chi\kappa} - \lambda_6^{\chi\kappa})+ v_{\phi }^2 (2 \lambda_1^{\chi\phi} - \lambda_2^{\chi\phi} - \lambda_6^{\chi\phi}))=0,
\end{align}

\begin{align}
&\frac{1}{2} v_{\kappa } (2 (v_H^2 \lambda^{H\kappa} - 2 \mu_\kappa^2  + 3 \lambda_1^{\psi \kappa} v_{\psi }^2 + \lambda^{\kappa \xi} v_{\xi }^2 + \lambda^{\kappa \rho} v_{\rho }^2 + \lambda^{\kappa \sigma} v_{\sigma }^2 ) 
+ v_{\eta }^2 (2 \lambda_1^{\eta \kappa} - \lambda_2^{\eta \kappa} - \lambda_6^{\eta \kappa} ) 
+ 4 \nonumber\\ &(\lambda_1^{\kappa} + \lambda_2^\kappa) v_{\kappa }^2 
+ v_{\chi }^2 (2 \lambda_1^{\chi \kappa} - \lambda_2^{\chi \kappa} - \lambda_6^{\chi \kappa} ) 
+ 2 v_{\phi }^2 (\lambda_1^{\phi \kappa} + \lambda_2^{\phi \kappa} + \lambda_6^{\phi \kappa}))=0,
\end{align}

\begin{align}
&\frac{1}{2} v_{\phi} (2 ( v_H^2 \lambda^{H\phi} - \,2 \mu_\phi^2 
+ 3 \lambda_1^{\psi \phi} v_{\psi}^2 \,+ \lambda^{\phi \xi} v_\xi^2 + \lambda^{ \phi \rho }v_\rho^2 +\, \lambda^{ \phi \sigma} v_\sigma^2) + v_{\eta}^2 (2\, \lambda_1^{\eta \phi} - \lambda_2^{\eta \phi}- \lambda_6^{\eta \phi})
+\nonumber\\ & 2 v_{\kappa}^2\, (\lambda_1^{\phi \kappa} + \lambda_2^{\phi \kappa} + \lambda_6^{\phi \kappa})+ v_{\chi}^2 (2 \lambda_1^{\chi \phi} - \lambda_2^{\chi \phi}- \lambda_6^{\chi \phi}) + 4 (\lambda_1^{\phi} + \lambda_2^\phi) v_{\phi}^2)=0,
\end{align}

\begin{align}
&v_\xi( v_H^2 \lambda^{H \xi} - 2 \mu_\xi^2 + \lambda^{\eta \xi} v_\eta^2 + \lambda^{\kappa \xi} v_\kappa^2+\lambda^{\phi \xi} v_\phi^2 + 2 \lambda^\xi v_\xi^2 
 + \lambda^{\xi \sigma} v_\sigma^2 +\lambda^{\xi \rho} v_\rho^2+ \lambda^{\chi \xi} v_\chi^2 
+ \nonumber\\ &3 \lambda^{\psi \xi} v_\psi^2)=0,
\end{align}

\begin{align}
&v_\sigma (v_H^2 \lambda^{H \sigma} - 2 \mu_\sigma^2
+ \lambda^{\eta \sigma} v_\eta^2 + \lambda^{\kappa \sigma} v_\kappa^2
+ \lambda^{\phi \sigma} v_\phi^2 +\lambda^{\xi \sigma} v_\xi^2 +\lambda^{\sigma \rho} v_\rho^2+ 2 \lambda^{\sigma} v \sigma^2
+ \lambda^{\chi \sigma} v_\chi^2+\nonumber\\ & 3 \lambda^{\psi \sigma} v_\psi^2
)=0.
\end{align}

\begin{align}
&v_\rho ( v_H^2 \lambda^{ H \rho} - 2 \mu_\rho^2 + \lambda^{ \eta \rho} v_{\eta}^2 + \lambda^{ \kappa \rho} v_{\kappa}^2 + \lambda^{ \chi \rho} v_{\chi}^2 + 3 \lambda^{ \psi \rho} v_{\psi}^2 + \lambda^{ \phi \rho} v_{\phi}^2 + \lambda^{ \xi \rho} v_{\xi}^2 + 2 \lambda ^{\rho} v_{\rho}^2 +\nonumber\\ & \lambda^{ \sigma \rho} v_{\sigma}^2)=0.
\end{align}

 \bibliography{ref.bib}

@article{Cabibbo:1963yz,
    author = "Cabibbo, Nicola",
    title = "{Unitary Symmetry and Leptonic Decays}",
    doi = "10.1103/PhysRevLett.10.531",
    journal = "Phys. Rev. Lett.",
    volume = "10",
    pages = "531--533",
    year = "1963"
}

@article{Kobayashi:1973fv,
    author = "Kobayashi, Makoto and Maskawa, Toshihide",
    title = "{CP Violation in the Renormalizable Theory of Weak Interaction}",
    reportNumber = "KUNS-242",
    doi = "10.1143/PTP.49.652",
    journal = "Prog. Theor. Phys.",
    volume = "49",
    pages = "652--657",
    year = "1973"
}

@article{ParticleDataGroup:2024cfk,
    author = "Navas, S. and others",
    collaboration = "Particle Data Group",
    title = "{Review of particle physics}",
    doi = "10.1103/PhysRevD.110.030001",
    journal = "Phys. Rev. D",
    volume = "110",
    number = "3",
    pages = "030001",
    year = "2024"
}

@misc{JUNO:2025gmd,
    author = "Abusleme, Angel and others",
    collaboration = "JUNO",
    title = "{First measurement of reactor neutrino oscillations at JUNO}",
    eprint = "2511.14593",
    archivePrefix = "arXiv",
    primaryClass = "hep-ex",
    month = "11",
    year = "2025"
}

@article{Maki:1962mu,
    author = "Maki, Ziro and Nakagawa, Masami and Sakata, Shoichi",
    title = "{Remarks on the unified model of elementary particles}",
    doi = "10.1143/PTP.28.870",
    journal = "Prog. Theor. Phys.",
    volume = "28",
    pages = "870--880",
    year = "1962"
}

@misc{nufit,
  author       = "NuFIT Collaboration",
  title        = "{NuFIT 6.1 (2025)}",
  howpublished = "\url{http://www.nu-fit.org}",
  note         = "Available at http://www.nu-fit.org"
}

@article{Georgi:1974sy,
    author = "Georgi, H. and Glashow, S. L.",
    title = "{Unity of All Elementary Particle Forces}",
    doi = "10.1103/PhysRevLett.32.438",
    journal = "Phys. Rev. Lett.",
    volume = "32",
    pages = "438--441",
    year = "1974"
}

@article{Carena:1994bv,
    author = "Carena, Marcela and Olechowski, M. and Pokorski, S. and Wagner, C. E. M.",
    title = "{Electroweak symmetry breaking and bottom - top Yukawa unification}",
    eprint = "hep-ph/9402253",
    archivePrefix = "arXiv",
    reportNumber = "MPI-PH-93-103, CERN-TH-7163-94",
    doi = "10.1016/0550-3213(94)90313-1",
    journal = "Nucl. Phys. B",
    volume = "426",
    pages = "269--300",
    year = "1994"
}

@article{Ananthanarayan:1994qt,
    author = "Ananthanarayan, B. and Shafi, Q. and Wang, X. M.",
    title = "{Improved predictions for top quark, lightest supersymmetric particle, and Higgs scalar masses}",
    eprint = "hep-ph/9311225",
    archivePrefix = "arXiv",
    reportNumber = "BA-93-25-REV, PRL-TH-93-6-REV, BA-93-25, PRL-TH-93-6",
    doi = "10.1103/PhysRevD.50.5980",
    journal = "Phys. Rev. D",
    volume = "50",
    pages = "5980--5984",
    year = "1994"
}

@article{Leontaris:1987nq,
    author = "Leontaris, G. K. and Vergados, J. D.",
    title = "{The Neutrino Masses in SO(10) Grand Unified Theory}",
    reportNumber = "CERN-TH-4639/87",
    doi = "10.1016/0370-2693(87)91647-9",
    journal = "Phys. Lett. B",
    volume = "188",
    pages = "455--461",
    year = "1987"
}

@article{Roy:2020vtm,
    author = "Roy, Subhankar and Sashikanta Singh, K. and Borah, Jyotirmoi",
    title = "{Revamped Bi-Large neutrino mixing with Gatto-Sartori-Tonin like relation}",
    eprint = "2001.07401",
    archivePrefix = "arXiv",
    primaryClass = "hep-ph",
    doi = "10.1016/j.nuclphysb.2020.115204",
    journal = "Nucl. Phys. B",
    volume = "960",
    pages = "115204",
    year = "2020"
}

@article{Roy:2014nua,
    author = "Roy, S. and Morisi, S. and Singh, N. N. and Valle, J. W. F.",
    title = "{The Cabibbo angle as a universal seed for quark and lepton mixings}",
    eprint = "1410.3658",
    archivePrefix = "arXiv",
    primaryClass = "hep-ph",
    reportNumber = "IFIC-14-XX",
    doi = "10.1016/j.physletb.2015.06.052",
    journal = "Phys. Lett. B",
    volume = "748",
    pages = "1--4",
    year = "2015"
}

@article{Roy:2015cza,
    author = "Roy, S. and Singh, N. N.",
    title = "{Mixing angle as a function of neutrino mass ratio}",
    eprint = "1603.07474",
    archivePrefix = "arXiv",
    primaryClass = "hep-ph",
    doi = "10.1103/PhysRevD.91.096003",
    journal = "Phys. Rev. D",
    volume = "91",
    number = "9",
    pages = "096003",
    year = "2015"
}

@article{Ding:2012wh,
    author = "Ding, Gui-Jun and Morisi, S. and Valle, J. W. F.",
    title = "{Bilarge neutrino mixing and Abelian flavor symmetry}",
    eprint = "1211.6506",
    archivePrefix = "arXiv",
    primaryClass = "hep-ph",
    doi = "10.1103/PhysRevD.87.053013",
    journal = "Phys. Rev. D",
    volume = "87",
    number = "5",
    pages = "053013",
    year = "2013"
}

@article{Boucenna:2012xb,
    author = "Boucenna, S. M. and Morisi, S. and Tortola, M. and Valle, J. W. F.",
    title = "{Bi-large neutrino mixing and the Cabibbo angle}",
    eprint = "1206.2555",
    archivePrefix = "arXiv",
    primaryClass = "hep-ph",
    reportNumber = "IFIC-12-42",
    doi = "10.1103/PhysRevD.86.051301",
    journal = "Phys. Rev. D",
    volume = "86",
    pages = "051301",
    year = "2012"
}

@article{Branco:2014zza,
    author = "Branco, G. C. and Rebelo, M. N. and Silva-Marcos, J. I. and Wegman, Daniel",
    title = "{Quasidegeneracy of Majorana Neutrinos and the Origin of Large Leptonic Mixing}",
    eprint = "1405.5120",
    archivePrefix = "arXiv",
    primaryClass = "hep-ph",
    doi = "10.1103/PhysRevD.91.013001",
    journal = "Phys. Rev. D",
    volume = "91",
    number = "1",
    pages = "013001",
    year = "2015"
}

@article{Pati:1974yy,
    author = "Pati, Jogesh C. and Salam, Abdus",
    title = "{Lepton Number as the Fourth Color}",
    reportNumber = "IC-74-7",
    doi = "10.1103/PhysRevD.10.275",
    journal = "Phys. Rev. D",
    volume = "10",
    pages = "275--289",
    year = "1974",
    note = "[Erratum: Phys.Rev.D 11, 703--703 (1975)]"
}

@article{Elias:1975kf,
    author = "Elias, Victor and Swift, Arthur R.",
    title = "{Generalization of the Pati-Salam Model}",
    reportNumber = "Print-75-0824 (MASS.U.,AMHERST)",
    doi = "10.1103/PhysRevD.13.2083",
    journal = "Phys. Rev. D",
    volume = "13",
    pages = "2083",
    year = "1976"
}

@article{Elias:1977bv,
    author = "Elias, Victor",
    title = "{Gauge Coupling Constant Magnitudes in the Pati-Salam Model}",
    reportNumber = "MdDP-TR-77-049, MdDP-PP-77-169",
    doi = "10.1103/PhysRevD.16.1586",
    journal = "Phys. Rev. D",
    volume = "16",
    pages = "1586",
    year = "1977"
}

@article{Blazek:2003wz,
    author = "Blazek, T. and King, S. F. and Parry, J. K.",
    title = "{Global analysis of a supersymmetric Pati-Salam model}",
    eprint = "hep-ph/0303192",
    archivePrefix = "arXiv",
    doi = "10.1088/1126-6708/2003/05/016",
    journal = "JHEP",
    volume = "05",
    pages = "016",
    year = "2003"
}

@article{Dent:2007eu,
    author = "Dent, James B. and Kephart, Thomas W.",
    title = "{Minimal Pati-Salam model from string theory unification}",
    eprint = "0705.1995",
    archivePrefix = "arXiv",
    primaryClass = "hep-ph",
    doi = "10.1103/PhysRevD.77.115008",
    journal = "Phys. Rev. D",
    volume = "77",
    pages = "115008",
    year = "2008"
}

@article{Antusch:2009gu,
    author = "Antusch, Stefan and Spinrath, Martin",
    title = "{New GUT predictions for quark and lepton mass ratios confronted with phenomenology}",
    eprint = "0902.4644",
    archivePrefix = "arXiv",
    primaryClass = "hep-ph",
    reportNumber = "MPP-2009-26",
    doi = "10.1103/PhysRevD.79.095004",
    journal = "Phys. Rev. D",
    volume = "79",
    pages = "095004",
    year = "2009"
}

@article{Marzocca:2011dh,
    author = "Marzocca, David and Petcov, Serguey T. and Romanino, Andrea and Spinrath, Martin",
    title = "{Sizeable $\theta_{13}$ from the Charged Lepton Sector in SU(5), (Tri-)Bimaximal Neutrino Mixing and Dirac CP Violation}",
    eprint = "1108.0614",
    archivePrefix = "arXiv",
    primaryClass = "hep-ph",
    reportNumber = "SISSA-40-2011-EP",
    doi = "10.1007/JHEP11(2011)009",
    journal = "JHEP",
    volume = "11",
    pages = "009",
    year = "2011"
}

@article{Antusch:2012fb,
    author = "Antusch, Stefan and Gross, Christian and Maurer, Vinzenz and Sluka, Constantin",
    title = "{$\theta^{PMNS}_13$ = $\theta_C / \sqrt{2}$ from GUTs}",
    eprint = "1205.1051",
    archivePrefix = "arXiv",
    primaryClass = "hep-ph",
    doi = "10.1016/j.nuclphysb.2012.09.002",
    journal = "Nucl. Phys. B",
    volume = "866",
    pages = "255--269",
    year = "2013"
}

@article{Antusch:2013ti,
    author = "Antusch, Stefan",
    editor = "Kobayashi, Takashi and Nakahata, Masayuki and Nakaya, Tsuyoshi",
    title = "{Models for Neutrino Masses and Mixings}",
    eprint = "1301.5511",
    archivePrefix = "arXiv",
    primaryClass = "hep-ph",
    doi = "10.1016/j.nuclphysbps.2013.04.026",
    journal = "Nucl. Phys. B Proc. Suppl.",
    volume = "235-236",
    pages = "303--309",
    year = "2013"
}

@article{Antusch:2013rxa,
    author = "Antusch, Stefan and King, Stephen F. and Spinrath, Martin",
    title = "{GUT predictions for quark-lepton Yukawa coupling ratios with messenger masses from non-singlets}",
    eprint = "1311.0877",
    archivePrefix = "arXiv",
    primaryClass = "hep-ph",
    reportNumber = "MPP-2013-282, SISSA-49-2013-FISI, TTP13-035",
    doi = "10.1103/PhysRevD.89.055027",
    journal = "Phys. Rev. D",
    volume = "89",
    number = "5",
    pages = "055027",
    year = "2014"
}

@article{Antusch:2013jca,
    author = "Antusch, Stefan and Maurer, Vinzenz",
    title = "{Running quark and lepton parameters at various scales}",
    eprint = "1306.6879",
    archivePrefix = "arXiv",
    primaryClass = "hep-ph",
    doi = "10.1007/JHEP11(2013)115",
    journal = "JHEP",
    volume = "11",
    pages = "115",
    year = "2013"
}

@article{Charles:2004jd,
    author = "Charles, J. and Hocker, Andreas and Lacker, H. and Laplace, S. and Le Diberder, F. R. and Malcles, J. and Ocariz, J. and Pivk, M. and Roos, L.",
    collaboration = "CKMfitter Group",
    title = "{CP violation and the CKM matrix: Assessing the impact of the asymmetric $B$ factories}",
    eprint = "hep-ph/0406184",
    archivePrefix = "arXiv",
    reportNumber = "CPT-2004-P-030, LAL-04-21, LAPP-EXP-2004-01, LPNHE-2004-01",
    doi = "10.1140/epjc/s2005-02169-1",
    journal = "Eur. Phys. J. C",
    volume = "41",
    number = "1",
    pages = "1--131",
    year = "2005"
}

@article{Barranco:2010we,
    author = "Barranco, J. and Gonzalez Canales, F. and Mondragon, A.",
    title = "{Universal Mass Texture, CP violation and Quark-Lepton Complementarity}",
    eprint = "1004.3781",
    archivePrefix = "arXiv",
    primaryClass = "hep-ph",
    doi = "10.1103/PhysRevD.82.073010",
    journal = "Phys. Rev. D",
    volume = "82",
    pages = "073010",
    year = "2010"
}

@article{GonzalezCanales:2009zz,
    author = "Gonzalez Canales, F. and Mondragon, A.",
    editor = "Bernabeu, J. and Botella, F. J. and Mavromatos, N. E. and Mitsou, V. A.",
    title = "{Universal mass texture and quark-lepton complementarity}",
    doi = "10.1088/1742-6596/171/1/012063",
    journal = "J. Phys. Conf. Ser.",
    volume = "171",
    pages = "012063",
    year = "2009"
}

@article{deMedeirosVarzielas:2017sdv,
    author = "de Medeiros Varzielas, Ivo and Ross, Graham G. and Talbert, Jim",
    title = "{A Unified Model of Quarks and Leptons with a Universal Texture Zero}",
    eprint = "1710.01741",
    archivePrefix = "arXiv",
    primaryClass = "hep-ph",
    reportNumber = "OUTP-17-13P, DESY-17-146",
    doi = "10.1007/JHEP03(2018)007",
    journal = "JHEP",
    volume = "03",
    pages = "007",
    year = "2018"
}

@article{Koide:2003rx,
    author = "Koide, Yoshio",
    title = "{Universal texture of quark and lepton mass matrices with an extended flavor 2 \ensuremath{<}---\ensuremath{>} 3 symmetry}",
    eprint = "hep-ph/0312207",
    archivePrefix = "arXiv",
    reportNumber = "US-03-10",
    doi = "10.1103/PhysRevD.69.093001",
    journal = "Phys. Rev. D",
    volume = "69",
    pages = "093001",
    year = "2004"
}

@article{Vien:2023xyq,
    author = "Vien, V. V.",
    title = "{Realistic fermion mass and mixing in $\mathbf {U(1)_L}$ model with $\mathbf {A_4}$ flavor symmetry for majorana neutrino}",
    eprint = "2303.04687",
    archivePrefix = "arXiv",
    primaryClass = "hep-ph",
    doi = "10.1007/s12648-024-03375-1",
    journal = "Indian J. Phys.",
    volume = "99",
    number = "3",
    pages = "1165--1183",
    year = "2025"
}

@article{Vien:2023zid,
    author = "Vien, V. V. and Long, H. N. and C\'arcamo Hern\'andez, A. E. and Marchant Gonz\'alez, Juan",
    title = "{Fermion masses and mixings and g \ensuremath{-} 2 muon anomaly in a Q6 flavored 2HDM}",
    eprint = "2301.07811",
    archivePrefix = "arXiv",
    primaryClass = "hep-ph",
    doi = "10.1016/j.nuclphysb.2024.116722",
    journal = "Nucl. Phys. B",
    volume = "1008",
    pages = "116722",
    year = "2024"
}

@article{Vien:2022sxh,
    author = "Vien, V. V.",
    title = "{Fermion mass hierarchies and mixings in B \ensuremath{-} L model with $\Delta$(27) \texttimes{} Z4 symmetry}",
    doi = "10.1142/S0217732322500948",
    journal = "Mod. Phys. Lett. A",
    volume = "37",
    number = "15",
    pages = "2250094",
    year = "2022"
}

@article{Vien:2021eog,
    author = "Vien, V. V.",
    title = "{Multiscalar $B-L$ extension with $A_4$ symmetry for fermion mass and mixing with co-bimaximal scheme}",
    doi = "10.1016/j.physletb.2021.136296",
    journal = "Phys. Lett. B",
    volume = "817",
    pages = "136296",
    year = "2021"
}

@article{Hernandez:2021mxo,
    author = "Hern\'andez, A. E. C\'arcamo and Long, Hoang Ngoc and Mora-Urrutia, M. L. and Thao, N. H. and Vien, V. V.",
    title = "{Fermion masses and mixings and $g-2$ muon anomaly in a 3-3-1 model with $D_4$ family symmetry}",
    eprint = "2104.04559",
    archivePrefix = "arXiv",
    primaryClass = "hep-ph",
    doi = "10.1140/epjc/s10052-022-10639-9",
    journal = "Eur. Phys. J. C",
    volume = "82",
    number = "8",
    pages = "769",
    year = "2022"
}

@article{Vien:2021xfp,
    author = "Vien, V. V.",
    title = "{$B−L$ model based on $Q_4$ symmetry for fermion spectrum with normal neutrino mass ordering}",
    doi = "10.1142/S0217732321500474",
    journal = "Mod. Phys. Lett. A",
    volume = "36",
    number = "07",
    pages = "2150047",
    year = "2021"
}

@article{Vien:2021ciw,
    author = "Vien, V. V. and Hern\'andez, A. E. C\'arcamo and Long, H. N.",
    title = "{Fermion masses and mixings in a U(1)X model based on the \ensuremath{\Sigma}(18) discrete symmetry}",
    eprint = "2101.03506",
    archivePrefix = "arXiv",
    primaryClass = "hep-ph",
    doi = "10.1093/ptep/ptab078",
    journal = "PTEP",
    volume = "2021",
    number = "8",
    pages = "083B02",
    year = "2021"
}

@article{Vien:2020uzf,
    author = "Vien, V. V.",
    title = "{Fermion mass and mixing in the $U(1)_{B-L}$ extension of the standard model with $D_4$ symmetry}",
    doi = "10.1088/1361-6471/ab7ec0",
    journal = "J. Phys. G",
    volume = "47",
    number = "5",
    pages = "055007",
    year = "2020"
}

@article{Vien:2019eju,
    author = "Vien, V. V. and Soi, N. V.",
    title = "{Fermion mass and mixing in an extension of the standard model with $D_5$ symmetry}",
    doi = "10.1142/S0217732320500030",
    journal = "Mod. Phys. Lett. A",
    volume = "35",
    number = "04",
    pages = "2050003",
    year = "2019"
}

@article{Vien:2019zhs,
    author = "Vien, V. V. and Long, H. N. and C\'arcamo Hern\'andez, A. E.",
    title = "{Fermion Mass and Mixing in a Low-Scale Seesaw Model based on the $S_4$ Flavor Symmetry}",
    eprint = "1909.09532",
    archivePrefix = "arXiv",
    primaryClass = "hep-ph",
    doi = "10.1093/ptep/ptz119",
    journal = "PTEP",
    volume = "2019",
    number = "11",
    pages = "113B04",
    year = "2019"
}

@article{Vien:2019lso,
    author = "Vien, V. V. and Khoi, D. P.",
    title = "{Fermion masses and mixings in a 3-3-1 model with $Q_4$ symmetry}",
    doi = "10.1142/S0217732319501980",
    journal = "Mod. Phys. Lett. A",
    volume = "34",
    number = "25",
    pages = "1950198",
    year = "2019"
}

@article{Vien:2016qbb,
    author = "Vien, V. V. and Long, H. N.",
    title = "{Fermion Mass and Mixing in a Simple Extension of the Standard Model Based on T$_{7}$ Flavor Symmetry}",
    eprint = "1609.03895",
    archivePrefix = "arXiv",
    primaryClass = "hep-ph",
    doi = "10.1134/S1063778819020133",
    journal = "Phys. Atom. Nucl.",
    volume = "82",
    number = "2",
    pages = "168--182",
    year = "2019"
}

@article{Gupta:2015iku,
    author = "Gupta, Manmohan and Fakay, Priyanka and Sharma, Samandeep and Ahuja, Gulsheen",
    title = "{Fermion mass matrices, textures and beyond}",
    eprint = "1604.03335",
    archivePrefix = "arXiv",
    primaryClass = "hep-ph",
    doi = "10.1142/S0217732315300244",
    journal = "Mod. Phys. Lett. A",
    volume = "30",
    pages = "1530024",
    year = "2015"
}

@article{Gupta:2011zzg,
    author = "Gupta, Manmohan and Ahuja, Gulsheen",
    title = "{Possible textures of the fermion mass matrices}",
    eprint = "1206.3844",
    archivePrefix = "arXiv",
    primaryClass = "hep-ph",
    doi = "10.1142/S0217751X11053754",
    journal = "Int. J. Mod. Phys. A",
    volume = "26",
    pages = "2973--2995",
    year = "2011"
}

@article{Randhawa:1999hi,
    author = "Randhawa, Monika and Bhatnagar, V. and Gill, P. S. and Gupta, M.",
    title = "{Unique mass texture for quarks and leptons}",
    eprint = "hep-ph/9903428",
    archivePrefix = "arXiv",
    doi = "10.1103/PhysRevD.60.051301",
    journal = "Phys. Rev. D",
    volume = "60",
    pages = "051301",
    year = "1999"
}

@article{Gupta:2012fsl,
    author = "Gupta, Manmohan and Ahuja, Gulsheen",
    title = "{Flavor mixings and textures of the fermion mass matrices}",
    eprint = "1302.4823",
    archivePrefix = "arXiv",
    primaryClass = "hep-ph",
    doi = "10.1142/S0217751X12300335",
    journal = "Int. J. Mod. Phys. A",
    volume = "27",
    pages = "1230033",
    year = "2012"
}

@article{Berezhiani:2024fsw,
    author = "Berezhiani, Zurab and Belfatto, Benedetta",
    title = "{Towards understanding fermion masses and mixings}",
    eprint = "2405.07923",
    archivePrefix = "arXiv",
    primaryClass = "hep-ph",
    reportNumber = "TTP24-012, P3H-24-029",
    doi = "10.1142/S0217751X24410100",
    journal = "Int. J. Mod. Phys. A",
    volume = "39",
    number = "09n10",
    pages = "2441010",
    year = "2024"
}

@article{Garces:2018nar,
    author = "Garc\'es, E. A. and G\'omez-Izquierdo, Juan Carlos and Gonzalez-Canales, F.",
    title = "{Flavored non-minimal left\textendash{}right symmetric model fermion masses and mixings}",
    eprint = "1807.02727",
    archivePrefix = "arXiv",
    primaryClass = "hep-ph",
    doi = "10.1140/epjc/s10052-018-6271-5",
    journal = "Eur. Phys. J. C",
    volume = "78",
    number = "10",
    pages = "812",
    year = "2018"
}

@article{King:2006np,
    author = "King, Stephen F. and Malinsky, Michal",
    title = "{A(4) family symmetry and quark-lepton unification}",
    eprint = "hep-ph/0610250",
    archivePrefix = "arXiv",
    doi = "10.1016/j.physletb.2006.12.006",
    journal = "Phys. Lett. B",
    volume = "645",
    pages = "351--357",
    year = "2007"
}

@article{King:2013hoa,
    author = "King, Stephen F.",
    title = "{A model of quark and lepton mixing}",
    eprint = "1311.3295",
    archivePrefix = "arXiv",
    primaryClass = "hep-ph",
    doi = "10.1007/JHEP01(2014)119",
    journal = "JHEP",
    volume = "01",
    pages = "119",
    year = "2014"
}

@article{King:2013hj,
    author = "King, S. F. and Morisi, S. and Peinado, E. and Valle, J. W. F.",
    title = "{Quark-Lepton Mass Relation in a Realistic $A_4$ Extension of the Standard Model}",
    eprint = "1301.7065",
    archivePrefix = "arXiv",
    primaryClass = "hep-ph",
    reportNumber = "IFIC-13-XX, IFIC-13-04",
    doi = "10.1016/j.physletb.2013.05.067",
    journal = "Phys. Lett. B",
    volume = "724",
    pages = "68--72",
    year = "2013"
}

@article{Chen:2023mwt,
    author = "Chen, Mu-Chun and King, Stephen F. and Medina, Omar and Valle, Jos\'e W. F.",
    title = "{Quark-lepton mass relations from modular flavor symmetry}",
    eprint = "2312.09255",
    archivePrefix = "arXiv",
    primaryClass = "hep-ph",
    doi = "10.1007/JHEP02(2024)160",
    journal = "JHEP",
    volume = "02",
    pages = "160",
    year = "2024"
}

@article{Bonilla:2014xla,
    author = "Bonilla, Cesar and Morisi, Stefano and Peinado, Eduardo and Valle, Jose W. F.",
    title = "{Relating quarks and leptons with the $T_7$ flavour group}",
    eprint = "1411.4883",
    archivePrefix = "arXiv",
    primaryClass = "hep-ph",
    reportNumber = "IFIC-14-76",
    doi = "10.1016/j.physletb.2015.01.017",
    journal = "Phys. Lett. B",
    volume = "742",
    pages = "99--106",
    year = "2015"
}

@article{Morisi:2011pt,
    author = "Morisi, S. and Peinado, E. and Shimizu, Yusuke and Valle, J. W. F.",
    title = "{Relating quarks and leptons without grand-unification}",
    eprint = "1104.1633",
    archivePrefix = "arXiv",
    primaryClass = "hep-ph",
    reportNumber = "IFIC-11-16",
    doi = "10.1103/PhysRevD.84.036003",
    journal = "Phys. Rev. D",
    volume = "84",
    pages = "036003",
    year = "2011"
}

@article{Wolfenstein:1983yz,
    author = "Wolfenstein, Lincoln",
    title = "{Parametrization of the Kobayashi-Maskawa Matrix}",
    reportNumber = "CMU-HEG83-9",
    doi = "10.1103/PhysRevLett.51.1945",
    journal = "Phys. Rev. Lett.",
    volume = "51",
    pages = "1945",
    year = "1983"
}

@article{CentellesChulia:2017koy,
    author = "Centelles Chuli\'a, Salvador and Srivastava, Rahul and Valle, Jos\'e W. F.",
    title = "{Generalized Bottom-Tau unification, neutrino oscillations and dark matter: predictions from a lepton quarticity flavor approach}",
    eprint = "1706.00210",
    archivePrefix = "arXiv",
    primaryClass = "hep-ph",
    doi = "10.1016/j.physletb.2017.07.065",
    journal = "Phys. Lett. B",
    volume = "773",
    pages = "26--33",
    year = "2017"
}

@misc{Borboruah:2024lli,
    author = "Borboruah, Zafri Ahmed and Borah, Debasish and Malhotra, Lekhika and Patel, Utkarsh",
    title = "{Minimal Dirac seesaw dark matter}",
    eprint = "2412.12267",
    archivePrefix = "arXiv",
    primaryClass = "hep-ph",
    month = "12",
    year = "2024"
}

@article{Borah:2024gql,
    author = "Borah, Debasish and Das, Pritam and Karmakar, Biswajit and Mahapatra, Satyabrata",
    title = "{Discrete dark matter with light Dirac neutrinos}",
    eprint = "2406.17861",
    archivePrefix = "arXiv",
    primaryClass = "hep-ph",
    doi = "10.1103/PhysRevD.111.035032",
    journal = "Phys. Rev. D",
    volume = "111",
    number = "3",
    pages = "035032",
    year = "2025"
}

@article{Borah:2017dmk,
    author = "Borah, Debasish and Karmakar, Biswajit",
    title = "{$A_4$ flavour model for Dirac neutrinos: Type I and inverse seesaw}",
    eprint = "1712.06407",
    archivePrefix = "arXiv",
    primaryClass = "hep-ph",
    doi = "10.1016/j.physletb.2018.03.047",
    journal = "Phys. Lett. B",
    volume = "780",
    pages = "461--470",
    year = "2018"
}

@article{Singh:2024imk,
    author = "Singh, Labh and Kashav, Monal and Verma, Surender",
    title = "{Minimal type-I Dirac seesaw and leptogenesis under A4 modular invariance}",
    eprint = "2405.07165",
    archivePrefix = "arXiv",
    primaryClass = "hep-ph",
    doi = "10.1016/j.nuclphysb.2024.116666",
    journal = "Nucl. Phys. B",
    volume = "1007",
    pages = "116666",
    year = "2024"
}

@article{Mahapatra:2023oyh,
    author = "Mahapatra, Satyabrata and Sahoo, Sujit Kumar and Sahu, Narendra and Thounaojam, Vicky Singh",
    title = "{Self-interacting dark matter and Dirac neutrinos via lepton quarticity}",
    eprint = "2312.12322",
    archivePrefix = "arXiv",
    primaryClass = "hep-ph",
    doi = "10.1103/PhysRevD.109.055036",
    journal = "Phys. Rev. D",
    volume = "109",
    number = "5",
    pages = "055036",
    year = "2024"
}

@article{Chen:2022bjb,
    author = "Chen, Su-Ping and Gu, Pei-Hong",
    title = "{Undemocratic Dirac seesaw}",
    eprint = "2210.05307",
    archivePrefix = "arXiv",
    primaryClass = "hep-ph",
    doi = "10.1016/j.nuclphysb.2022.116028",
    journal = "Nucl. Phys. B",
    volume = "985",
    pages = "116028",
    year = "2022"
}

@misc{Goswami:2025jde,
    author = "Goswami, Sagar Tirtha and Roy, Subhankar",
    title = "{Permuted Charged Lepton Correction in the Framework of Dirac Seesaw}",
    eprint = "2501.18181",
    archivePrefix = "arXiv",
    primaryClass = "hep-ph",
    month = "1",
    year = "2025"
}

@article{Planck:2018vyg,
    author = "Aghanim, N. and others",
    collaboration = "Planck",
    title = "{Planck 2018 results. VI. Cosmological parameters}",
    eprint = "1807.06209",
    archivePrefix = "arXiv",
    primaryClass = "astro-ph.CO",
    doi = "10.1051/0004-6361/201833910",
    journal = "Astron. Astrophys.",
    volume = "641",
    pages = "A6",
    year = "2020",
    note = "[Erratum: Astron.Astrophys. 652, C4 (2021)]"
}

@article{Higgs:1964pj,
    author = "Higgs, Peter W.",
    editor = "Taylor, J. C.",
    title = "{Broken Symmetries and the Masses of Gauge Bosons}",
    doi = "10.1103/PhysRevLett.13.508",
    journal = "Phys. Rev. Lett.",
    volume = "13",
    pages = "508--509",
    year = "1964"
}

@article{Englert:1964et,
    author = "Englert, F. and Brout, R.",
    editor = "Taylor, J. C.",
    title = "{Broken Symmetry and the Mass of Gauge Vector Mesons}",
    doi = "10.1103/PhysRevLett.13.321",
    journal = "Phys. Rev. Lett.",
    volume = "13",
    pages = "321--323",
    year = "1964"
}

@article{Guralnik:1964eu,
    author = "Guralnik, G. S. and Hagen, C. R. and Kibble, T. W. B.",
    editor = "Taylor, J. C.",
    title = "{Global Conservation Laws and Massless Particles}",
    doi = "10.1103/PhysRevLett.13.585",
    journal = "Phys. Rev. Lett.",
    volume = "13",
    pages = "585--587",
    year = "1964"
}

@article{Glashow:1961tr,
    author = "Glashow, S. L.",
    title = "{Partial Symmetries of Weak Interactions}",
    doi = "10.1016/0029-5582(61)90469-2",
    journal = "Nucl. Phys.",
    volume = "22",
    pages = "579--588",
    year = "1961"
}

@article{Weinberg:1967tq,
    author = "Weinberg, Steven",
    title = "{A Model of Leptons}",
    doi = "10.1103/PhysRevLett.19.1264",
    journal = "Phys. Rev. Lett.",
    volume = "19",
    pages = "1264--1266",
    year = "1967"
}

@article{Salam:1968rm,
    author = "Salam, Abdus",
    title = "{Weak and Electromagnetic Interactions}",
    doi = "10.1142/9789812795915_0034",
    journal = "Conf. Proc. C",
    volume = "680519",
    pages = "367--377",
    year = "1968"
}

@article{Froggatt:1978nt,
    author = "Froggatt, C. D. and Nielsen, Holger Bech",
    title = "{Hierarchy of Quark Masses, Cabibbo Angles and CP Violation}",
    reportNumber = "CERN-TH-2519",
    doi = "10.1016/0550-3213(79)90316-X",
    journal = "Nucl. Phys. B",
    volume = "147",
    pages = "277--298",
    year = "1979"
}

@article{Chakraborty:2024hhq,
    author = "Chakraborty, Pralay and Roy, Subhankar",
    title = "{A unique neutrino mass matrix texture under exponential parametrization}",
    eprint = "2407.02550",
    archivePrefix = "arXiv",
    primaryClass = "hep-ph",
    doi = "10.1007/JHEP07(2025)246",
    journal = "JHEP",
    volume = "07",
    pages = "246",
    year = "2025"
}

@article{Branco:2011iw,
    author = "Branco, G. C. and Ferreira, P. M. and Lavoura, L. and Rebelo, M. N. and Sher, Marc and Silva, Joao P.",
    title = "{Theory and phenomenology of two-Higgs-doublet models}",
    eprint = "1106.0034",
    archivePrefix = "arXiv",
    primaryClass = "hep-ph",
    doi = "10.1016/j.physrep.2012.02.002",
    journal = "Phys. Rept.",
    volume = "516",
    pages = "1--102",
    year = "2012"
}

@article{Antusch:2006vwa,
    author = "Antusch, S. and Biggio, C. and Fernandez-Martinez, E. and Gavela, M. B. and Lopez-Pavon, J.",
    title = "{Unitarity of the Leptonic Mixing Matrix}",
    eprint = "hep-ph/0607020",
    archivePrefix = "arXiv",
    reportNumber = "FTUAM-06-8, IFT-UAM-CSIC-06-30",
    doi = "10.1088/1126-6708/2006/10/084",
    journal = "JHEP",
    volume = "10",
    pages = "084",
    year = "2006"
}

@article{MEG:2016leq,
    author = "Baldini, A. M. and others",
    collaboration = "MEG",
    title = "{Search for the lepton flavour violating decay $\mu^+ \rightarrow \mathrm {e}^+ \gamma $ with the full dataset of the MEG experiment}",
    eprint = "1605.05081",
    archivePrefix = "arXiv",
    primaryClass = "hep-ex",
    doi = "10.1140/epjc/s10052-016-4271-x",
    journal = "Eur. Phys. J. C",
    volume = "76",
    number = "8",
    pages = "434",
    year = "2016"
}

@article{BaBar:2009hkt,
    author = "Aubert, Bernard and others",
    collaboration = "BaBar",
    title = "{Searches for Lepton Flavor Violation in the Decays $\tau^\pm \to e^\pm \gamma$ and $\tau^\pm \to \mu^\pm \gamma$}",
    eprint = "0908.2381",
    archivePrefix = "arXiv",
    primaryClass = "hep-ex",
    reportNumber = "SLAC-PUB-13753, BABAR-PUB-09-026",
    doi = "10.1103/PhysRevLett.104.021802",
    journal = "Phys. Rev. Lett.",
    volume = "104",
    pages = "021802",
    year = "2010"
}

@article{MEGII:2023ltw,
    author = "Afanaciev, K. and others",
    collaboration = "MEG II",
    title = "{A search for $\mu^+ \rightarrow e^+ \gamma $ with the first dataset of the MEG~II experiment}",
    eprint = "2310.12614",
    archivePrefix = "arXiv",
    primaryClass = "hep-ex",
    doi = "10.1140/epjc/s10052-024-12416-2",
    journal = "Eur. Phys. J. C",
    volume = "84",
    number = "3",
    pages = "216",
    year = "2024",
    note = "[Erratum: Eur.Phys.J.C 84, 1042 (2024)]"
}

@article{Fernandez-Martinez:2007iaa,
    author = "Fernandez-Martinez, E. and Gavela, M. B. and Lopez-Pavon, J. and Yasuda, O.",
    title = "{CP-violation from non-unitary leptonic mixing}",
    eprint = "hep-ph/0703098",
    archivePrefix = "arXiv",
    reportNumber = "FTUAM-07-4, IFT-UAM-CSIC-07-10",
    doi = "10.1016/j.physletb.2007.03.069",
    journal = "Phys. Lett. B",
    volume = "649",
    pages = "427--435",
    year = "2007"
}

@article{Antusch:2005gp,
    author = {Antusch, Stefan and Kersten, J{\"o}rn and Lindner, Manfred and Ratz, Michael and Schmidt, Michael Andreas},
    title = "{Running neutrino mass parameters in see-saw scenarios}",
    eprint = "hep-ph/0501272",
    archivePrefix = "arXiv",
    reportNumber = "DESY-05-013, TUM-HEP-576-05, SHEP-0504",
    doi = "10.1088/1126-6708/2005/03/024",
    journal = "JHEP",
    volume = "03",
    pages = "024",
    year = "2005"
}

@article{Tanimoto:1995bf,
    author = "Tanimoto, Morimitsu",
    title = "{Renormalization effect on large neutrino flavor mixing in the minimal supersymmetric standard model}",
    eprint = "hep-ph/9508247",
    archivePrefix = "arXiv",
    reportNumber = "EHU-1995-08",
    doi = "10.1016/0370-2693(95)01107-2",
    journal = "Phys. Lett. B",
    volume = "360",
    pages = "41--46",
    year = "1995"
}

@article{Casas:1999tp,
    author = "Casas, J. A. and Espinosa, J. R. and Ibarra, A. and Navarro, I.",
    title = "{Naturalness of nearly degenerate neutrinos}",
    eprint = "hep-ph/9904395",
    archivePrefix = "arXiv",
    reportNumber = "IEM-FT-191-99, CERN-TH-99-103, IFT-UAM-CSIC-99-15",
    doi = "10.1016/S0550-3213(99)00383-1",
    journal = "Nucl. Phys. B",
    volume = "556",
    pages = "3--22",
    year = "1999"
}

@article{Casas:1999ac,
    author = "Casas, J. A. and Espinosa, J. R. and Ibarra, A. and Navarro, I.",
    title = "{Nearly degenerate neutrinos, supersymmetry and radiative corrections}",
    eprint = "hep-ph/9905381",
    archivePrefix = "arXiv",
    reportNumber = "IEM-FT-193-99, CERN-TH-99-142",
    doi = "10.1016/S0550-3213(99)00605-7",
    journal = "Nucl. Phys. B",
    volume = "569",
    pages = "82--106",
    year = "2000"
}

@article{King:2000hk,
    author = "King, S. F. and Singh, N. Nimai",
    title = "{Renormalization group analysis of single right-handed neutrino dominance}",
    eprint = "hep-ph/0006229",
    archivePrefix = "arXiv",
    doi = "10.1016/S0550-3213(00)00545-9",
    journal = "Nucl. Phys. B",
    volume = "591",
    pages = "3--25",
    year = "2000"
}

@article{He:2006dk,
    author = "He, Xiao-Gang and Keum, Yong-Yeon and Volkas, Raymond R.",
    title = "{A(4) flavor symmetry breaking scheme for understanding quark and neutrino mixing angles}",
    eprint = "hep-ph/0601001",
    archivePrefix = "arXiv",
    doi = "10.1088/1126-6708/2006/04/039",
    journal = "JHEP",
    volume = "04",
    pages = "039",
    year = "2006"
}

@article{Feruglio:2008ht,
    author = "Feruglio, Ferruccio and Hagedorn, Claudia and Lin, Yin and Merlo, Luca",
    title = "{Lepton Flavour Violation in Models with A(4) Flavour Symmetry}",
    eprint = "0807.3160",
    archivePrefix = "arXiv",
    primaryClass = "hep-ph",
    reportNumber = "DFPD-08-TH-09",
    doi = "10.1016/j.nuclphysb.2008.10.002",
    journal = "Nucl. Phys. B",
    volume = "809",
    pages = "218--243",
    year = "2009"
}

@article{deMedeirosVarzielas:2010ppv,
    author = "de Medeiros Varzielas, Ivo and Merlo, Luca",
    title = "{Ultraviolet Completion of Flavour Models}",
    eprint = "1011.6662",
    archivePrefix = "arXiv",
    primaryClass = "hep-ph",
    reportNumber = "DO-TH-10-21, TUM-HEP-776-10",
    doi = "10.1007/JHEP02(2011)062",
    journal = "JHEP",
    volume = "02",
    pages = "062",
    year = "2011"
}

@article{Altarelli:2012bn,
    author = "Altarelli, Guido and Feruglio, Ferruccio and Merlo, Luca and Stamou, Emmanuel",
    title = "{Discrete Flavour Groups, $theta_{13}$ and Lepton Flavour Violation}",
    eprint = "1205.4670",
    archivePrefix = "arXiv",
    primaryClass = "hep-ph",
    reportNumber = "CERN-PH-TH-2012-137, RM3-TH-12-7, DFPD-2012-TH-4, TUM-HEP-833-12",
    doi = "10.1007/JHEP08(2012)021",
    journal = "JHEP",
    volume = "08",
    pages = "021",
    year = "2012"
}

@article{Ahn:2012tv,
    author = "Ahn, Y. H. and Kang, Sin Kyu",
    title = "{Non-zero $\theta_{13}$ and CP violation in a model with $A_4$ flavor symmetry}",
    eprint = "1203.4185",
    archivePrefix = "arXiv",
    primaryClass = "hep-ph",
    reportNumber = "KIAS-P12022",
    doi = "10.1103/PhysRevD.86.093003",
    journal = "Phys. Rev. D",
    volume = "86",
    pages = "093003",
    year = "2012"
}

@article{Memenga:2013vc,
    author = "Memenga, Nina and Rodejohann, Werner and Zhang, He",
    title = "{$A_4$ flavor symmetry model for Dirac neutrinos and sizable $U_{e3}$}",
    eprint = "1301.2963",
    archivePrefix = "arXiv",
    primaryClass = "hep-ph",
    doi = "10.1103/PhysRevD.87.053021",
    journal = "Phys. Rev. D",
    volume = "87",
    number = "5",
    pages = "053021",
    year = "2013"
}

@article{GonzalezFelipe:2013yhh,
    author = "Gonzalez Felipe, R. and Serodio, H. and Silva, Joao P.",
    title = "{Neutrino masses and mixing in A4 models with three Higgs doublets}",
    eprint = "1304.3468",
    archivePrefix = "arXiv",
    primaryClass = "hep-ph",
    reportNumber = "CFTP-13-011",
    doi = "10.1103/PhysRevD.88.015015",
    journal = "Phys. Rev. D",
    volume = "88",
    number = "1",
    pages = "015015",
    year = "2013"
}

@article{deMedeirosVarzielas:2012cet,
    author = "de Medeiros Varzielas, Ivo and Pidt, Daniel",
    title = "{UV completions of flavour models and large $\theta_{13}$}",
    eprint = "1211.5370",
    archivePrefix = "arXiv",
    primaryClass = "hep-ph",
    reportNumber = "DO-TH-12-36",
    doi = "10.1007/JHEP03(2013)065",
    journal = "JHEP",
    volume = "03",
    pages = "065",
    year = "2013"
}

@article{Ishimori:2012fg,
    author = "Ishimori, Hajime and Ma, Ernest",
    title = "{New Simple $A_4$ Neutrino Model for Nonzero $\theta_{13}$ and Large $\delta_{CP}$}",
    eprint = "1205.0075",
    archivePrefix = "arXiv",
    primaryClass = "hep-ph",
    reportNumber = "UCRHEP-T520",
    doi = "10.1103/PhysRevD.86.045030",
    journal = "Phys. Rev. D",
    volume = "86",
    pages = "045030",
    year = "2012"
}

@article{CarcamoHernandez:2013yiy,
    author = {Carcamo Hernandez, Antonio Enrique and de Medeiros Varzielas, Ivo and Kovalenko, S. G. and P{\"a}s, H. and Schmidt, Ivan},
    title = "{Lepton masses and mixings in an $A_4$ multi-Higgs model with a radiative seesaw mechanism}",
    eprint = "1307.6499",
    archivePrefix = "arXiv",
    primaryClass = "hep-ph",
    doi = "10.1103/PhysRevD.88.076014",
    journal = "Phys. Rev. D",
    volume = "88",
    number = "7",
    pages = "076014",
    year = "2013"
}

@article{Babu:2002dz,
    author = "Babu, K. S. and Ma, Ernest and Valle, J. W. F.",
    title = "{Underlying A(4) symmetry for the neutrino mass matrix and the quark mixing matrix}",
    eprint = "hep-ph/0206292",
    archivePrefix = "arXiv",
    reportNumber = "UCRHEP-T341, OSU-HEP-02-07, IFIC-02-26",
    doi = "10.1016/S0370-2693(02)03153-2",
    journal = "Phys. Lett. B",
    volume = "552",
    pages = "207--213",
    year = "2003"
}

@article{Altarelli:2005yx,
    author = "Altarelli, Guido and Feruglio, Ferruccio",
    title = "{Tri-bimaximal neutrino mixing, A(4) and the modular symmetry}",
    eprint = "hep-ph/0512103",
    archivePrefix = "arXiv",
    reportNumber = "CERN-PH-TH-2005-226",
    doi = "10.1016/j.nuclphysb.2006.02.015",
    journal = "Nucl. Phys. B",
    volume = "741",
    pages = "215--235",
    year = "2006"
}

@article{Gupta:2011ct,
    author = "Gupta, Shivani and Joshipura, Anjan S. and Patel, Ketan M.",
    title = "{Minimal extension of tri-bimaximal mixing and generalized $Z_2 \to Z_2$ symmetries}",
    eprint = "1112.6113",
    archivePrefix = "arXiv",
    primaryClass = "hep-ph",
    doi = "10.1103/PhysRevD.85.031903",
    journal = "Phys. Rev. D",
    volume = "85",
    pages = "031903",
    year = "2012"
}

@article{Morisi:2013eca,
    author = "Morisi, S. and Nebot, M. and Patel, Ketan M. and Peinado, E. and Valle, J. W. F.",
    title = "{Quark-Lepton Mass Relation and CKM mixing in an A4 Extension of the Minimal Supersymmetric Standard Model}",
    eprint = "1303.4394",
    archivePrefix = "arXiv",
    primaryClass = "hep-ph",
    reportNumber = "IFIC-13-17, TIFR-TH-13-07",
    doi = "10.1103/PhysRevD.88.036001",
    journal = "Phys. Rev. D",
    volume = "88",
    pages = "036001",
    year = "2013"
}

@article{Altarelli:2005yp,
    author = "Altarelli, Guido and Feruglio, Ferruccio",
    title = "{Tri-bimaximal neutrino mixing from discrete symmetry in extra dimensions}",
    eprint = "hep-ph/0504165",
    archivePrefix = "arXiv",
    reportNumber = "DFPD-05-TH-14, CERN-PH-TH-2005-067",
    doi = "10.1016/j.nuclphysb.2005.05.005",
    journal = "Nucl. Phys. B",
    volume = "720",
    pages = "64--88",
    year = "2005"
}

@article{Kadosh:2010rm,
    author = "Kadosh, A. and Pallante, E.",
    title = "{An A(4) flavor model for quarks and leptons in warped geometry}",
    eprint = "1004.0321",
    archivePrefix = "arXiv",
    primaryClass = "hep-ph",
    doi = "10.1007/JHEP08(2010)115",
    journal = "JHEP",
    volume = "08",
    pages = "115",
    year = "2010"
}

@article{Kadosh:2013nra,
    author = "Kadosh, Avihay",
    title = "{$\Theta_13$ and charged Lepton Flavor Violation in ''warped'' $A_4$ models}",
    eprint = "1303.2645",
    archivePrefix = "arXiv",
    primaryClass = "hep-ph",
    doi = "10.1007/JHEP06(2013)114",
    journal = "JHEP",
    volume = "06",
    pages = "114",
    year = "2013"
}

@article{delAguila:2010vg,
    author = "del Aguila, Francisco and Carmona, Adrian and Santiago, Jose",
    title = "{Neutrino Masses from an A4 Symmetry in Holographic Composite Higgs Models}",
    eprint = "1001.5151",
    archivePrefix = "arXiv",
    primaryClass = "hep-ph",
    reportNumber = "CAFPE-128-10, UGFT-258-10",
    doi = "10.1007/JHEP08(2010)127",
    journal = "JHEP",
    volume = "08",
    pages = "127",
    year = "2010"
}

@article{Campos:2014lla,
    author = "Campos, Miguel D. and C{\'a}rcamo Hern{\'a}ndez, A. E. and Kovalenko, S. and Schmidt, Iv{\'a}n and Schumacher, Erik",
    title = "{Fermion masses and mixings in an $SU(5)$ grand unified model with an extra flavor symmetry}",
    eprint = "1403.2525",
    archivePrefix = "arXiv",
    primaryClass = "hep-ph",
    doi = "10.1103/PhysRevD.90.016006",
    journal = "Phys. Rev. D",
    volume = "90",
    number = "1",
    pages = "016006",
    year = "2014"
}

@article{Vien:2014pta,
    author = "Vien, Vo Van and Long, Hoang Ngoc",
    title = "{Neutrino mixing with nonzero $\theta_{13}$ and CP violation in the 3-3-1 model based on $A_4$ flavor symmetry}",
    eprint = "1405.4665",
    archivePrefix = "arXiv",
    primaryClass = "hep-ph",
    doi = "10.1142/S0217751X15501171",
    journal = "Int. J. Mod. Phys. A",
    volume = "30",
    number = "21",
    pages = "1550117",
    year = "2015"
}

@article{Karmakar:2016cvb,
    author = "Karmakar, Biswajit and Sil, Arunansu",
    title = "{An $A_4$ realization of inverse seesaw: neutrino masses, $\theta_{13}$ and leptonic non-unitarity}",
    eprint = "1610.01909",
    archivePrefix = "arXiv",
    primaryClass = "hep-ph",
    doi = "10.1103/PhysRevD.96.015007",
    journal = "Phys. Rev. D",
    volume = "96",
    number = "1",
    pages = "015007",
    year = "2017"
}

@article{Chattopadhyay:2017zvs,
    author = "Chattopadhyay, Pratik and Patel, Ketan M.",
    title = "{Discrete symmetries for electroweak natural type-I seesaw mechanism}",
    eprint = "1703.09541",
    archivePrefix = "arXiv",
    primaryClass = "hep-ph",
    doi = "10.1016/j.nuclphysb.2017.06.008",
    journal = "Nucl. Phys. B",
    volume = "921",
    pages = "487--506",
    year = "2017"
}

@article{Srivastava:2017sno,
    author = "Srivastava, Rahul and Ternes, C. A. and T{\'o}rtola, M. and Valle, J. W. F.",
    title = "{Testing a lepton quarticity flavor theory of neutrino oscillations with the DUNE experiment}",
    eprint = "1711.10318",
    archivePrefix = "arXiv",
    primaryClass = "hep-ph",
    reportNumber = "IFIC-17-XX",
    doi = "10.1016/j.physletb.2018.01.014",
    journal = "Phys. Lett. B",
    volume = "778",
    pages = "459--463",
    year = "2018"
}

@article{Srivastava:2018ser,
    author = "Srivastava, Rahul and Ternes, Christoph A. and T{\'o}rtola, Mariam and Valle, Jos{\'e} W. F.",
    title = "{Zooming in on neutrino oscillations with DUNE}",
    eprint = "1803.10247",
    archivePrefix = "arXiv",
    primaryClass = "hep-ph",
    reportNumber = "IFIC/18-xxx, IFIC-18-XXX",
    doi = "10.1103/PhysRevD.97.095025",
    journal = "Phys. Rev. D",
    volume = "97",
    number = "9",
    pages = "095025",
    year = "2018"
}

@article{Borah:2018nvu,
    author = "Borah, Debasish and Karmakar, Biswajit",
    title = "{Linear seesaw for Dirac neutrinos with $A_4$ flavour symmetry}",
    eprint = "1806.10685",
    archivePrefix = "arXiv",
    primaryClass = "hep-ph",
    doi = "10.1016/j.physletb.2018.12.006",
    journal = "Phys. Lett. B",
    volume = "789",
    pages = "59--70",
    year = "2019"
}

@article{Pramanick:2019qpg,
    author = "Pramanick, Soumita",
    title = "{Radiative generation of realistic neutrino mixing with $A4$}",
    eprint = "1903.04208",
    archivePrefix = "arXiv",
    primaryClass = "hep-ph",
    doi = "10.1016/j.nuclphysb.2020.115282",
    journal = "Nucl. Phys. B",
    volume = "963",
    pages = "115282",
    year = "2021"
}

@article{Vien:2022euq,
    author = "Vien, V. V.",
    title = "{Renormalizable standard model extension with S4 symmetry for neutrino mass and mixing}",
    doi = "10.1139/cjp-2022-0009",
    journal = "Can. J. Phys.",
    volume = "101",
    number = "3",
    pages = "106--119",
    year = "2023"
}

@article{Vien:2020aya,
    author = "Vien, V. V. and Long, H. N.",
    title = "{Multiscalar $B-L$ extension based on $S_4$ flavor symmetry for neutrino masses and mixing}",
    eprint = "2012.01715",
    archivePrefix = "arXiv",
    primaryClass = "hep-ph",
    doi = "10.1088/1674-1137/abe1c7",
    journal = "Chin. Phys. C",
    volume = "45",
    number = "4",
    pages = "043112",
    year = "2021"
}

@article{Vien:2016jkz,
    author = "Vien, V. V.",
    title = "{Lepton mass and mixing in a neutrino mass model based on $S_4$ flavor symmetry}",
    eprint = "1603.03933",
    archivePrefix = "arXiv",
    primaryClass = "hep-ph",
    doi = "10.1142/S0217751X16500391",
    journal = "Int. J. Mod. Phys. A",
    volume = "31",
    number = "09",
    pages = "1650039",
    year = "2016"
}

@article{Dong:2010zu,
    author = "Dong, P. V. and Long, H. N. and Soa, D. V. and Vien, V. V.",
    title = "{The 3-3-1 model with $S_4$ flavor symmetry}",
    eprint = "1009.2328",
    archivePrefix = "arXiv",
    primaryClass = "hep-ph",
    doi = "10.1140/epjc/s10052-011-1544-2",
    journal = "Eur. Phys. J. C",
    volume = "71",
    pages = "1544",
    year = "2011"
}

@article{Pakvasa:1978tx,
    author = "Pakvasa, Sandip and Sugawara, Hirotaka",
    title = "{Mass of the t Quark in SU(2) x U(1)}",
    reportNumber = "COO-881-66",
    doi = "10.1016/0370-2693(79)90436-2",
    journal = "Phys. Lett. B",
    volume = "82",
    pages = "105--107",
    year = "1979"
}

@article{Derman:1979nf,
    author = "Derman, Emanuel and Tsao, Hung-Sheng",
    title = "{SU(2) X U(1) X S($n$) Flavor Dynamics and a Bound on the Number of Flavors}",
    reportNumber = "COO-2232B-185",
    doi = "10.1103/PhysRevD.20.1207",
    journal = "Phys. Rev. D",
    volume = "20",
    pages = "1207",
    year = "1979"
}

@article{Yamanaka:1981pa,
    author = "Yamanaka, Y. and Sugawara, H. and Pakvasa, S.",
    title = "{Permutation Symmetries and the Fermion Mass Matrix}",
    reportNumber = "UH-511-453-81",
    doi = "10.1103/PhysRevD.25.1895",
    journal = "Phys. Rev. D",
    volume = "25",
    pages = "1895",
    year = "1982",
    note = "[Erratum: Phys.Rev.D 29, 2135 (1984)]"
}

@article{Brown:1984mq,
    author = "Brown, Tim and Deshpande, Nilendra and Pakvasa, Sandip and Sugawara, Hirotaka",
    title = "{{CP} Nonconservation and Rare Processes in S(4) Model of Permutation Symmetry}",
    reportNumber = "UH-511-522-84, OITS-240",
    doi = "10.1016/0370-2693(84)90568-9",
    journal = "Phys. Lett. B",
    volume = "141",
    pages = "95--99",
    year = "1984"
}

@article{Brown:1984dk,
    author = "Brown, T. and Pakvasa, S. and Sugawara, H. and Yamanaka, Y.",
    title = "{Neutrino Masses, Mixing and Oscillations in S(4) Model of Permutation Symmetry}",
    reportNumber = "UH-511-523-84",
    doi = "10.1103/PhysRevD.30.255",
    journal = "Phys. Rev. D",
    volume = "30",
    pages = "255",
    year = "1984"
}

@article{Lee:1994qx,
    author = "Lee, Dae-Gyu and Mohapatra, R. N.",
    title = "{An SO(10) x S(4) scenario for naturally degenerate neutrinos}",
    eprint = "hep-ph/9403201",
    archivePrefix = "arXiv",
    reportNumber = "UMD-PP-94-95",
    doi = "10.1016/0370-2693(94)91091-X",
    journal = "Phys. Lett. B",
    volume = "329",
    pages = "463--468",
    year = "1994"
}

@article{Mohapatra:2003tw,
    author = "Mohapatra, R. N. and Parida, M. K. and Rajasekaran, G.",
    title = "{High scale mixing unification and large neutrino mixing angles}",
    eprint = "hep-ph/0301234",
    archivePrefix = "arXiv",
    reportNumber = "UMD-PP-03-038, NEHU-PHY-MP-02-03, IMSC-2003-01-02",
    doi = "10.1103/PhysRevD.69.053007",
    journal = "Phys. Rev. D",
    volume = "69",
    pages = "053007",
    year = "2004"
}

@article{Ma:2005pd,
    author = "Ma, Ernest",
    title = "{Neutrino mass matrix from S(4) symmetry}",
    eprint = "hep-ph/0508231",
    archivePrefix = "arXiv",
    reportNumber = "UCRHEP-T397",
    doi = "10.1016/j.physletb.2005.10.019",
    journal = "Phys. Lett. B",
    volume = "632",
    pages = "352--356",
    year = "2006"
}

@article{Cai:2006mf,
    author = "Cai, Yi and Yu, Hai-Bo",
    title = "{A SO(10) GUT Model with S4 Flavor Symmetry}",
    eprint = "hep-ph/0608022",
    archivePrefix = "arXiv",
    doi = "10.1103/PhysRevD.74.115005",
    journal = "Phys. Rev. D",
    volume = "74",
    pages = "115005",
    year = "2006"
}

@article{Caravaglios:2006aq,
    author = "Caravaglios, Francesco and Morisi, Stefano",
    title = "{Gauge boson families in grand unified theories of fermion masses: $E^4_6 \rtimes S_{4}$}",
    eprint = "hep-ph/0611078",
    archivePrefix = "arXiv",
    reportNumber = "IFUM-880-FT",
    doi = "10.1142/S0217751X07036646",
    journal = "Int. J. Mod. Phys. A",
    volume = "22",
    pages = "2469--2492",
    year = "2007"
}

@article{Zhang:2006fv,
    author = "Zhang, He",
    title = "{Flavor S(4) x Z(2) symmetry and neutrino mixing}",
    eprint = "hep-ph/0612214",
    archivePrefix = "arXiv",
    doi = "10.1016/j.physletb.2007.09.003",
    journal = "Phys. Lett. B",
    volume = "655",
    pages = "132--140",
    year = "2007"
}

@article{CarcamoHernandez:2018iel,
    author = "C{\'a}rcamo Hern{\'a}ndez, A. E. and Long, H. N. and Vien, V. V.",
    title = "{The first $\Delta(27)$ flavor 3-3-1 model with low scale seesaw mechanism}",
    eprint = "1803.01636",
    archivePrefix = "arXiv",
    primaryClass = "hep-ph",
    doi = "10.1140/epjc/s10052-018-6284-0",
    journal = "Eur. Phys. J. C",
    volume = "78",
    number = "10",
    pages = "804",
    year = "2018"
}

@article{Vien:2020hzy,
    author = "Vien, V. V. and Khoi, D. P.",
    title = "{$U(1)_B−L$ extension based on $\Delta(27)$ symmetry for lepton masses and mixings}",
    doi = "10.1142/S0217732320501813",
    journal = "Mod. Phys. Lett. A",
    volume = "35",
    number = "22",
    pages = "2050181",
    year = "2020"
}

@article{Branco:1983tn,
    author = "Branco, G. C. and Gerard, J. M. and Grimus, W.",
    title = "{GEOMETRICAL T VIOLATION}",
    reportNumber = "CFMC E-9/83",
    doi = "10.1016/0370-2693(84)92024-0",
    journal = "Phys. Lett. B",
    volume = "136",
    pages = "383--386",
    year = "1984"
}

@article{deMedeirosVarzielas:2006fc,
    author = "de Medeiros Varzielas, I. and King, S. F. and Ross, G. G.",
    title = "{Neutrino tri-bi-maximal mixing from a non-Abelian discrete family symmetry}",
    eprint = "hep-ph/0607045",
    archivePrefix = "arXiv",
    reportNumber = "OUTP-0614P, CERN-PH-TH-2006-124",
    doi = "10.1016/j.physletb.2007.03.009",
    journal = "Phys. Lett. B",
    volume = "648",
    pages = "201--206",
    year = "2007"
}

@article{Ma:2007wu,
    author = "Ma, Ernest",
    title = "{Near tribimaximal neutrino mixing with Delta(27) symmetry}",
    eprint = "0709.0507",
    archivePrefix = "arXiv",
    primaryClass = "hep-ph",
    reportNumber = "UCRHEP-T439",
    doi = "10.1016/j.physletb.2007.12.060",
    journal = "Phys. Lett. B",
    volume = "660",
    pages = "505--507",
    year = "2008"
}

@article{deMedeirosVarzielas:2012ylr,
    author = "de Medeiros Varzielas, Ivo and Emmanuel-Costa, David and Leser, Philipp",
    title = "{Geometrical CP Violation from Non-Renormalisable Scalar Potentials}",
    eprint = "1204.3633",
    archivePrefix = "arXiv",
    primaryClass = "hep-ph",
    reportNumber = "CFTP-12-005, DO-TH-12-12",
    doi = "10.1016/j.physletb.2012.08.008",
    journal = "Phys. Lett. B",
    volume = "716",
    pages = "193--196",
    year = "2012"
}

@article{Bhattacharyya:2012pi,
    author = "Bhattacharyya, Gautam and de Medeiros Varzielas, Ivo and Leser, Philipp",
    title = "{A common origin of fermion mixing and geometrical CP violation, and its test through Higgs physics at the LHC}",
    eprint = "1210.0545",
    archivePrefix = "arXiv",
    primaryClass = "hep-ph",
    reportNumber = "SINP-TNP-2012-12, DO-TH-12-28",
    doi = "10.1103/PhysRevLett.109.241603",
    journal = "Phys. Rev. Lett.",
    volume = "109",
    pages = "241603",
    year = "2012"
}

@article{Ferreira:2012ri,
    author = "Ferreira, P. M. and Grimus, W. and Lavoura, L. and Ludl, P. O.",
    title = "{Maximal CP Violation in Lepton Mixing from a Model with Delta(27) flavour Symmetry}",
    eprint = "1206.7072",
    archivePrefix = "arXiv",
    primaryClass = "hep-ph",
    reportNumber = "UWTHPH-2012-24, CFTP-12-009",
    doi = "10.1007/JHEP09(2012)128",
    journal = "JHEP",
    volume = "09",
    pages = "128",
    year = "2012"
}

@article{Ma:2013xqa,
    author = "Ma, Ernest",
    title = "{Neutrino Mixing and Geometric CP Violation with Delta(27) Symmetry}",
    eprint = "1304.1603",
    archivePrefix = "arXiv",
    primaryClass = "hep-ph",
    reportNumber = "UCRHEP-T527-(APR-2013)",
    doi = "10.1016/j.physletb.2013.05.011",
    journal = "Phys. Lett. B",
    volume = "723",
    pages = "161--163",
    year = "2013"
}

@article{Nishi:2013jqa,
    author = "Nishi, C. C.",
    title = "{Generalized $CP$ symmetries in $\Delta(27)$ flavor models}",
    eprint = "1306.0877",
    archivePrefix = "arXiv",
    primaryClass = "hep-ph",
    doi = "10.1103/PhysRevD.88.033010",
    journal = "Phys. Rev. D",
    volume = "88",
    number = "3",
    pages = "033010",
    year = "2013"
}

@article{deMedeirosVarzielas:2013xas,
    author = "de Medeiros Varzielas, Ivo and Pidt, Daniel",
    title = "{Towards realistic models of quark masses with geometrical CP violation}",
    eprint = "1307.0711",
    archivePrefix = "arXiv",
    primaryClass = "hep-ph",
    reportNumber = "DO-TH-13-18",
    doi = "10.1088/0954-3899/41/2/025004",
    journal = "J. Phys. G",
    volume = "41",
    pages = "025004",
    year = "2014"
}

@article{Harrison:2014jqa,
    author = "Harrison, P. F. and Krishnan, R. and Scott, W. G.",
    title = "{Deviations from tribimaximal neutrino mixing using a model with $\Delta(27)$ symmetry}",
    eprint = "1406.2025",
    archivePrefix = "arXiv",
    primaryClass = "hep-ph",
    doi = "10.1142/S0217751X1450095X",
    journal = "Int. J. Mod. Phys. A",
    volume = "29",
    number = "18",
    pages = "1450095",
    year = "2014"
}

@article{CarcamoHernandez:2018djj,
    author = "C{\'a}rcamo Hern{\'a}ndez, A. E. and G{\'o}mez-Izquierdo, Juan Carlos and Kovalenko, Sergey and Mondrag{\'o}n, Myriam",
    title = "{$\Delta \left( 27\right)$ flavor singlet-triplet Higgs model for fermion masses and mixings}",
    eprint = "1810.01764",
    archivePrefix = "arXiv",
    primaryClass = "hep-ph",
    doi = "10.1016/j.nuclphysb.2019.114688",
    journal = "Nucl. Phys. B",
    volume = "946",
    pages = "114688",
    year = "2019"
}

@article{Bjorkeroth:2019csz,
    author = {Bj{\"o}rkeroth, Fredrik and de Medeiros Varzielas, Ivo and L{\'o}pez-Ib{\'a}{\~n}ez, M. L. and Melis, Aurora and Vives, {\'O}scar},
    title = "{Leptogenesis in $\Delta(27)$ with a Universal Texture Zero}",
    eprint = "1904.10545",
    archivePrefix = "arXiv",
    primaryClass = "hep-ph",
    reportNumber = "IFIC/19-21,FTUV-19-0416, IFIC-19-21",
    doi = "10.1007/JHEP09(2019)050",
    journal = "JHEP",
    volume = "09",
    pages = "050",
    year = "2019"
}

@article{Chakraborty:2023msb,
    author = "Chakraborty, Pralay and Roy, Subhankar",
    title = "{The other variants of mixed {\ensuremath{\mu}}-{\ensuremath{\tau}} symmetry}",
    eprint = "2304.06737",
    archivePrefix = "arXiv",
    primaryClass = "hep-ph",
    doi = "10.1016/j.nuclphysb.2023.116252",
    journal = "Nucl. Phys. B",
    volume = "992",
    pages = "116252",
    year = "2023"
}

@article{CarcamoHernandez:2016piw,
    author = "C{\'a}rcamo Hern{\'a}ndez, A. E. and Long, H. N. and Vien, V. V.",
    title = "{A 3-3-1 model with right-handed neutrinos based on the $\varDelta \left( 27\right) $ family symmetry}",
    eprint = "1601.05062",
    archivePrefix = "arXiv",
    primaryClass = "hep-ph",
    doi = "10.1140/epjc/s10052-016-4074-0",
    journal = "Eur. Phys. J. C",
    volume = "76",
    number = "5",
    pages = "242",
    year = "2016"
}

@article{Vien:2016tmh,
    author = "Vien, V. V. and C{\'a}rcamo Hern{\'a}ndez, A. E. and Long, H. N.",
    title = "{The $\Delta(27)$ flavor 3-3-1 model with neutral leptons}",
    eprint = "1601.03300",
    archivePrefix = "arXiv",
    primaryClass = "hep-ph",
    doi = "10.1016/j.nuclphysb.2016.10.010",
    journal = "Nucl. Phys. B",
    volume = "913",
    pages = "792--814",
    year = "2016"
}

@article{Vien:2025cak,
    author = "Vien, V. V. and Singh, Mayengbam Kishan",
    title = "{Dirac neutrino mass in a U(1)L model with T7{\texttimes}Z4 symmetry}",
    doi = "10.1016/j.cjph.2025.06.031",
    journal = "Chin. J. Phys.",
    volume = "97",
    pages = "239--246",
    year = "2025"
}

@article{Bazzocchi:2009qg,
    author = "Bazzocchi, Federica and de Medeiros Varzielas, Ivo",
    title = "{Tri-bi-maximal mixing in viable family symmetry unified model with extended seesaw}",
    eprint = "0902.3250",
    archivePrefix = "arXiv",
    primaryClass = "hep-ph",
    doi = "10.1103/PhysRevD.79.093001",
    journal = "Phys. Rev. D",
    volume = "79",
    pages = "093001",
    year = "2009"
}

@article{Varzielas:2012nn,
    author = "de Medeiros Varzielas, Ivo and Emmanuel-Costa, David and Leser, Philipp",
    title = "{Geometrical CP Violation from Non-Renormalisable Scalar Potentials}",
    eprint = "1204.3633",
    archivePrefix = "arXiv",
    primaryClass = "hep-ph",
    reportNumber = "CFTP-12-005, DO-TH-12-12",
    doi = "10.1016/j.physletb.2012.08.008",
    journal = "Phys. Lett. B",
    volume = "716",
    pages = "193--196",
    year = "2012"
}

@article{Aranda:2013gga,
    author = "Aranda, Alfredo and Bonilla, Cesar and Morisi, S. and Peinado, E. and Valle, J. W. F.",
    title = "{Dirac neutrinos from flavor symmetry}",
    eprint = "1307.3553",
    archivePrefix = "arXiv",
    primaryClass = "hep-ph",
    reportNumber = "DCP-13-03, IFIC-13-XX",
    doi = "10.1103/PhysRevD.89.033001",
    journal = "Phys. Rev. D",
    volume = "89",
    number = "3",
    pages = "033001",
    year = "2014"
}

@article{Abbas:2014ewa,
    author = "Abbas, Mohammed and Khalil, Shaaban",
    title = "{Fermion masses and mixing in $\Delta(27)$ flavour model}",
    eprint = "1406.6716",
    archivePrefix = "arXiv",
    primaryClass = "hep-ph",
    doi = "10.1103/PhysRevD.91.053003",
    journal = "Phys. Rev. D",
    volume = "91",
    number = "5",
    pages = "053003",
    year = "2015"
}

@article{Abbas:2015zna,
    author = "Abbas, Mohammed and Khalil, Shaaban and Rashed, Ahmed and Sil, Arunansu",
    title = "{Neutrino masses and deviation from tribimaximal mixing in \ensuremath{\Delta}(27) model with inverse seesaw mechanism}",
    eprint = "1508.03727",
    archivePrefix = "arXiv",
    primaryClass = "hep-ph",
    doi = "10.1103/PhysRevD.93.013018",
    journal = "Phys. Rev. D",
    volume = "93",
    number = "1",
    pages = "013018",
    year = "2016"
}

@article{Bjorkeroth:2015uou,
    author = {Bj\"orkeroth, Fredrik and de Anda, Francisco J. and de Medeiros Varzielas, Ivo and King, Stephen F.},
    title = "{Towards a complete $\Delta(27) \times SO(10)$ SUSY GUT}",
    eprint = "1512.00850",
    archivePrefix = "arXiv",
    primaryClass = "hep-ph",
    doi = "10.1103/PhysRevD.94.016006",
    journal = "Phys. Rev. D",
    volume = "94",
    number = "1",
    pages = "016006",
    year = "2016"
}

\end{document}